\documentclass[11pt]{article}
\usepackage[utf8]{inputenc}

 \usepackage[T1]{fontenc}

\usepackage[margin=1in]{geometry}
\usepackage{authblk}

\usepackage{amsmath}
\usepackage{amsthm}
\usepackage{amssymb}
\usepackage{thmtools}
\usepackage{thm-restate}
\usepackage{enumitem}
\usepackage{dsfont}
\usepackage{xspace}
\usepackage{comment}
\usepackage{xparse}
\usepackage{tikz}
\usepackage{fixme}
\usepackage[colorlinks]{hyperref}
\usepackage[capitalize]{cleveref}
\usepackage{subcaption}

\setlist[enumerate]{nosep, topsep=0ex}
\setlist[itemize]{nosep, topsep=0ex}

\usepackage{titlesec}
\titlespacing*{\section} {0pt}{2ex}{2ex}
\titlespacing*{\subsection} {0pt}{2ex}{2ex}
\titlespacing*{\subsubsection} {0pt}{2ex}{2ex}

\newtheorem{theorem}{Theorem}[]

\newtheorem{note}{Note}[]
\newtheorem{definition}{Definition}[]
\newtheorem{remark}{Remark}[]
\newtheorem{corollary}{Corollary}[]
\newtheorem{lemma}{Lemma}[]

\usepackage{algorithm}
\usepackage{algpseudocode}
\usepackage{algorithmicx}  
\usepackage{float}
\usepackage{microtype}
\usepackage{algorithm}
\usepackage{algpseudocode}
\usepackage{algorithmicx}   
\usepackage{floatflt}
\usepackage{graphics}
\usepackage{amsthm}

\title{
The connectivity carcass of a vertex subset in a graph:\\
both odd and even case \footnote{Preliminary version of this article appeared in the proceedings of the SIAM Symposium on Simplicity in Algorithms (SOSA) 2025.}
}

\author[1]{Surender Baswana}
\affil[1]{Indian Institute of Technology Kanpur, India}
\affil[1]{\texttt{sbaswana@cse.iitk.ac.in}}
\date{}

\author[2]{Abhyuday Pandey}
\affil[2]{Tower Research Capital LLC, New York, USA}
\affil[2]{\texttt{pandey.abhyuday07@gmail.com}}
\date{}

\date{}

\begin{document}

\begin{titlepage}

\maketitle




\pagenumbering{arabic}
\begin{abstract}
    Let $G=(V,E)$ be an undirected unweighted multi-graph and $S\subseteq V$ be a subset of vertices. A set of edges with the least cardinality whose removal disconnects $S$, that is, there is no path between at least one pair of vertices from $S$, is called a Steiner mincut for $S$ or simply an $S$-mincut. Connectivity Carcass is a compact data structure storing all $S$-mincuts in $G$ announced by Dinitz and Vainshtein in an extended abstract \cite{DBLP:conf/stoc/DinitzV94} in 1994. The complete proof of various results of this data structure for the simpler case when the capacity of $S$-mincut is odd appeared in the year 2000 \cite{DBLP:journals/siamcomp/DinitzV00}. Over the last couple of decades, there have been attempts towards the proof for the case when the capacity of $S$-mincut is even \cite{MS_thesis_Alex_Belous}, but none of them met a logical end. We present the following results.
    \begin{enumerate}
        \item We present the first complete, self-contained exposition of the connectivity carcass which covers both even and odd cases of the capacity of $S$-mincut.
        \item We derive the results using an alternate and much simpler approach. In particular, we derive the results using \textit{submodularity of cuts} -- a well-known property of graphs expressed using a simple inequality.
        \item We also show how the connectivity carcass can be helpful in efficiently answering some basic queries related to $S$-mincuts using some additional insights.
    \end{enumerate}
\end{abstract}

\end{titlepage}

\tableofcontents{}

\section{Introduction}
\label{sec: intro-and-overview}
Minimum cuts are fundamental structures in graph theory with numerous applications in the real world as well. Let $G=(V,E)$ be an undirected, unweighted, and connected multi-graph on $n=|V|$ vertices and $m=|E|$ edges. Henceforth, for the sake of brevity, a {\em mincut} will refer to a minimum cut.
Two well-known types of mincuts are global mincuts and pairwise mincuts. A smallest set of edges whose removal disconnects $G$, that is, breaks it into 2 connected components, is called a global mincut. For any pair of vertices $s,t\in V$, a $(s,t)$-mincut is a smallest set of edges whose removal breaks $G$ into 2 connected components -- one containing $s$ and another containing $t$.

%

Recently, there has been significant research towards designing efficient algorithms for computing $(s,t)$-mincuts and global mincuts; see \cite{DBLP:conf/focs/Brand0PKLGSS23, DBLP:conf/soda/HenzingerLRW24, DBLP:conf/focs/Abboud0PS23} and references therein. On the other hand, the data structural and graph theoretical aspects of mincuts have also been well-researched, though not recently. These results are few, but very fundamental and seminal. However, these are not as widely known as the algorithmic results. 
We therefore begin by motivating these structural perspectives.

Consider the following fundamental query on $(s,t)$-mincuts. Given any two disjoint subsets of vertices $A$ and $B$, does there exist a $(s,t)$-mincut that keeps $A$ on the side of $s$ and $B$ on the side of $t$? 
We can also frame similar query on global mincuts as follows. Given any two disjoint subsets of vertices $A$ and $B$, does there exist a global mincut that keeps $A$ and $B$ on the different sides of the cut? These queries can be answered trivially by invoking an algorithm for $(s,t)$-mincut or global mincut after a suitable modification of $G$. However, 
classical results show that
$G$ can be preprocessed to build another {\em simpler} graph ${\cal G}$ such that these queries and many other queries on mincuts in $G$ can be answered efficiently using elementary queries (e.g. reachability or connectivity) on ${\cal G}$. The reason ${\cal G}$ turns out to be an efficient data structure is that 
${\cal G}$ offers an easy way to {\em characterize} all mincuts using a simple graph property. We briefly describe these graphs for $(s,t)$-mincuts and global mincuts as follows.

For $(s,t)$-mincuts, Picard and Queyranne \cite{DBLP:journals/mp/PicardQ82} showed that there is a directed acyclic graph (DAG) of ${\cal O}(m)$ size that stores and characterizes all $(s,t)$-mincuts. This is remarkable because there exist graphs with $\Omega(2^n)$ $(s,t)$-mincuts. So while it is just impossible to store all $(s,t)$-mincuts explicitly, this DAG serves as a compact structure to store, characterize, and even enumerate all of them efficiently. 
Similarly, there exist graphs with $\Omega(n^2)$ global mincuts. However, Dinitz, Karzanov, and Lomonosov \cite{DL76}, in a seminal work, showed that all global mincuts can be stored and characterized by a {\em cactus} of ${\cal O}(n)$ size. A cactus is a connected graph in which no edge is shared by 2 or more cycles. 

There exists a family of mincuts that subsumes the family of $(s,t)$-mincuts and the family of global mincuts. This is the family of Steiner mincuts defined as follows. Suppose $S$ is any subset of 2 or more vertices. A smallest set of edges whose removal disconnects a given set $S$, that is, there is no path between at least one pair of vertices from $S$ in the resulting graph, is called a Steiner mincut for $S$ or simply an $S$-mincut. 
%
Recently, there has been a growing interest in the algorithmic aspects of $S$-mincuts \cite{DBLP:conf/soda/HeHS24, DBLP:conf/focs/LiP20}. 
Observe that the global mincuts are the $S$-mincuts for $S = V$; and $(s, t)$-mincuts are the $S$-mincuts for $S=\{s,t\}$. So it is a natural and important research objective to design a compact structure that stores and characterizes all $S$-mincuts for any given set $S$. However, this objective is quite challenging because the structures at the two extremes of this spectrum are fundamentally very different – a DAG when $|S| = 2$ and a cactus when $S = V$. 

 In a seminal work in 1994, Dinitz and Vainshtein \cite{DBLP:conf/stoc/DinitzV94} designed a compact structure, called connectivity carcass, that stores and characterizes all $S$-mincuts of a graph. This structure turns out to be an elegant hybrid of a cactus and a DAG. It consists of three parts, namely, a cactus ${\cal H}$, a {\em multi-terminal} DAG ${\cal F}$, and a mapping from the vertices of ${\cal F}$ to the paths in ${\cal H}$. While the characterization of the $S$-mincuts using these ingredients is quite intricate, 
 all of them together occupy only ${\cal O}(m)$ space.
 Dinitz and Vainshtein \cite{DBLP:conf/stoc/DinitzV94} also designed an incremental algorithm for the connectivity carcass, albeit for a very restricted case -- when the insertion of an edge does not increase the capacity of $S$-mincut.
 
 These results on connectivity carcass were announced by Dinitz and Vainshtein in an extended abstract that appeared in the Proceedings of ACM STOC, 716--725, 1994 \cite{DBLP:conf/stoc/DinitzV94}. 
 The article doesn't give the proofs of most of the lemmas and theorems. Moreover, many sophisticated tools and terminologies are developed specifically to arrive at these results.
 A subsequent article by Dinitz and Vainshtein, that appeared in SIAM J. Comput. 30(3): 753-808, 2000 \cite{DBLP:journals/siamcomp/DinitzV00}, provides details of the proofs only for the case when the number of edges of an $S$-mincut is odd. This is a simpler case since the cactus ${\cal H}$ turns out to be a tree. We came to know about some efforts towards the even-case proof of the connectivity carcass \cite{MS_thesis_Alex_Belous} through private communication with Dinitz and Vainshtein.  However, these efforts did not meet a logical end. So, there does not exist any peer-reviewed proof for the even case till date, as noted by other researchers as well (e.g. refer to the footnote on page 4 of article \cite{DBLP:conf/icalp/PettieY21}).



We now present an overview of the compact structures that store and characterize $(s,t)$-mincuts \cite{DBLP:journals/mp/PicardQ82} and global mincuts \cite{DL76}. Thereafter, we present an overview of the connectivity carcass \cite{DBLP:conf/stoc/DinitzV94, DBLP:conf/soda/DinitzV95, DBLP:journals/siamcomp/DinitzV00}. In order to fully grasp this overview, the reader is advised to develop familiarity with some basic graph structures in the following section. These structures also pave the way to a significantly simplified and complete analysis of the connectivity carcass in this article. 



\subsection{Basic graphs}\label{sec:basic-graphs}

We provide a gentle introduction to the following graphs.~$(i)$ single-source single-sink balanced DAG, ~$(ii)$ $t$-cactus and ~$(iii)$ quotient graph induced by a family of cuts.

\begin{figure}[ht]
 \centering  
 \includegraphics[width=160pt]{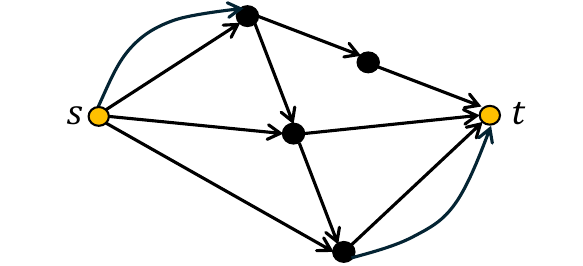} 
  \caption{Each vertex, other than $s$ and $t$, has the same number of incoming and outgoing edges.}
  \label{fig:special-DAG-overview}
\end{figure}

The DAG invented by Picard and Queyranne 
\cite{DBLP:journals/mp/PicardQ82} stores and
characterizes all $(s,t)$-mincuts of a directed graph. However, if the graph is undirected, this DAG turns out to be a {\em special} DAG defined as follows. Refer to Figure \ref{fig:special-DAG-overview} for its illustration. 

\begin{definition}[Single-source single-sink balanced DAG]
A single-source single-sink balanced DAG is a DAG with the following additional properties:
\begin{enumerate}
    \item[$(a)$] 
    There is exactly one vertex called {\em source} with indegree 0, and exactly one vertex called {\em sink} with outdegree 0.
    \item[$(b)$] 
    For each vertex, other than the source and the sink, the number of incoming edges is the same as the number of outgoing edges.
\end{enumerate}
\label{def:balanced-DAG-property}
\end{definition}
%
%

\begin{figure}[ht]
 \centering  
 \includegraphics[width=480pt]{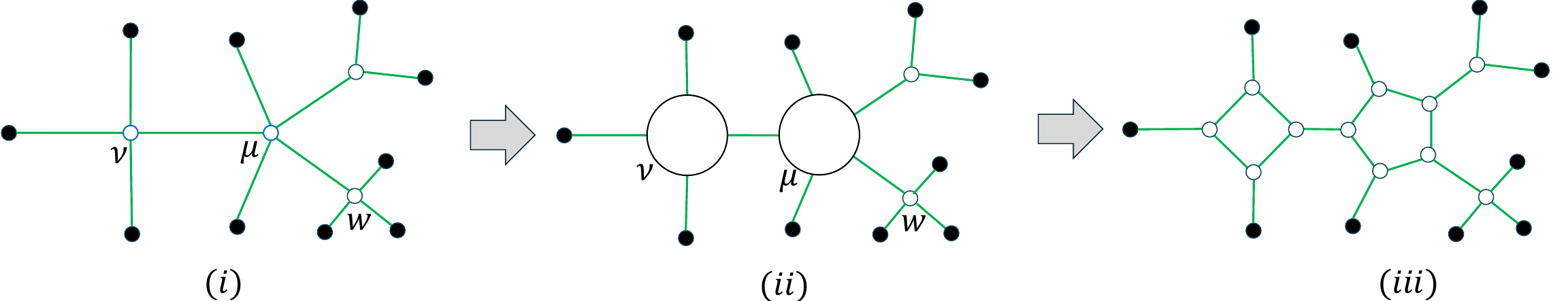} 
  \caption{($i$) $\nu,\mu$, and $w$ have degrees $\ge 4$, ($ii$) inflating $\nu$ and $\mu$, $(iii)$ implanting cycles at $\nu$ and $\mu$.}
  \label{fig:special-Cactus-overview}
\end{figure}

The second graph type can be best explained through its construction as follows. 
Consider any tree, and focus on its nodes of degree at least 4. We pick some of these nodes, inflate them, and {\em implant} a cycle of length equal to their degrees. Figure \ref{fig:special-Cactus-overview} illustrates one such transformation. Obviously, no edge in the resulting graph is shared by multiple cycles; hence, it is a cactus. In fact, it satisfies an even stronger property as follows. No vertex in this cactus is shared by multiple cycles. Henceforth, we shall call it a $t$-cactus (where $t$ stands for tree). 
An edge belonging to a cycle in a $t$-cactus is called a cycle edge, otherwise it is called a tree edge.  
 We shall study the {\em minimal} cuts of a $t$-cactus in this article. A cut is a minimal cut if no proper subset of it is a cut. It is easy to observe that a minimal cut of a $t$-cactus is either a tree edge or a pair of edges belonging to the same cycle.


%

The third graph type is a graph defined using a family of cuts. 
\begin{definition}[Quotient graph induced by a family of cuts]
Let ${\mathcal X}$ be any family of cuts. 
We build a quotient graph of $G$ using ${\cal X}$ as follows. Each pair of vertices $u,v\in V$ are mapped to the same vertex in the quotient graph if there is no mincut from ${\cal X}$ that separates $u$ and $v$.   
\label{def:quotient-graph-induced-by-cuts}    
\end{definition}

It is easy to observe that the quotient graph induced by a family ${\cal X}$ of cuts preserves all cuts of ${\mathcal X}$. So, it would be equally good to work with this quotient graph instead of the original graph. It turns out that this quotient graph often helps in the design of a compact structure that stores and characterizes some cut families.  

\subsection{Overview of the compact structures for \texorpdfstring{$(s,t)$}{(s,t)}-mincuts and global mincuts} 

Let us consider the family of $(s,t)$-mincuts.
Let ${\cal D}_{s,t}$ denote the quotient graph of $G$ induced by all $(s,t)$-mincuts. There exists a unique way of assigning direction to the edges of ${\cal D}_{s,t}$ such that the resulting graph, denoted by $\overrightarrow{\cal D}_{s,t}$, is a single-source single-sink balanced DAG \cite{DBLP:conf/stoc/DinitzV94,DBLP:journals/mp/PicardQ82}. 
Let $\omega$ be any vertex other than source or sink in ${\cal D}_{s,t}$, and let $E(\omega)$ be the set of edges incident on $\omega$. 
Observe that $\omega$ has exactly the same number of incoming edges as the outgoing edges in $\overrightarrow{\cal D}_{s,t}$.
In this way, $\overrightarrow{\cal D}_{s,t}$ defines a unique partition of the set $E(\omega)$ into two subsets of equal size. This partition is called the {\em inherent} partition of $E(\omega)$.
It is because of this inherent partition that the DAG $\overrightarrow{\cal D}_{s,t}$  
is able to characterize all $(s,t)$-mincuts of $G$ as follows.  
Consider any $(s,t)$-cut, and let $A$ be the set of vertices lying on the side of $s$ of this cut. This cut is a $(s,t)$-mincut if and only if all edges of this cut are emanating from $A$ and entering into $V\backslash A$ in $\overrightarrow{\cal D}_{s,t}$. Henceforth, we refer to this DAG as a {\em strip}. The vertices containing $s$ and $t$ in the strip are called terminals, and every other vertex is called a non-terminal.

Let us consider the family of global mincuts of $G$.  
The quotient graph induced by all global mincuts does not seem to have any remarkable structure to characterize all global mincuts. However, it follows from the seminal work of Dinitz, Karzanov, and Lomonosov \cite{DL76} that there exists a $t$-cactus ${\cal H}$ that stores and characterizes all global mincuts of $G$ as follows. 
Each vertex of the quotient graph is mapped to a unique node in ${\cal H}$. 
Observe that each cut of ${\cal H}$ defines a partition of $V$ as well. All global mincuts are characterized using ${\cal H}$ as follows \cite{DL76}. A cut of $G$ is a global mincut if and only if the partition of $V$ defined by it can be defined by a minimal cut in ${\cal H}$.

In the following section, we present an overview of the connectivity carcass designed by Dinitz and Vainshtein \cite{DBLP:conf/stoc/DinitzV94}. To appreciate its depth and elegance,  the reader is encouraged to reflect on the ways of building it using the structures for $(s,t)$-mincuts and global mincuts sketched above.
\begin{remark}
Dinitz and Vainshtein, in a subsequent article \cite{DBLP:conf/soda/DinitzV95} on incremental maintenance of connectivity carcass, presented slightly modified and more refined concepts compared to \cite{DBLP:conf/stoc/DinitzV94}. In the overview that follows and the rest of this article, we have used these refined concepts only.
\end{remark}

\subsection{Overview of the connectivity carcass designed by Dinitz and Vainshtein }

Let $x$ be any arbitrary vertex from the set $S$. Observe that any $S$-mincut is also a $(x,y)$-mincut for some suitably selected vertex $y\in S\setminus \{x\}$, and hence is present in the strip constructed with source $x$ and sink $y$. Therefore, a set of $O(|S|)$ strips, each with source vertex $x$, is sufficient to store and characterize all $S$-mincuts. However, not only this structure requires $O(m|S|)$ space which is huge, but also provides no insight into the relation among these $S$-mincuts.

Let us consider the quotient graph of $G$ induced by all $S$-mincuts. Let us denote it by ${\cal F}$. It follows from the construction that any two vertices from $S$ are mapped to the same vertex in ${\cal F}$ if there is no $S$-mincut that separates them. The following is a natural classification of the vertices of ${\cal F}$. A vertex in ${\cal F}$ is called a $S$-unit if it contains at least one vertex from $S$, otherwise it is called a non-$S$-unit. Unfortunately, unlike a strip storing all $(s,t)$-mincuts, apparently ${\cal F}$ does not seem to possess any special structure for characterizing all $S$-mincuts.

\begin{figure}[ht]
 \centering  
 \includegraphics[width=500pt]{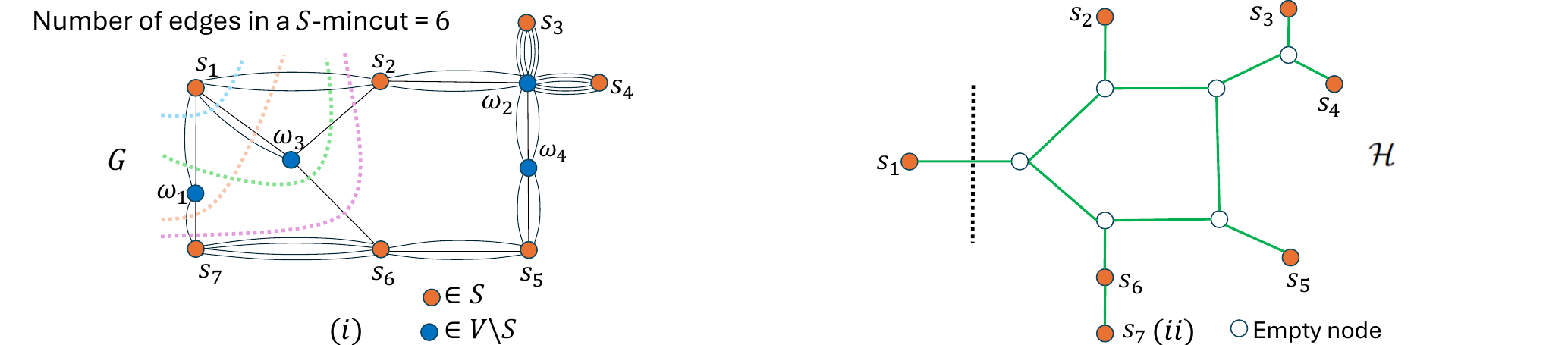} 
  \caption{There may be many $S$-mincuts associated with one minimal cuts in ${\cal H}$. 
  }
  \label{fig:flesh-and-cactus-overview}
\end{figure}

Observe that each $S$-mincut, by definition, partitions set $S$ into two non-empty subsets.  
%
%
Dinitz and Vainshtein 
note (and credit D. Naor and Westbrook \cite{Westbrook93}) that there exists a $t$-cactus ${\cal H}$ storing the vertices of $S$ as follows. 
Each node of ${\cal H}$ stores a (possibly empty) subset of vertices from $S$. 
This $t$-cactus has ${\mathcal O}(|S|)$ size, and it satisfies the following property. 
For each $S$-mincut, the corresponding partition of $S$ is defined by at least one minimal cut in ${\cal H}$, and vice versa. Figure \ref{fig:flesh-and-cactus-overview}($ii$) represents the $t$-cactus ${\cal H}$ for the graph shown in Figure \ref{fig:flesh-and-cactus-overview}($i$). For the sake of simplicity, each pair of vertices belonging to $S$ is separated by at least one $S$-mincut in $G$. 
Observe that there may be multiple $S$-mincuts that partition $S$ in the same way, and ${\cal H}$ does not provide any way to distinguish them. 
As an example, there is a minimal cut in ${\cal H}$, shown using a black dotted line in Figure \ref{fig:flesh-and-cactus-overview}($ii$), that separates $s_1$ from the rest of the vertices of set $S$.
However, there are 4 $S$-mincuts in $G$, shown using dotted curves of different colors, that keep $s_1$ on one side and the remaining vertices of set $S$ on the other side. 
So ${\cal H}$, by itself, is not adequate for storing and characterizing all $S$-mincuts. It should not appear surprising since ${\cal H}$ does not provide {\em any} information about the vertices belonging to set $V\backslash S$.

In the pursuit of a compact structure that stores and characterizes all $S$-mincuts, it seems quite natural to explore the relationship between the units of ${\cal F}$ and the nodes of ${\cal H}$. Observe that two vertices of set $S$ belong to the same node of ${\cal H}$ if and only if they belong to the same unit in ${\cal F}$. 
So we can define an injective mapping from the set of all $S$-units of ${\cal F}$ to the set of nodes of ${\cal H}$ as follows -- each $S$-unit $\omega$ is mapped to the node in ${\cal H}$ that stores all vertices of the set $S$ present in $\omega$. 
But, apparently ${\cal H}$ does not reveal any information about the non-$S$-units. 
However, an obvious way to use ${\cal H}$ for storing all $S$-mincuts is the following. Consider any minimal cut in ${\cal H}$. Suppose this minimal cut defines a partition of $S$ into two subsets $S_1$ and $S_2$. Observe that a strip with source $S_1$ and sink $S_2$ stores all $S$-mincuts associated with the partition $(S_1,S_2)$. Therefore, the set of all such strips together stores and characterizes all $S$-mincuts. 
The first insight made by Dinitz and Vainshtein 
is that each strip is closely related to graph ${\cal F}$ and a pair of strips are not necessarily totally different from each other. 
We introduce a terminology here to explain this insight.
Let $\omega$ be any non-$S$-unit in ${\cal F}$. A minimal cut of ${\cal H}$ is said to {\em distinguish} $\omega$ if $\omega$ belongs to a non-terminal in the strip corresponding to the minimal cut. 
Dinitz and Vainshtein  
showed that $\omega$ satisfies the following two properties.

\begin{enumerate} 
\item [${\cal Q}_1$:] 
If $\omega$ is distinguished by a minimal cut in ${\cal H}$, $\omega$ appears as a distinct non-terminal in the corresponding strip, along with all edges incident on it in ${\cal F}$. In other words, if there is any other non-$S$-unit $\omega'$ distinguished by the same minimal cut, it cannot happen that both $\omega$ and $\omega'$ belong to the same non-terminal in the strip. 
\item [${\cal Q}_2$:]
Consider the strips associated with all the minimal cuts that distinguish $\omega$. The inherent partition of $E(\omega)$ is the same in each of them. So a single partition of $E(\omega)$ will serve as an inherent partition of $E(\omega)$ in all these strips.
\end{enumerate}

It follows from these properties that a strip associated with any minimal cut in ${\cal H}$ is a quotient graph of ${\cal F}$ obtained as follows: keep all the non-$S$-units distinguished by the minimal cut intact and merge the remaining ones with the source or sink suitably. 
This insight suggests the following alternate and implicit way of storing the strips associated with all the minimal cuts of ${\cal H}$ -- For each non-$S$-unit $\omega$, store the set of all minimal cuts of ${\cal H}$ that distinguish $\omega$. 
It turns out that there is an amazing structure underlying all these minimal cuts as stated in the following theorem. 
\begin{theorem}[Theorem 4.4 in \cite{DBLP:conf/soda/DinitzV95}]
Let $\omega$ be any non-$S$-unit in ${\cal F}$. There exists a path $P(\omega)$ in the $t$-cactus ${\cal H}$ such that $\omega$ is distinguished by those and exactly those minimal cuts in ${\cal H}$ that share an edge with $P(\omega)$. If there is no minimal cut that distinguishes $\omega$, $P(\omega)$ is a node in ${\cal H}$. 
\label{thm: main-theorem-DV}
\end{theorem}
%
%
%
%
%
%
%

\begin{figure}[ht]
 \centering  
 \includegraphics[width=500pt]{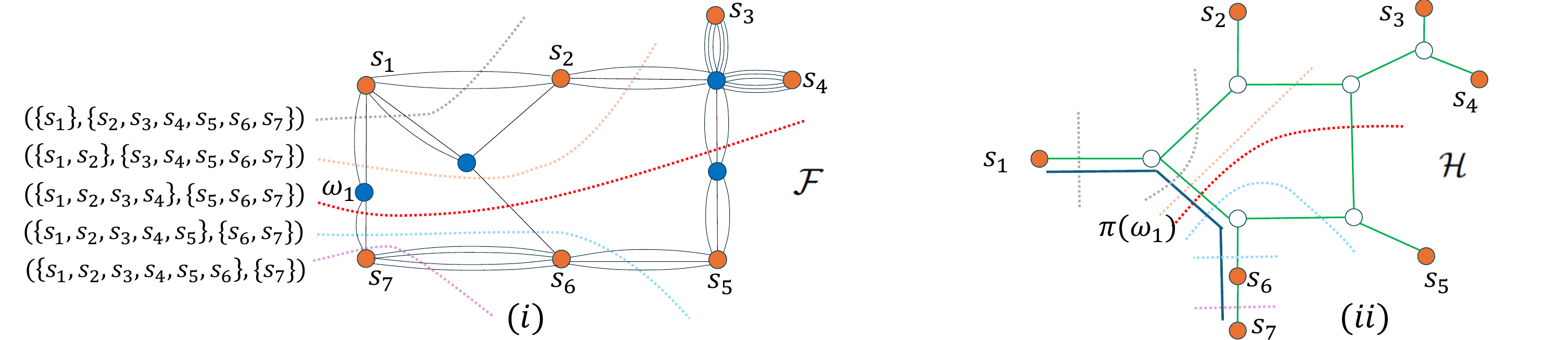} 
  \caption{($i$) the partitions of $S$ that distinguish $\omega_1$ shown in distinct colors, ($ii$) the corresponding cuts in ${\cal H}$.  
  }
  \label{fig:projection-mapping-of-omega_1-in-cactus-overview}
\end{figure}

 We illustrate Theorem \ref{thm: main-theorem-DV} by an example shown in Figure \ref{fig:projection-mapping-of-omega_1-in-cactus-overview}.  
 For the sake of simplicity in the illustration, we have chosen $G$ and $S$ so that ${\cal F}$, as shown in Figure \ref{fig:projection-mapping-of-omega_1-in-cactus-overview}($i$), is identical to $G$. $\omega_1$ is a non-$S$-unit in ${\cal F}$. All those partitions of $S$ that distinguish  $\omega_1$ are shown using dotted curves of different colors. Figure \ref{fig:projection-mapping-of-omega_1-in-cactus-overview}($ii$) shows the minimal cuts in ${\cal H}$ corresponding to these partitions. Observe that each of these cuts is defined by an edge belonging to the thick blue path joining $s_1$ and $s_7$ in ${\cal H}$. So, this path captures all minimal cuts in ${\cal H}$ that distinguish $\omega_1$. 

\begin{figure}[ht]
 \centering  
 \includegraphics[width=500pt]{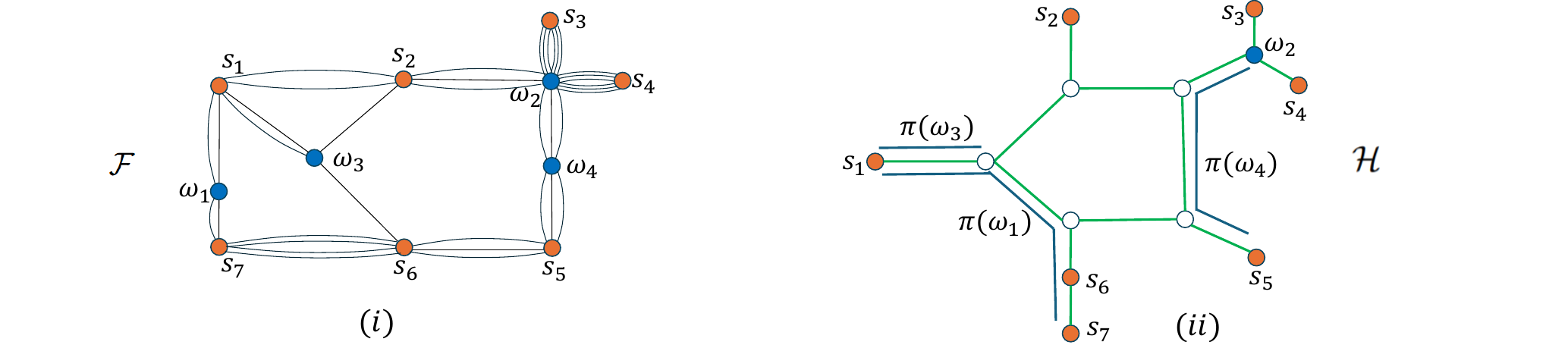} 
  \caption{($i$) ${\cal F}$, ($ii$) Projection mapping of all non-$S$-units in skeleton ${\cal H}$. 
  }
  \label{fig:projection-mapping-of-all-units-overview}
\end{figure}

It follows from Theorem \ref{thm: main-theorem-DV} that 
each non-$S$-unit of ${\cal F}$ can be {\em mapped} to a path in ${\cal H}$. Interestingly, this path turns out to share exactly one edge with each cycle that it passes through. Such a path is called a {\em proper} path, and it is uniquely defined by its endpoints in ${\cal H}$.  
Hence the path to which a non-$S$-unit of ${\cal F}$ is mapped can be stored compactly in ${\cal O}(1)$ space. 
This mapping $\pi$, called projection mapping, of all the non-$S$-units of ${\cal F}$ can thus be stored in just ${\cal O}(n)$ space. The projection mapping $\pi$ acts as the important bridge between ${\cal F}$ and ${\cal H}$. Refer to Figure \ref{fig:projection-mapping-of-all-units-overview} for ${\cal H}$ along with the projection mapping of all the non-$S$-units of ${\cal F}$. While $\omega_1, \omega_3$ and $\omega_4$ are mapped to paths of length 1 or more, $\omega_2$ is mapped to an empty node in ${\cal H}$ since it is not distinguished by any minimal cut of ${\cal H}$.

The connectivity carcass for a given set $S$ has three components -- ${\cal F}$ as its flesh, ${\cal H}$ as its skeleton, and $\pi$ the projection mapping. 
The space occupied by ${\cal H}$ and $\pi$ 
is ${\cal O}(n)$ only, and explicitly storing ${\cal F}$ requires ${\cal O}(m)$ space.

The proofs of Properties ${\cal Q}_1$ and ${\cal Q}_2$ given in \cite{DBLP:journals/siamcomp/DinitzV00} hold irrespective of whether the number of edges present in a $S$-mincut is odd or even. As described above, it is Theorem \ref{thm: main-theorem-DV} that leads to the ${\cal O}(n)$ size projection mapping, and hence plays the most crucial role in achieving the ${\cal O}(m)$ space for the connectivity carcass. 
However, the proof of Theorem \ref{thm: main-theorem-DV}, as given in \cite{DBLP:journals/siamcomp/DinitzV00}, is specific for the case when ${\cal H}$ is a tree and thus holds only for the case when the number of edges present in a $S$-mincut is odd. When the number of edges in a $S$-mincut is even, ${\cal H}$ may have cycles leading to the following challenges.  Firstly, the number of partitions of $S$ defined by all $S$-mincuts may grow from $O(|S|)$ to $O(|S|^2)$.
Secondly, it turns out that there may be multiple minimal cuts in ${\cal H}$ that represent the same partition of $S$. 
Hence, the mapping from the set of partitions of $S$ defined by all $S$-mincuts to the set of minimal cuts in ${\cal H}$ may no longer be injective when the number of edges in a $S$-mincut is even. We now state the contributions of this article.

\subsection{Our Contribution}
1.~    {\bf The first complete exposition of the connectivity carcass:}~\\
    We present the first complete and self-contained exposition of the foundational results of the connectivity carcass announced by Dinitz and Vainshtein in the extended abstract \cite{DBLP:conf/stoc/DinitzV94}.\\

\noindent
2. {\bf A Simpler and Alternate approach:}~\\
To derive their results, Dinitz and Vainshtein \cite{DBLP:conf/stoc/DinitzV94,DBLP:journals/siamcomp/DinitzV00} introduce and use the concept of a {\em locally orientable} graph which can be viewed as a generalization of a directed graph. This concept is quite sophisticated. Moreover, one of the main tools they use is the ${\textsc{3-Star Lemma}}$ which has two assertions $(i)$ and $(ii)$. This lemma has a nontrivial and involved proof. 
In this article, we pursue an alternate approach which is much simpler. Our approach uses  elementary graph properties or the {\em submodularity} of cuts \cite{Lovasz-book} -- a well-known property of cuts in graphs expressed only using a short and simple inequality. Even the proof of the ${\textsc{3-Star Lemma}}(i)$ turns out to be its immediate application.
We do not require the ${\textsc{3-Star Lemma}}(ii)$. 
The alternate approach allows us to generalize the results uniformly for both even and odd case, treating them without distinction. Additionally, avoiding complex concepts makes it much more accessible to a broader audience.\\

\noindent
3.~{\bf An Additional bridge between ${\cal F}$ and ${\cal H}$:}~\\
Dinitz and Vainshtein \cite{DBLP:conf/stoc/DinitzV94} established a mapping from the units of ${\cal F}$ to proper paths in ${\cal H}$.
We extend this mapping to the edges of ${\cal F}$ in a seamless manner as follows. Let $e$ be any edge in ${\cal F}$. We show that there exists a proper path in ${\cal H}$ such that $e$ appears in the strip of those and exactly those minimal cuts of ${\cal H}$ that share an edge with this path. This leads to establishing a relationship between the projection of an edge and the projection of its endpoints in ${\cal H}$. These insights serve as an additional bridge between ${\cal F}$ and ${\cal H}$.

We only require ${\cal O}(n)$ additional space to store the connectivity carcass if we already have access to the original graph $G$. 
Essentially, we need not store ${\cal F}$ explicitly as a graph. Instead, we just need to store the mapping from the set of vertices of $G$ to the set of units of ${\cal F}$.
%
The following theorem summarizes our results.

\begin{theorem}
Let $G=(V,E)$ be an undirected multigraph stored in the form of adjacency lists, and let $S\subseteq V$ be any set of 2 or more vertices. There exists an ${\cal O}(n)$ size data structure comprising of (implicit) flesh ${\cal F}$, skeleton ${\cal H}$, and projection mapping ${\cal \pi}$ satisfying the following properties.
\begin{enumerate}
    \item 
    A cut in $G$ is a $S$-mincut if and only if it appears as an $(s,t)$-mincut in the strip corresponding to a minimal cut in ${\cal H}$. 
    \item For any $e\in E$, we can report a $S$-mincut separating its endpoints, if exists, in ${\cal O}(m+n)$ time.
    \item For any pair of vertices $u,v\in S$ separated by at least one $S$-mincut, we can compute strip $\overrightarrow{\cal D}_{u,v}$ that stores all $S$-mincuts separating $u$ and $v$ in ${\cal O}(m+n)$ time.
\end{enumerate}
It takes ${\cal O}(|S|)$ max-flow computations to build the data structure.
\label{thm: data-structure-result}
\end{theorem}

\noindent 

\subsection{Organization of the manuscript}
Section \ref{sec:preliminaries} presents notations and basic lemmas on cuts including the submodularity of cuts. Section \ref{sec:balaned_DAG_and_Cactus} states properties of two well-known graph structures, namely, single-source single-sink balanced DAG and $t$-cactus. These structures underlie most of the compact structures for various mincuts. 
Section \ref{sec:existing-structures-for-various-mincuts} presents the classical structure for storing and characterizing all $(s,t)$-mincuts followed by a $t$-cactus that stores all partitions of $S$ induced by all $S$-mincuts. 
Section \ref{sec:CompactStructure_for_S-mincuts} and Section \ref{sec:projection-of-a-stretched-unit} together establish the mapping of a non-$S$-unit of ${\cal F}$
to a path in ${\cal H}$, and establish the proof of Theorem \ref{thm: main-theorem-DV} for the general case.
Section \ref{sec:carcass_as_an_efficient_datastructure} highlights the data structural aspects of the connectivity carcass. Section \ref{sec:projection_of_an_edge} presents additional insights into the relation between ${\cal F}$ and ${\cal H}$, including the projection mapping of an edge of ${\cal F}$ to a proper path in ${\cal H}$.

The length of this manuscript may initially seem extensive. However, more than half of the content focuses on essential background topics: (1) properties of a $t$-cactus and a single-source single-sink balanced DAG, (2) existing structure for $(s, t)$-mincuts, and (3) $t$-cactus structure representing all partitions of $S$ induced by all $S$-mincuts. The connectivity carcass, a complex structure, builds on these foundational concepts. To ensure the manuscript is self-contained and serves as a standard reference, it is necessary to include these topics, which are otherwise dispersed across multiple sources.

\section{Preliminaries} \label{sec:preliminaries}
We begin with a generic definition of a cut. This will assist in expressing our results more compactly and formally.

\begin{definition}[Cut]
Let $Z$ be any set, and $A$ be any subset of $Z$ such that $\emptyset\not= A\not=Z$. The ordered partition $(A,\overline{A})$ is called the cut of $Z$ {\em defined} by $A$. The sets $A$ and $\overline{A}$ are called the two {\em sides} of this cut. Any element $x\in Z$ is said to lie {\em inside} the cut $(A,\overline{A})$ if $x\in A$, and {\em outside} otherwise.
\label{def: cut}
\end{definition}

We now define the intersection and union of two cuts.
\begin{definition}[Intersection and union of cuts] Let $(A,\overline{A})$ and $(B,\overline{B})$ be any two cuts of a set $Z$. If $A\cap B\not= \emptyset$, $(A\cap B,\overline{A\cap B})$ is also a cut of $Z$, called the {\em intersection} of the two cuts. If $A\cup B\not= Z$, $(A\cup B,\overline{A\cup B})$ is also a cut of $Z$, called the {\em union} of the two cuts. 
\label{def: intersection-and-union-of-cuts}
\end{definition}

The following definition captures the dominance of one cut over another.
{ \begin{definition}[Dominance] Let $(A,\overline{A})$ and $(A',\overline{A'})$ be any two cuts of a set $Z$. $(A',\overline{A'})$ is said to dominate $(A,\overline{A})$ if $A\subsetneq A'$.
\label{def: dominance}
\end{definition}
}

We assume, without loss of generality, that the given multi-graph $G=(V,E)$ is connected.
For any given $A,B\subseteq V$ such that $A\cap B=\emptyset$, we use $c(A,B)$ to denote the number of edges with one endpoint in $A$ and another in $B$. 

When $Z=V$, the {\em edge-set} of a cut $(A,\overline{A})$ is defined as the set of all those edges from $E$ having one endpoint in $A$ and another endpoint in $\overline{A}$. $c(A)$ refers to 
$c(A,\overline{A})$, and is called the {\em capacity} of the cut $(A,\overline{A})$. In this article, a cut will refer to a cut of $V$ unless explicitly mentioned otherwise. 

For each cut $(A,\overline{A})$, the {\em opposite} cut is $(\overline{A},A)$. Notice that both of them have identical edge-sets since $G$ is undirected. Using the fact that $G$ is connected, it can be easily shown that any two cuts have identical edge-sets if and only if they are opposite to each other.  
In other words, the edge-set of a cut  reveals the set of vertices that defines the cut (or the opposite cut). 

\begin{definition}[Minimum cut]
Let ${\cal X}$ be a family of cuts, and let $(A,\overline{A})$ 
be any cut belonging to ${\mathcal X}$. 
$(A,\overline{A})$ is said to be a {\em minimum cut} in ${\mathcal X}$ if its capacity is the least among all cuts in ${\mathcal X}$.
\label{def: minimum-cut}
\end{definition}

\begin{definition}[Minimal cut]
Let ${\cal X}$ be a family of cuts, and let
$(A,\overline{A})$ be any cut belonging to ${\mathcal X}$. $(A,\overline{A})$ is said to be a {\em minimal cut} in ${\mathcal X}$ if there is no cut in ${\cal X}$ whose edge-set is a proper subset of the edge-set of $(A,\overline{A})$.
\label{def: minimal-cut}
\end{definition}

\noindent
Observe that a minimum cut is always a minimal cut; but a minimal cut needs not be a minimum cut.

\begin{definition}[($s,t$)-cut] 
A cut $(A,\overline{A})$ 
is said to be an $(s,t)$-cut for any 
given $s,t\in V$ if $s\in A$ and $t\in \overline{A}$. An
($s,t$)-cut is said to be an 
$(s,t)$-mincut if its capacity is the least among all $(s,t)$-cuts.
\label{def: (s,t)-cuts}
\end{definition}

The following lemma states an important property that holds for any pair of $(s,t)$-mincuts. 

\begin{lemma}(\cite{DBLP:journals/siamcomp/DinitzV00,DBLP:journals/talg/BaswanaBP23})
    If $(A,\overline{A})$ and $(A',\overline{A'})$ are any two $(s,t)$-mincuts, there does not exist any edge between $A\backslash A'$ and $A'\backslash A$.
\label{lem: no edge between AA' and A'A}
\end{lemma}

A cut $(A,\overline{A})$ is said to {\em divide}  a set $B$ if $A\cap B \not= \emptyset \not= \overline{A}\cap B$. Let $S$ be an arbitrary subset of $V$ of size at least~$2$.

\begin{definition}[$S$-cuts and $S$-mincuts]
A cut $(A,\overline{A})$ is said to be a Steiner cut or $S$-cut if it divides $S$.
We use $\lambda$ to denote the minimum capacity of any $S$-cut. An $S$-cut is called an $S$-mincut if its capacity is $\lambda$. 
\label{Def: Steiner-cuts}
\end{definition}

The following definition extends the notion of $(s,t)$-cuts.

\begin{definition}[$(S_1,S_2)$-cut]
    Let $S_1$ and $S_2$ be any two disjoint subsets of $S$.
    A cut $(A,\overline{A})$ is said to be an $(S_1,S_2)$-cut if $S_1\subseteq A$ and 
$S_2\subseteq \overline{A}$.
\label{def: (S1,S2)-cut}
\end{definition}

Given any two disjoint subsets $S_1,S_2$ of vertices,
let $G'$ be the graph obtained from $G$ by contracting $S_1$ into a vertex $s$ and $S_2$ into a vertex $t$. 
 Observe that there is a bijection between the set of $(S_1,S_2)$-cuts in $G$ and the set of $(s,t)$-cuts in $G'$. Therefore, the notion of $(s,t)$-mincut gets extended seamlessly to $(S_1,S_2)$-mincut. 

Observe that each $S$-cut $(A,\overline{A})$ defines a unique cut $(S\cap A, S\cap \overline{A})$ of the
set $S$. 
There may be multiple $S$-mincuts that define the same cut of $S$. However, for any arbitrary cut of $S$, there might not exist any $S$-mincut that defines it. Henceforth, a cut of $S$ is said to be a {\em valid} cut of $S$ if there exists at least one $S$-mincut that defines it.
A valid cut $(S_1,\overline{S_1})$ is said to be a {\em indivisible} valid cut if there is no $S$-mincut that divides $S_1$.

\begin{definition}[Bunch]
Two $S$-mincuts are said to be $S$-equivalent if they define the same cut of $S$. Each of the corresponding equivalence classes of $S$-mincuts is called a {\em bunch}.
\label{def: bunches-and-units}
\end{definition}

In the following section, we state a well-known property satisfied by the cuts of graphs. This property will be used crucially in deriving a majority of results in this article.

\subsection{Submodularity condition}

Let $A$ and $B$ be any two subsets of vertices.
It is well-known (Problem 6.48($i$) in \cite{Lovasz-book}) that they satisfy the submodularity condition.

    \begin{center} 
    $c(A)+c(B)\ge c(A\cap B)+c(A\cup B)$
    \end{center}

The following lemma, an immediate corollary of the above condition, 
highlights its important application.

\begin{lemma}[Submodularity of cuts]
Let ${\mathcal X}$ be a family of cuts in $G$. Let $(A,\overline{A})$ and $(B,\overline{B})$ be any two minimum cuts in ${\mathcal X}$. If cuts defined by $A\cap B$ and $A\cup B$ also belong to ${\mathcal X}$, then $A\cap B$ and $A\cup B$ also define minimum cuts in ${\mathcal X}$.
\label{lem: cor-submodularity-of-cuts}
\end{lemma}

We now state a lemma which can be viewed as a generalization of the submodularity condition. 
\begin{lemma}[Problem 6.48($iii$) in \cite{Lovasz-book}]
For any three subsets $C_1, C_2,C_3\subseteq V$,
\begin{center}
$c(C_1) + c(C_2) + c(C_3) $~~$\ge$~~ $c(C_1\cap\overline{C_2}\cap \overline{C_3}) ~+~ c(\overline{C_1}\cap {C_2} \cap\overline{C_3}) ~+~ $ \\
\hspace{1.5in}$c(\overline{C_1}\cap \overline{C_2} \cap C_3) ~+~ c(C_1\cap C_2\cap C_3)$
\end{center}
\label{lem:four point lemma}
\end{lemma}

The following lemma states a property satisfied by any three $S$-mincuts. It is equivalent to {\textsc{3-Star Lemma}($i$)} in \cite{DBLP:journals/siamcomp/DinitzV00}, albeit its proof is much shorter and lighter.
\begin{lemma}
Let $C_1,  C_2,  C_3\subset V$ define three $S$-mincuts in $G$. If no vertex from $S$ belongs to more than one of the subsets $C_1,C_2$, and $C_3$, then $C_1\cap C_2\cap C_3=\emptyset$.
\label{lem: corollary of four point lemma}
\end{lemma}
\begin{proof}
Given that any vertex from $S$ belongs to at most one of $C_1,C_2$, and $C_3$, it is easy to
observe that the cuts defined by each of the following three subsets are also $S$-cuts:
\[
C_1\cap\overline{C_2}\cap \overline{C_3},~~~~~~~~
\overline{C_1}\cap {C_2} \cap\overline{C_3},~~~~~~~~ 
\overline{C_1}\cap \overline{C_2} \cap C_3.
\] 
Hence the capacity of the cut defined by each of them is at least $\lambda$. 
So, Lemma \ref{lem:four point lemma} implies that $c(C_1 \cap C_2 \cap C_3)=0$. Since $G$ is a connected graph, this equality implies that $C_1 \cap C_2 \cap C_3 = \emptyset$.
\end{proof}

\subsection{Tight and Loose mincuts}
We now introduce the notion of tight and loose mincuts that will be used crucially in this article. 

\begin{definition}[Tight and loose mincut from $s$ to $t$]
Let $(A,\overline{A})$ be any $(s,t)$-mincut. 
It is said to be a tight mincut from $s$ to $t$ if
it does not dominate any other $(s,t)$-mincut, and 
a loose mincut from $s$ to $t$ if 
it is not dominated by any other $(s,t)$-mincut.
\label{def:tight-s-t-mincut}
\end{definition}

Observe that the intersection (as well as the union) of any two $(s,t)$-mincuts is always a $(s,t)$-cut. So it follows from the submodularity of cuts (Lemma \ref{lem: cor-submodularity-of-cuts}) that 
the family of $(s,t)$-mincuts is closed under intersection and union operations. 
Therefore, the tight (likewise the loose) mincut from $s$ to $t$ exists and is unique --
the intersection (likewise the union) of all $(s,t)$-mincuts is the {\em tight} (likewise the {\em loose}) $(s,t)$-mincut. 
The tight and the loose mincuts are sometimes called the nearest and the farthest mincuts respectively. 
We use $N(s,t)$ and $F(s,t)$ to denote respectively the tight and the loose mincut from $s$ to $t$. The following lemma, whose proof is an easy exercise, states the relationship between tight and loose mincuts.
\begin{lemma}
$N(s,t)$ is the opposite cut of $F(t,s)$. 
\label{lem: N(s,t)-is-opposite-of-F(t,s)}
\end{lemma}

Let $S_1$ and $S_2$ be any pair of disjoint subsets of $S$. The tight (likewise the loose) mincut from $s$ to $t$ gets extended, in a seamless manner, to the tight (likewise the loose) mincut from $S_1$ to $S_2$ (refer to Definition \ref{def: (S1,S2)-cut} and the paragraph that follows it). We now state a couple of lemmas about the tight and the loose $S$-mincuts. 

\begin{lemma}
Let $(S_1,\overline{S_1})$ be a valid  cut of $S$, and let $(B,\overline{B})$ be the tight  mincut from $S_1$ to $\overline{S_1}$. If $(A,\overline{A})$ is any $S$-mincut that divides $B$, $(A,\overline{A})$ divides $S_1$ as well. 
\label{lem: S-crossing-tight-Steiner-mincut}
\end{lemma}
\begin{proof}
We present a proof by contradiction. 
Suppose $(A,\overline{A})$ does not divide $S_1$.
Without loss of generality, assume that $S_1\subseteq A\cap S$ (otherwise swap $A$ and $\overline{A}$). So the cut defined by $A\cap B$ is an $S$-cut. Also, observe that 
$(A\cup B)\cap S=A\cap S$ because $B\cap S=S_1$ and $S_1\subseteq A\cap S$. 
Hence, the cut defined by $A\cup B$ is also an $S$-cut. So it follows from the submodularity of cuts (Lemma \ref{lem: cor-submodularity-of-cuts})  that $A\cap B$ and $A\cup B$ define $S$-mincuts. Notice that the $S$-cut defined by $A\cap B$ is $(S_1,\overline{S_1})$.
However, $A\cap B$ is a proper subset of $B$ since the cut defined by $A$ divides $B$. This would contradict that $(B,\overline{B})$ is the tight mincut from $S_1$ to $\overline{S_1}$. 
\end{proof}

\noindent
Lemma \ref{lem: S-crossing-tight-Steiner-mincut} provides important an insight on the tight and the loose $S$-mincuts as stated in the following lemma. 

\begin{lemma}
Let $(S_1,\overline{S_1})$ be any valid cut of $S$, and let $B\subsetneq V$ define $N(S_1,\overline{S_1})$. The following assertions hold. 
\begin{enumerate}
\item     
If $(S_2,\overline{S_2})$ is any valid cut of $S$ that dominates the cut 
$(S_1,\overline{S_1})$, $N(S_2,\overline{S_2})$ dominates $N(S_1,\overline{S_1})$, 
and $F(S_2,\overline{S_2})$ dominates
$F(S_1,\overline{S_1})$.
\item 
 If $(S_1,\overline{S_1})$ is an indivisible valid cut of $S$, then no $S$-mincut can divide $B$.
\end{enumerate}
\label{lemma: loose-tight-mincut-subset-property}
\end{lemma}
\begin{proof}
We first prove assertion (1). 
Let $A\subsetneq V$ 
define $N(S_2,\overline{S_2})$.
Observe that $(A,\overline{A})$ does not divide $S_1$ since $(S_2,\overline{S_2})$ dominates $(S_1,\overline{S_1})$ and $S_2\subseteq A$. Hence, it follows from Lemma \ref{lem: S-crossing-tight-Steiner-mincut} that $(A,\overline{A})$ does not divide $B$. However, $A\cap B\neq \emptyset$ since
$S_1\subseteq B$ and $S_1\subsetneq S_2\subseteq A$. So $B\subseteq A$. But $B=A$ is ruled out because $S_2\subseteq A$ and $S_2\nsubseteq B$. Therefore, $B\subsetneq A$. So
$N(S_2,\overline{S_2})$ dominates 
$N(S_1,\overline{S_1})$. Similarly, using the relationship between tight and loose mincuts from Lemma \ref{lem: N(s,t)-is-opposite-of-F(t,s)}, we can prove that $F(S_2,\overline{S_2})$ dominates
$F(S_1,\overline{S_1})$. This completes the proof of assertion (1). 

We now prove assertion $(2)$. Let $(A,\overline{A})$ be any $S$-mincut. Observe that $(A,\overline{A})$ cannot divide $S_1$ since $(S_1,\overline{S_1})$ is an indivisible valid cut of $S$. So, Lemma \ref{lem: S-crossing-tight-Steiner-mincut} implies that $(A,\overline{A})$ cannot divide $B$.
\end{proof} 

\begin{definition}[Laminar set family]
Let ${\cal Y}$ be a family of subsets of a set $Y$. ${\cal Y}$ is called a laminar set family if for each $A,A'\in {\cal Y}$, either $A$ and $A'$ are disjoint or one of them is contained inside the other.
\label{def:laminar}
\end{definition}
 It is a folklore that any laminar set family defined over a set $Y$ has size ${\cal O}(|Y|)$. The following section presents a few well-known families of cuts.

\subsection{Parallel, Laminar, and Crossing cuts}

\begin{definition}[Pair of crossing cuts]
Let ${\mathcal X}$ be any family of cuts. 
Two cuts $(A,\overline{A})$ and $(A',\overline{A'})$ from ${\mathcal X}$ are said to {\em cross} each other if each of the 4 corner sets, that is, $A\cap A'$, $A\setminus A'$, $A'\setminus A$, $\overline{A}\cap \overline{A'}$ are non-empty. If two cuts do not cross each other, they are said to be parallel to each other. 
\label{def: pair-of-crossing-cut}
\end{definition}
An alternative way to define a pair of parallel cuts is the following.
Two cuts $(A,\overline{A})$ and $(A',\overline{A'})$ are parallel to each other if 
one of $A$ and $\overline{A}$ is a subset of one of $A'$ and $\overline{A'}$. We can use Definition \ref{def: pair-of-crossing-cut} to classify all cuts of a family as follows.

\begin{definition}[Laminar and crossing cuts]
    Let ${\mathcal X}$ be a family of cuts. A cut in ${\mathcal X}$ is said to be a \emph{laminar} cut if every other cut in ${\mathcal X}$ is parallel to it. A cut that is not laminar is called a \emph{crossing} cut.
\label{def: laminar-and-crossing}
\end{definition}

A family of cuts is said to be a laminar cut family if each cut in the family is a laminar cut. 
Let ${\cal X}$ be a laminar family of cuts of a set $Z$. Let ${\cal Z}({\cal X})$ be the family of subsets of $Z$ that define the cuts of ${\cal X}$. Note that the subset of vertices that define a cut can also be used to implicitly represent the opposite cut. So if a cut and its opposite cut are both present in ${\cal X}$, we keep exactly one of the corresponding subsets of vertices defining them in ${\cal Z}({\cal X})$. 
It follows easily from the discussion preceding Definition \ref{def: laminar-and-crossing} that 
${\cal Z}({\cal X})$ is a laminar set family.  So, the following lemma provides an upper bound on the size of a laminar cut family. 
\begin{lemma}
A laminar cut family defined over a set $Z$ has size ${\cal O}(|Z|)$.
\label{lem: laminar-valid-cuts}
\end{lemma}

The following lemma can be easily proved using the fact that $(s,t)$-mincuts are closed under intersection and union. 
\begin{lemma}
    $N(s,t)$ (likewise $F(s,t)$) is a laminar cut in the family of $(s,t)$-mincuts.
\label{lem: s^N_t and s^F_t are parallel}
\end{lemma}

We now define another family of cuts called a {\em crossing family}. 

\begin{definition}[Crossing family of cuts]
A family 
of cuts ${\mathcal X}$ is said to be a crossing family if for any pair $(A,\overline{A})$ and $(A',\overline{A'})$ of crossing cuts in ${\mathcal X}$,  
the following properties hold:
\begin{enumerate}
    \item Each of the 4 corner sets defined by $(A,\overline{A})$ and $(A',\overline{A'})$
    also define a cut belonging to ${\mathcal X}$.
    \item The cut defined by 
    $(A\setminus A') \cup (A'\setminus A)$ does not belong to ${\mathcal X}$.
\end{enumerate}
\label{def:crossing family}
\end{definition}
Observe that a laminar cut family is vacuously a crossing cut family. 
The set of all $(s,t)$-mincuts does not form a crossing family unless there is no crossing pair of $(s,t)$-mincuts in $G$. 
However, the set of all global mincuts is a crossing family. For this, an interested  reader may refer to Lemma \ref{lem:global-mincuts-are-crossing-family} in Appendix \ref{app: global-mincuts-is-a-crossing-family}; its proof uses Lemma \ref{lem: no edge between AA' and A'A} and the submodularity of cuts (Lemma \ref{lem: cor-submodularity-of-cuts}). 
Two $S$-mincuts $(A,\overline{A})$ and $(A',\overline{A'})$ are said to $S$-cross each other if each of the four corner subsets defined by $A$ and $A'$ also defines an $S$-cut. We now state an important property of valid cuts of $S$. 

\begin{lemma}
The set of all valid cuts of $S$ is a crossing family.
\label{lem: Valid-cuts-form-crossing-family}
\end{lemma}
\begin{proof}
Let $(A,\overline{A})$ and $(A',\overline{A'})$ be any  two $S$-mincuts and suppose they $S$-cross each other.
So each of the four corner sets defined by these cuts define $S$-cuts. Let $\{ S_1,S_2,S_3,S_4\}$ be the partition of $S$ defined by these corner sets as shown in Figure \ref{fig: proof-of-crossing-family-for-valid-cuts}. 

\begin{figure}[ht]
 \centering  \includegraphics[width=200pt]{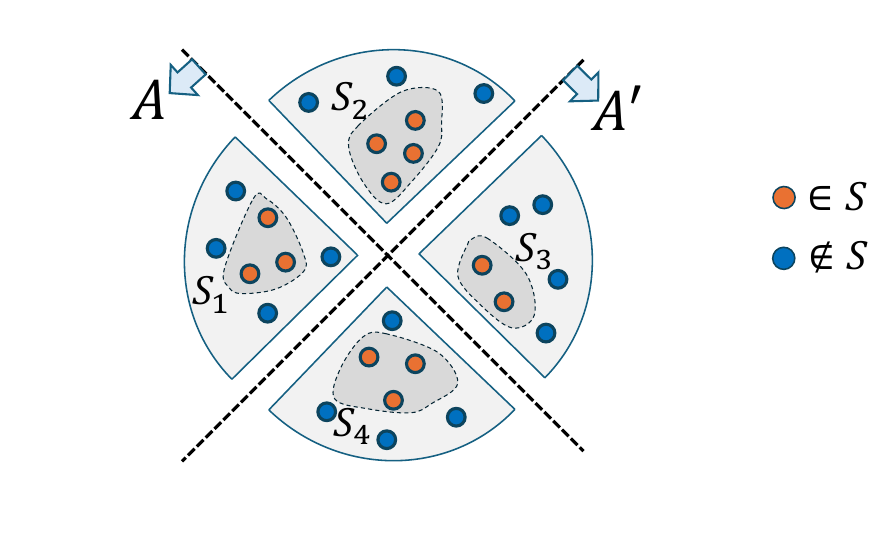} 
  \caption{$A$ and $A'$ are $S$-mincuts. They are also $S$-crossing.}
    \label{fig: proof-of-crossing-family-for-valid-cuts}
\end{figure}

Applying the submodularity of cuts (Lemma \ref{lem: cor-submodularity-of-cuts}), we can infer that each of the four corner sets also define a $S$-mincut. So each of $S_1$,$S_2$, $S_3$, and $S_4$ defines a valid cut of $S$. Thus the 1st condition of a crossing family (refer to Definition \ref{def:crossing family}) is satisfied. 
In order to show that the 2nd condition of a crossing family is also satisfied, we have to show that $S_2\cup S_4$ does not define a valid cut of $S$. We now provide a proof by contradiction for the same. 

Suppose $S_2\cup S_4$ defines a valid cut of $S$, so let $(B,\overline{B})$ be any $S$-mincut defined by $S_2\cup S_4$. 
Let $C_1=A, C_2=B$, and $C_3=A'$. Note that $C_1\cap C_2\cap C_3\not=\emptyset$ since $S_4\subseteq C_1\cap C_2\cap C_3$. 

Observe that each of the 3 sets, namely $C_1\cap \overline{C_2}\cap \overline{C_3}$, $\overline{C_1}\cap C_2\cap \overline{C_3}$, and $\overline{C_1}\cap \overline{C_2}\cap C_3$, defines a $S$-cut. The capacity of these cuts has to be at least $\lambda$ (the capacity of a $S$-mincut). Thus it follows from Lemma \ref{lem:four point lemma} that $c(C_1\cap C_2\cap C_3)=0$. Since the graph is connected, this would imply that $C_1\cap C_2\cap C_3=\emptyset$ -- a contradiction. Hence $S_2\cup S_4$ does not define a valid cut of $S$. 
\end{proof}

Consider any pair of vertices $u,v\in S$ separated by at least one valid cut of $S$. Let $(S_1,\overline{S_1})$ and $(S_2,\overline{S_2})$ be any two valid cuts of $S$ with $u\in S_1, S_2$ and $v\in \overline{S_1}, \overline{S_2}$. It follows from Lemma
\ref{lem: Valid-cuts-form-crossing-family}
that $S_1\cap S_2$ defines a valid cut of $S$ with $u\in S_1\cap S_2$ and $v\in \overline{S_1\cup S_2}$. Therefore, the notion of tight mincut from $u$ to $v$ gets extended to the tight valid cut from $u$ to $v$ in a seamless manner. So, along the lines of Lemma \ref{lem: s^N_t and s^F_t are parallel}, we can derive the following lemma.

\begin{lemma}
Let $u,v \in S$ be separated by at least one valid cut of $S$.
The tight (likewise loose) valid cut of $S$ from $u$ to $v$  exists, is unique, and is a laminar cut in the family of all valid $S$-cuts.
\label{lem: tight-S--mincut-is-laminar}
\end{lemma}

\noindent 
Lemma \ref{lem: tight-S--mincut-is-laminar} plays a crucial role in the construction of a $t$-cactus that stores all valid cuts of $S$ (refer to Appendix \ref{app: cactus-construction-for-valid-cuts}). The following lemma states an interesting property for certain pairs of $S$-mincuts.

\begin{lemma}[Lemma 2.1($i$)  \cite{DBLP:journals/siamcomp/DinitzV00}]
Let $(A,\overline{A})$ and $(B,\overline{B})$ be any two  $S$-mincuts such that $A\cap B$ and $A\cup B$ define $S$-cuts. There can not be any edge between $A\backslash B$ and $B\backslash A$.
\label{lem:no-edge-CminusC'-and-C'-minus-C}
\end{lemma}
\begin{proof}
   It follows from the submodularity of cuts (Lemma \ref{lem: cor-submodularity-of-cuts}) that  $A\cap B$ and $A\cup B$ define $S$-mincuts. Contract all those vertices of set $S$ that belong to $A\cap B$ to form the source vertex $s$, and contract all those vertices of set $S$ that belong to $\overline{A} \cap \overline{B}$ to form the sink vertex $t$. Observe that both $A\cap B$ and $A\cup B$ define $(s,t)$-mincuts in the resulting graph. 
   So it follows from Lemma 
   \ref{lem: no edge between AA' and A'A} that there is no edge between $A\backslash B$ and $B\backslash A$.
\end{proof}

\section{Properties and terminologies of a balanced DAG and a \texorpdfstring{$t$}{t}-Cactus} \label{sec:balaned_DAG_and_Cactus}
\subsection{A single-source single-sink balanced DAG}
%
We begin with the following lemma that states two properties of a single-source single-sink DAG.
%
%
\begin{lemma}
Let $\overrightarrow{\cal G}$ be any single-source single-sink balanced DAG. \begin{enumerate}
\item 
For each edge $(u,v)$ in $\overrightarrow{\cal G}$, there exists a path from the source to the sink passing through it. 
\item 
The indegree of the sink in $\overrightarrow{\cal G}$ is the same as the outdegree of the source. 
\end{enumerate}
\label{lem: flow-in-balanced-DAG}
\end{lemma}
\begin{proof}
    Proof of Assertion (1):~Let $s$ and $t$ denote respectively the source and the sink of $\overrightarrow{\cal G}$. Suppose we start traversing $\overrightarrow{\cal G}$ starting from $v$. For each newly visited vertex $y\not=t$, there must always be an outgoing edge due to Property $(b)$ of the single-source single-sink balanced DAG (see Definition \ref{def:balanced-DAG-property}). Therefore, we shall be able to leave $y$ along some edge. Moreover, we shall never visit a vertex again since $\overrightarrow{\cal G}$ is a DAG. So the traversal is bound to terminate at $t$ after a finite number of steps. The sequence of edges traversed constitutes a path, say $Q_v$ from $v$ to $t$. Along exactly similar lines, we can show that there is a path, say $Q_u$ from $s$ to $u$. $Q_u$ and $Q_v$ can not intersect since $\overrightarrow{\cal G}$ is a DAG. So 
    $Q_u$ concatenated with edge $(u,v)$ and $Q_v$ is a path that originates from $s$, passes through $(u,v)$, and terminates at $t$.

Proof of Assertion (2):~It follows from the proof of Assertion (1) that there always exists a path, say $Q$, in $\overrightarrow{\cal G}$ that originates from $s$ and terminates at $t$. If we remove each edge of this path, the indegree of the sink as well as the outdegree of the source decreases by 1; and the indegree and the outdegree of every other vertex on the path $Q$ decreases by 1. So the resulting graph is still a single-source single-sink balanced DAG. Using this observation, the validity of Assertion (2) can be established through an easy proof by induction on the number of edges.
\end{proof}

For any given single-source single-sink balanced DAG $\overrightarrow{\cal G}$, let ${\cal G}$ be the undirected graph obtained from $\overrightarrow{\cal G}$ by removing the direction of each of its edges. We call ${\cal G}$ the {\em undirected analogue} of $\overrightarrow{\cal G}$. Interestingly, the structure of $\overrightarrow{\cal G}$ is so rich that the direction of its edges is {\em redundant} as shown by the following lemma.

\begin{lemma}
There exists an ${\cal O}(m)$ time algorithm that can construct $\overrightarrow{\cal G}$ from ${\cal G}$.
\label{lem: G-from-Gu}
\end{lemma}
\begin{proof}
Recall the classical ${\cal O}(m)$ time algorithm by Kahn \cite{DBLP:journals/CommACM/Kahn} that computes a topological ordering of a DAG with $m$ edges. It maintains a queue ${\cal Q}$ storing vertices of indegree 0, and iterates the following step until ${\cal Q}$ becomes empty.

{\em Perform Dequeue operation on ${\cal Q}$; let $w$ be the vertex returned by this operation. Remove all outgoing edges of $w$, and decrement the indegree of each of the corresponding neighbors of $w$ by 1. As a result, if the indegree of any neighbor of $w$ becomes 0, we insert that neighbor into the queue ${\cal Q}$.}

Our algorithm {\em simulates} the algorithm by Kahn \cite{DBLP:journals/CommACM/Kahn} on ${\cal G}$. 
We initialize ${\cal Q}$ to store the source vertex $s$. Since ${\cal G}$ is undirected, the only hurdle in implementing the algorithm of Kahn \cite{DBLP:journals/CommACM/Kahn} is to detect the moment when the indegree of a vertex becomes 0. We overcome this hurdle by exploiting Property $(b)$ of the single-source single-sink balanced DAG (see Definition \ref{def:balanced-DAG-property}) as follows. 
For each vertex $v$ in ${\cal G}$, let $d(v)$ be the number of edges incident on it. We compute $d(v)$ for all $v\in {\cal G}$ in $O(m)$ time in the beginning.
We keep a variable $d'(v)$ for each vertex $v$ (other than $s$ and $t$) that stores the number of edges incident on it at any stage during the algorithm. We initialize $d'(v)$ to $d(v)$. Every time $v$ loses an edge incident on it, we decrement $d'(v)$ by 1. Note that each such edge is an incoming edge of $v$ in $\overrightarrow{\cal G}$. Therefore, once $d'(v)=d(v)/2$, $v$ has lost all incoming edges in $\overrightarrow{\cal G}$ and is ready for insertion into the queue ${\cal Q}$.  

Thus our algorithm, executed on ${\cal G}$, computes the topological ordering of $\overrightarrow{\cal G}$. 
Once we have the topological ordering, we can determine the direction of each edge of ${\cal G}$ in $\overrightarrow{\cal G}$ in ${\mathcal O}(1)$ time. 
\end{proof}

We shall use $\overleftarrow{\cal G}$ to denote the balanced DAG obtained by reversing the direction of each edge in $\overrightarrow{\cal G}$. Observe that ${\cal G}$ is also the undirected analogue of the DAG $\overleftarrow{\cal G}$ with the roles of the source and the sink reversed. 
The following concept, though coined by Dinitz and Vainshtein \cite{DBLP:conf/stoc/DinitzV94}  in the context of locally orientable graphs, was implicit in the work of  Picard and Queyranne \cite{DBLP:journals/mp/PicardQ82} on the characterization of $(s,t)$-mincuts.

\begin{figure}[ht]
 \centering  \includegraphics[width=350pt]{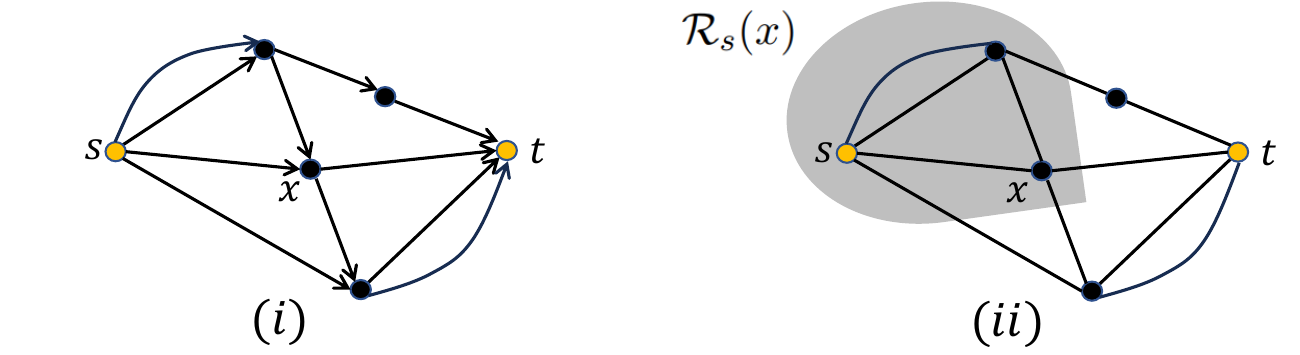} 
  \caption{($i$) A balanced DAG $(ii)$ the undirected analogue and a reachability cone.}
    \label{fig:balanced-dag-and-reach-cone}
\end{figure}

\begin{definition}[Reachability cone]
Let ${\mathcal G}$ be the undirected analogue of a single-source single-sink balanced DAG. Let $x$ be any vertex other than the source and the sink. The reachability cone of $x$ in ${\mathcal G}$ in the direction of the sink is defined as the set of all vertices that are reachable from $x$ in $\overrightarrow{\cal G}$. We denote it by ${\cal R}_t(x)$. The reachability cone of $x$ in the direction of the source, is defined similarly in $\overleftarrow{\cal G}$, and is denoted by ${\cal R}_s(x)$. 
\label{def:reachability-cone}
\end{definition}
Figure \ref{fig:balanced-dag-and-reach-cone} illustrates the undirected analogue of a balanced DAG and a reachability cone. 

The following lemma is immediate from the acyclicity of $\overrightarrow{\cal G}$ and the definition of ${\cal R}_s(x)$.
\begin{lemma} Let $\overrightarrow{\cal G}$ be any balanced DAG with source $s$ and sink $t$. 
\begin{enumerate} 
\item There is no edge entering ${\mathcal R}_s(x)$
in $\overrightarrow{\mathcal G}$.
\item For any two vertices $x$ and $y$ in ${\mathcal G}$, $y\in {\cal R}_{s}(x)$ and $x\in {\cal R}_{s}(y)$ can not hold simultaneously.
\end{enumerate}
\label{lem:Properties-of-R_s(x)}
\end{lemma}
%

\subsection{A \texorpdfstring{$t$}{t}-cactus}
\label{sec: cactus}
We described $t$-cactus in the Introduction. 
%
%
Henceforth, we shall use \emph{nodes} and \emph{structural edges}, respectively, for vertices and edges of a $t$-cactus. 
A path is said to {\em pass through a cycle} in a $t$-cactus if it shares at least one edge with the cycle. A node in a $t$-cactus is called a cycle node if it belongs to a cycle; otherwise it is called a tree node. A tree node with degree 1 is called a leaf node. 
\begin{figure}[ht]
  \begin{center}
    \includegraphics[width=0.5\textwidth]{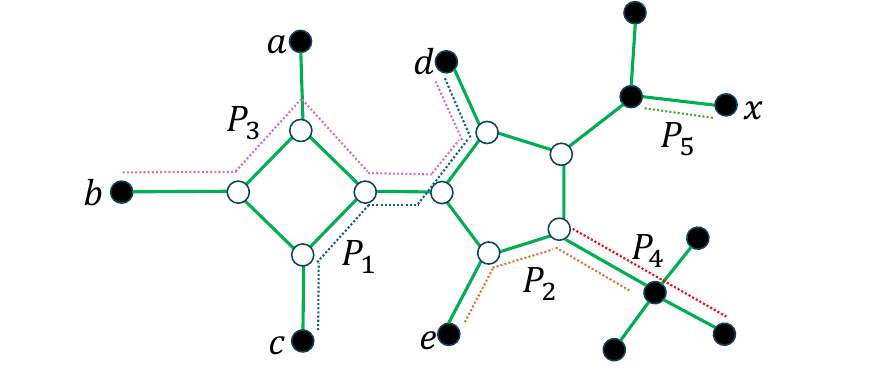}
  \end{center}    
  \caption{A $t$-cactus with various paths.}
    \label{fig:t-cactus-and-proper-paths-and-intersection}
\end{figure}

\begin{definition}[Proper path]
A path in a $t$-cactus is said to be a \textit{proper path} if it shares exactly one structural edge with each cycle it passes through. 
\label{def:proper-path}
\end{definition}
 Refer to Figure \ref{fig:t-cactus-and-proper-paths-and-intersection} that illustrates a $t$-cactus. Paths $P_1, P_2, P_4,$ and $P_5$ are proper paths, but $P_3$ is not a proper path. 
 There is a proper path between $c$ and $d$, but there is no proper path between $c$ and $x$. Observe that there can be at most one proper path between any pair of nodes in a $t$-cactus. 
 We now define the {\em intersection} of two paths in a $t$-cactus.

\begin{definition}[Intersection of two paths]
\label{def:intersection-of-two-paths}
Two paths in a $t$-cactus are said to intersect if they share a tree edge or if there is a cycle that shares edge(s) with each of them. In the former case, the paths are said to intersect at a tree edge, and in the latter case, the paths are said to intersect at a cycle.
\end{definition}

Refer to Figure \ref{fig:t-cactus-and-proper-paths-and-intersection}. Path $P_1$ intersects $P_2$, but does not intersect $P_4$ or $P_5$. $P_2$ intersects $P_4$.
Length of a path is the number of edges lying on the path. 
The distance between any pair of nodes $\nu$ and $\mu$ in a $t$-cactus is the length of the shortest path between $\nu$ and $\mu$. It is denoted by $\delta(\nu,\mu)$.

\begin{definition}[Distance between two structural edges] 
Let $e_1=(\nu_1,\nu_2)$ and $e_2=(\mu_1,\mu_2)$ be any pair of structural edges in a $t$-cactus. The distance between $e_1$ and $e_2$ is defined as $\min\{\delta(\nu_1,\mu_1),\delta(\nu_1,\mu_2),(\nu_2,\mu_1),(\nu_2,\mu_2)\}$. 
\label{def:distance-between-edges}
\end{definition}

\subsection{A tree data structure for storing a \texorpdfstring{$t$}{t}-cactus} 
\begin{figure}[ht]
  \begin{center}
    \includegraphics[width=0.67\textwidth]{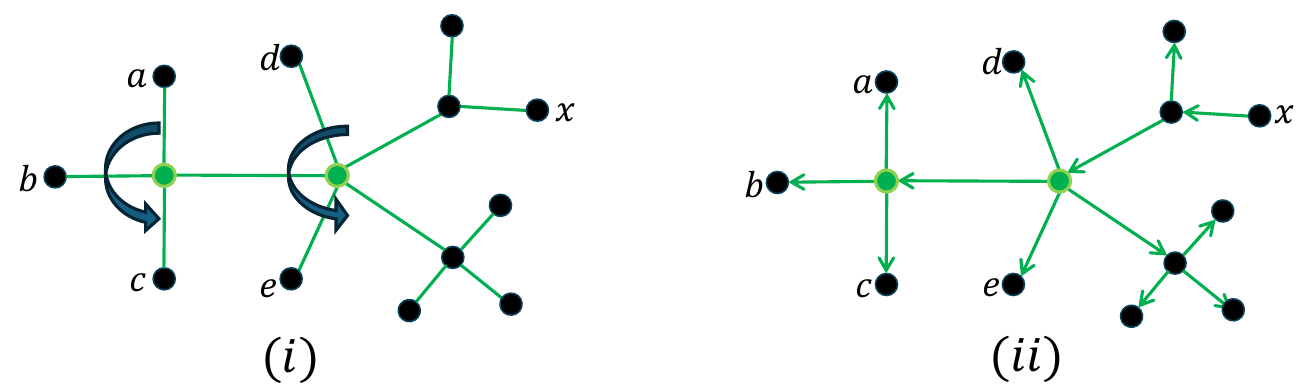}
  \end{center}    
  \caption{($i$) the tree
  $t({\cal H})$,~($ii$) tree rooted at $x$.}
    \label{fig:data-structure-for-t-cactus}
\end{figure}

 Dinitz and Westbrook \cite{DBLP:journals/algorithmica/DinitzW98} described a compact data structure for a cactus. We describe an even simpler data structure for a $t$-cactus. 
Let $t({\cal H})$ be the tree transformed into a $t$-cactus ${\cal H}$. Figure \ref{fig:data-structure-for-t-cactus}($i$) illustrates $t({\cal H})$ for the $t$-cactus ${\cal H}$ shown in Figure 
\ref{fig:t-cactus-and-proper-paths-and-intersection}.
%
Interestingly, $t({\cal H})$ with suitable augmentation serves as a data structure for  ${\cal H}$ as follows.

\begin{enumerate} 
\item  We root $t({\mathcal H})$ on any leaf vertex. 
\item Consider a vertex in $t({\mathcal H})$ at which a cycle is implanted during its transformation into ${\cal H}$. For each such vertex, we arrange its children in the rooted $t({\mathcal H})$ in the order defined by the cycle. This ordering captures the cycle completely. Refer to 
Figure \ref{fig:data-structure-for-t-cactus}.
\item 
We augment $t({\cal H})$ with the data structure that can answer any LCA query in ${\cal O}(1)$ time \cite{DBLP:journals/tcs/BenderF04}. 
\end{enumerate} 


We define a mapping $f$ from the set of nodes of ${\cal H}$ to the set of vertices of $t({\cal H})$ as follows. Let $\nu$ be any node in ${\cal H}$. If $\nu$ is a tree node, it is also present in ${\cal H}$, so $f(\nu)=\nu$. If $\nu$ belongs to a cycle, say $O$, $f(\nu)$ is the vertex in $t({\cal H})$ at which $O$ is implanted. Let $\nu$ and $\nu'$ be any pair of nodes in ${\cal H}$. Observe that all paths between $\nu$ and $\nu'$ in ${\cal H}$ are represented by the unique path between $f(\nu)$ and $f(\nu')$ in $t({\cal H})$. 
It is a simple exercise to show that it takes ${\cal O}(1)$ LCA queries to determine whether any two given paths in a tree share any vertex or any edge. Using this, we can efficiently determine if two paths in ${\cal H}$ intersect as follows.

Let $P_1$ and $P_2$ be any pair of paths  in ${\cal H}$, and let $f(P_1)$ and $f(P_2)$ be the corresponding paths in $t({\cal H})$. If $f(P_1)$ and $f(P_2)$ do not share any vertex, $P_1$ and $P_2$ do not intersect. If $f(P_1)$ and $f(P_2)$ share at least one edge, $P_1$ and $P_2$ surely intersect. These inferences exploit the fact that $t({\cal H})$ is a quotient graph of ${\cal H}$. The only case, that is left, is the case when $f(P_1)$ and 
$f(P_2)$ share exactly one vertex, say $v$. If a cycle is not implanted at $v$, surely $P_1$ and 
$P_2$ do not intersect. If a cycle is implanted at $v$, there are two subcases as follows. If $v$ is an internal vertex on $f(P_1)$ as well as 
$f(P_2)$, $P_1$ and 
$P_2$ intersect. If $v$ is an endpoint of one or both of $f(P_1)$ and 
$f(P_2)$, we need to determine whether $P_1$ and $P_2$ share an edge with the cycle implanted at $v$. The latter information can be determined in ${\cal O}(1)$ time. 
Considering all the cases, we can state the following lemma. 

\begin{lemma}
\label{lem:skeleton-tree-queries}
For any given $t$-cactus ${\cal H}$, there exists a data structure occupying space of the order of ${\cal H}$, that can determine in ${\cal O}(1)$ time for any given pairs of nodes  $(\nu_1,\nu_2)$ and $(\nu_3,\nu_4)$ in ${\cal H}$, whether there exist a path between $\nu_1$ and $\nu_2$, and a path between $\nu_3$ and $\nu_4$ that intersect each other. 
\end{lemma}

\subsection{Cuts in a \texorpdfstring{$t$}{t}-cactus}
 Let ${\cal H}$ be a $t$-cactus. The edge-set of a minimal cut of ${\cal H}$ consists of either a tree edge or any two edges belonging to the same cycle. Let $e$ be any structural edge in ${\cal H}$. $e$ is said to define a minimal cut in ${\cal H}$ if $e$ belongs to the edge-set of the cut. 
 If $e$ is a tree edge, there are 2 minimal cuts in ${\cal H}$ defined by $e$, and they are opposite to each other. 
 If $e$ is a cycle edge, $e$ defines $k-1$ pairs of opposite minimal cuts in ${\cal H}$,
 where $k$ is the number of edges in the cycle. 
The following lemma states an easy way to classify the minimal cuts of a $t$-cactus into laminar cuts and crossing cuts. Its proof is a simple exercise based on the structure of $t$-cactus and Definition \ref{def: laminar-and-crossing}. 
\begin{lemma}[Laminar and crossing cuts in $t$-cactus] A minimal cut defined by a tree edge or a pair of adjacent edges of a cycle in a $t$-cactus is a laminar cut; otherwise, it is a crossing cut.

\label{lem:parallel-crossing-cuts-in-cactus}
\end{lemma}

\begin{figure}[ht]
\begin{center}
\includegraphics[width=.8\textwidth]{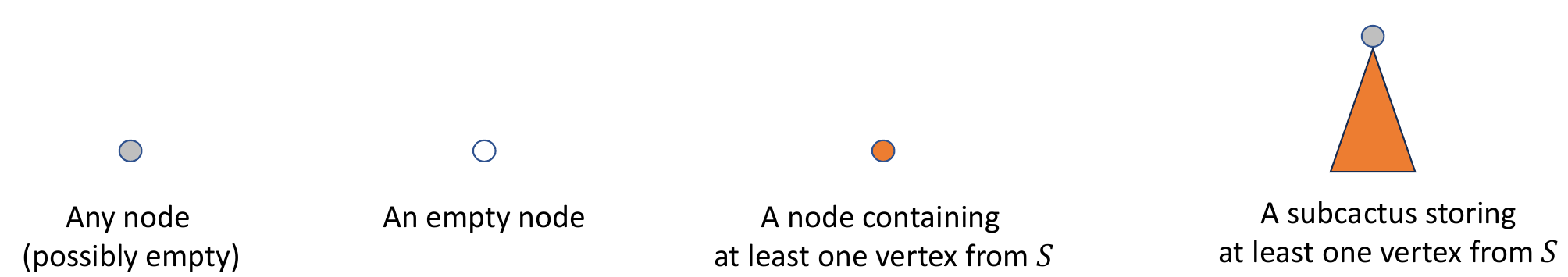}
\end{center}   
\caption{Notations used in the figures in this article.}
    \label{fig:subcactus}
\end{figure}

For better readability, we shall draw figures involving the $t$-cactus storing vertices from a given set $S$. Kindly refer to Figure \ref{fig:subcactus} for the notations used in the figures in the rest of the article.

\section{Existing data structures for various mincuts}
\label{sec:existing-structures-for-various-mincuts}
In this section, we describe structures that store and characterize all $(s,t)$-mincuts and all valid cuts of a set $S$.
On a few occasions, we shall use the inverse of a mapping stated formally as follows. 

\begin{definition}[Inverse of a mapping]
    Let $h:X \rightarrow Y$ be any mapping. For any subset $W\subseteq Y$, we define $h^{-1}(W)$ to be $\{x\in X ~|~ h(x)\in W\}$.
\label{def: inverse-of-mapping}
\end{definition}

\subsection{A balanced DAG that stores all {(s,t)}-mincuts}
\label{sec: D(s,t)}
For storing and characterizing all $(s,t)$-mincuts, Dinitz and Vainsthein \cite{DBLP:conf/stoc/DinitzV94, DBLP:journals/siamcomp/DinitzV00} introduced a structure, called {\em strip}, using the concept of locally orientable graphs. We present a more gentle exposure of a strip using the concept of a balanced DAG 
and maximum $(s,t)$-flow.

Let us consider the quotient graph of $G$ induced by all $(s,t)$-mincuts. We use bold letters to denote the vertices of this graph. The two vertices of the quotient graph, to which $s$ and $t$ are mapped, are called the terminals, denoted by ${\bf s}$ and ${\bf t}$ respectively. Every other vertex is called a non-terminal. Observe that
any $(s,t)$-mincut in $G$ appears as a $({\bf s},{\bf t})$-mincut in this quotient graph. 
This graph turns out to be the undirected analogue of a balanced DAG with source ${\bf s}$ and sink ${\bf t}$ as follows.

Let $f$ be any maximum $(s,t)$-flow in $G$. The strong duality between $(s,t)$-maxflow and $(s,t)$-mincut implies that each edge of any $(s,t)$-mincut is fully saturated in flow $f$. Let $u$ and $v$ be any two vertices joined by an edge in $G$ but separated by an $(s,t)$-mincut. Therefore, edge $(u,v)$ carries a unit flow in $f$ and must be present in the quotient graph (induced by all $(s,t)$-mincuts).  
Let $\overrightarrow{\cal D}_{s,t}$ be the graph obtained after directing each edge of the quotient graph in the direction of flow $f$ passing through it. Observe that $f$ also defines a maximum $({\bf s},{\bf t})$-flow in $\overrightarrow{\cal D}_{s,t}$ leading to the following lemma.
\begin{lemma}
 In every maximum $({\bf s},{\bf t})$-flow, each edge of $\overrightarrow{\cal D}_{s,t}$ carries a unit flow.
\label{lem: flow-on-edges-of-Dst}
\end{lemma}

The following lemma states an important property on the structure of $\overrightarrow{\cal D}_{s,t}$.

\begin{lemma}
$\overrightarrow{\cal D}_{s,t}$ is a balanced DAG with source ${\bf s}$ and sink ${\bf t}$.
\label{lem: balanced-DAG}
\end{lemma}
\begin{proof}
Consider any maximum $({\bf s},{\bf t})$-flow in $\overrightarrow{\cal D}_{s,t}$.
Suppose $\overrightarrow{\cal D}_{s,t}$ has a cycle. Reducing value of flow on each edge of the cycle keeps the value of flow unchanged but leads to zero flow on some edges of $\overrightarrow{\cal D}_{s,t}$ -- a contradiction to Lemma \ref{lem: flow-on-edges-of-Dst}. So $\overrightarrow{\cal D}_{s,t}$ must be acyclic. The conservation of flow and Lemma \ref{lem: flow-on-edges-of-Dst} implies that each vertex other than ${\bf s}$ or 
${\bf t}$ in $\overrightarrow{\cal D}_{s,t}$ will have exactly the same number of incoming edges as the number of outgoing edges. Hence $\overrightarrow{\cal D}_{s,t}$ is a balanced DAG. It is easy to observe that ${\bf s}$ is the only source and ${\bf t}$ is the only sink in this DAG.
\end{proof}

 It follows from Lemma \ref{lem: balanced-DAG} that the quotient graph induced by all $(s,t)$-mincuts can be viewed as the undirected analogue of the balanced DAG $\overrightarrow{\cal D}_{s,t}$ with source ${\bf s}$ and sink ${\bf t}$. 
 Let $E({\mathbf v})$ be the set of edges incident on a non-terminal ${\mathbf v}$ in the quotient graph. $\overrightarrow{{\cal D}}_{s,t}$ induces a unique balanced partition, called {\em inherent partition}, of $E({\mathbf v})$ into 2 subsets -- 
 the edges entering ${\mathbf v}$ and the edges leaving ${\mathbf v}$ in $\overrightarrow{{\cal D}}_{s,t}$. These subsets are called the two sides of the inherent partition of $E({\bf v})$. This leads us to define a {\em strip}. 

\begin{definition}[Strip] For any given pair of vertices $s$ and $t$, the quotient graph induced by all $(s,t)$-mincuts along with the inherent partition of the set of edges incident on each non-terminal is called a strip, denoted by ${\cal D}_{s,t}$. We use $\Phi$ to denote the mapping from $V$ to the vertices of ${\cal D}_{s,t}$. 
\label{def: Strip}
\end{definition}
 
 We now define certain (${\mathbf s},{\mathbf t}$)-cuts in ${\cal D}_{s,t}$ that will help in characterizing all $(s,t)$-mincuts of $G$.
 
  \begin{definition}[Transversal Cut]
     Let $({\bf U},\overline{\bf U})$ be a $({\mathbf s},{\mathbf t})$-cut in ${\cal D}_{s,t}$. $({\bf U},\overline{\bf U})$ is said to be a transversal cut if there is no edge entering ${\bf U}$ in $\overrightarrow{{\cal D}}_{s,t}$. Alternatively and equivalently, there is no edge entering $\overline{\bf U}$ in $\overleftarrow{{\cal D}}_{s,t}$. 
\label{def: transversal}
 \end{definition}

Using Lemma \ref{lem: flow-on-edges-of-Dst} and the strong duality between $(s,t)$-maxflow and $(s,t)$-mincut, it is easy to characterize all $(s,t)$-mincuts of $G$ using strip ${\cal D}_{s,t}$ as stated in the following lemma.
\begin{lemma}
$(A,\overline{A})$ is a $(s,t)$-mincut in $G$ if and only if $(\Phi(A),\overline{\Phi(A)})$ is a transversal cut in ${\cal D}_{s,t}$.
\label{lem:mincut-transversal}
\end{lemma}
\begin{proof}
Let $(s,t)$-mincut $(A,\overline{A})$ appear as $({\bf s},{\bf t})$-mincut $({\bf U},\overline{\bf U})$ in $\overrightarrow{\cal D}_{s,t}$. Let $f$ be any maximum $({\bf s},{\bf t})$-flow
in $\overrightarrow{\cal D}_{s,t}$. It follows from Lemma \ref{lem: flow-on-edges-of-Dst} that each edge of cut $({\bf U},\overline{\bf U})$ carries a unit flow. Suppose there is an edge entering ${\bf U}$ in 
$\overrightarrow{\cal D}_{s,t}$. Hence $f_{in}({\bf U})>0$. Applying conservation of flow, $f_{out}({\bf U})>$ value$(f)$. $f_{out}({\bf U})$ is upper bounded by the capacity of the cut $({\bf U},\overline{\bf U})$, so the capacity of $({\bf U},\overline{\bf U})$ exceeds value$(f)$. So $({\bf U},\overline{\bf U})$ is not a mincut in $\overrightarrow{\cal D}_{s,t}$ -- a contradiction. Hence there is no edge entering ${\bf U}$ in $\overrightarrow{\cal D}_{s,t}$. So $({\bf U},\overline{\bf U})$ is a transversal cut in ${\cal D}_{s,t}$. 

Let $({\bf U},\overline{\bf U})$ be a transversal cut in strip ${\cal D}_{s,t}$. So all edges of this cut are outgoing in $\overrightarrow{\cal D}_{s,t}$. 
Lemma \ref{lem: flow-on-edges-of-Dst} implies that each edge of the cut is fully saturated in every maximum $({\bf s},{\bf t})$-flow. Hence, $({\bf U},\overline{\bf U})$ is a $({\bf s},{\bf t})$-mincut in ${\cal D}_{s,t}$. So it follows from the construction of ${\cal D}_{s,t}$ that $(\Phi^{-1}({\bf U}),\Phi^{-1}(\overline{\bf U}))$ is a $(s,t)$-mincut in $G$.
\end{proof}

Observe that $\{\mathbf{s}\}$ and $V\setminus \{\mathbf{t}\}$ are transversal cuts, and hence define $(s,t)$-mincuts in $G$. Consider any non-terminal ${\bf x}$ in ${\cal D}_{s,t}$. The following lemma states a $(s,t)$-mincut defined using the reachability cone ${\cal R}_s({\bf x})$ (refer to Definition \ref{def:reachability-cone}).
\begin{lemma}
$\Phi^{-1}({\cal R}_s({\bf x}))$ defines a $(s,t)$-mincut in $G$.
\label{lem:R_s(x)_is_(s,t)-mincut}
\end{lemma}
\begin{proof}
 Lemma \ref{lem:Properties-of-R_s(x)}(1) implies that ${\cal R}_s({\bf x})$ defines a transversal cut in ${\cal D}_{s,t}$, and hence $\Phi^{-1}({\cal R}_s({\bf x}))$ defines a $(s,t)$-mincut due to Lemma \ref{lem:mincut-transversal}.
\end{proof}


\subsection{A \texorpdfstring{$t$}{t}-cactus that stores all valid cuts of \texorpdfstring{$S$}{set S}}
 Dinitz and Nutov \cite{DBLP:DN94}, by carrying out an elegant generalization of cactus for global mincuts \cite{DL76}, showed that there exists a $t$-cactus for every crossing family of cuts. 
The set of all valid cuts of $S$ is a crossing family as stated in Lemma \ref{lem: Valid-cuts-form-crossing-family}.  
So there exists a $t$-cactus that stores and characterizes all valid cuts of $S$. This result was also stated (without proof) independently by Westbrook \cite{Westbrook93} and D. Naor as reported in the articles \cite{DBLP:conf/stoc/DinitzV94, DBLP:journals/siamcomp/DinitzV00}. 

\begin{theorem}
For a given graph $G$ and a set $S$, there exists an 
${\mathcal O}(|S|)$ size $t$-cactus ${\cal H}=({\cal N},{\cal E})$ with a mapping $\varphi: S \rightarrow {\cal N}$ such that 
\begin{enumerate}
    \item Each minimal cut in ${\cal H}$ defines a valid cut of $S$ as follows. If ${\cal A}\subset {\cal N}$ defines a minimal cut in ${\cal H}$, then the set $\varphi^{-1}({\cal A})$ defines a valid cut of $S$.
    \item For each valid cut $(S_1,\overline{S_1})$ of $S$, there is at least one minimal cut in ${\cal H}$, say $({\cal A},{\overline{\cal A}})$, that represents it, that is, $\varphi^{-1}({\cal A})=S_1$.    
\end{enumerate}
\label{thm: skeleton-theorem}
\end{theorem}

There is an elegant 2-stage procedure to build 
the $t$-cactus storing all valid cuts of $S$ as stated in Theorem \ref{thm: skeleton-theorem}. In the first stage, a tree storing all vertices of $S$ is constructed. Any two vertices of $S$ belong to the same node in this tree if there is no valid cut that separates them. There may be some internal nodes in this tree that are empty. Each laminar valid cut is represented by a unique edge in this tree. 
In the second stage, a subset of empty nodes of degree at least 4 is picked suitably in this tree, and a cycle is implanted at each of these nodes. The resulting $t$-cactus is ${\cal H}$. 
%
%
Each crossing valid cut of $S$ is represented by a unique pair of non-adjacent edges of a cycle in ${\cal H}$. 
For the sake of completeness, we present this 2-stage construction of ${\cal H}$ in Appendix \ref{app: cactus-construction-for-valid-cuts}. 
%
%
%
%
However, one only needs to remember the three structural properties of ${\cal H}$, stated in the following lemma, to understand all the results of this article. 

\begin{lemma}
The cactus ${\cal H}$ as stated in Theorem \ref{thm: skeleton-theorem} satisfies the following structural properties.
\begin{enumerate}
\item[${\cal P}_1$:] 
The vertices of set $S$ mapped to any leaf node in ${\cal H}$ defines an indivisible valid cut of $S$. 
\item[${\cal P}_2$:] 
Each cycle in ${\cal H}$ has at least 4 edges. Each node of a cycle is empty and has exactly 3 edges incident on it -- two edges from the cycle and one tree edge.
\item[${\cal P}_3$:] 
 If a tree node in ${\cal H}$ is empty, its degree must be at least 3. In other words, a tree node with degree 2 must have one or more vertices from $S$ mapped to it.
\end{enumerate} 
\label{lem: structural-properties-of--cactus-H}
\end{lemma}

\begin{figure}[ht]
 \centering  \includegraphics[width=480pt]{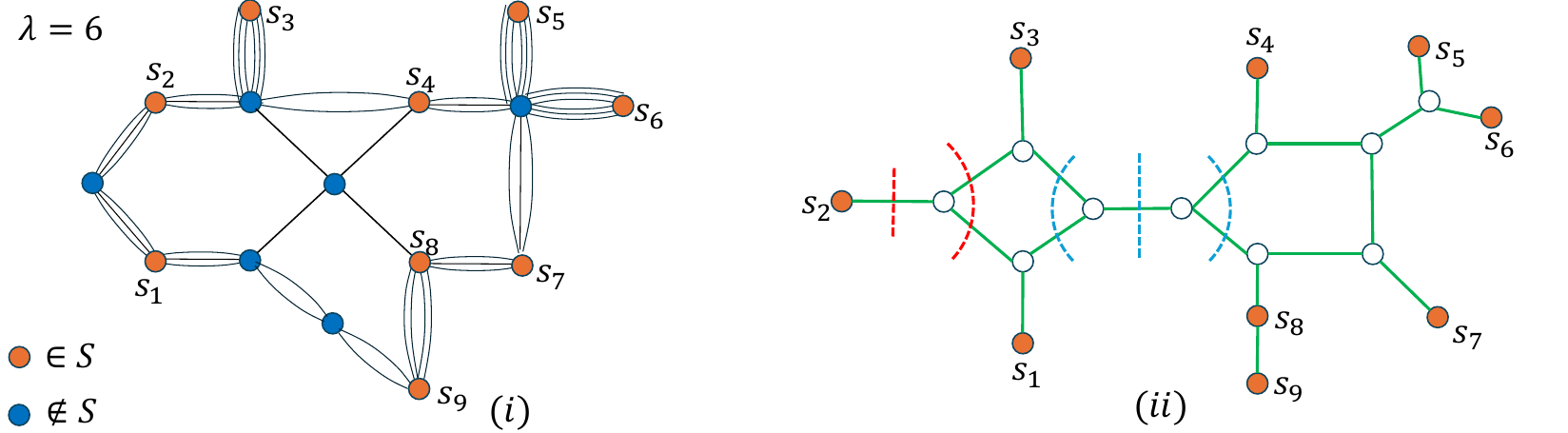} 
  \caption{$(i)$ Multigraph $G$,~($ii$) the $t$-cactus ${\cal H}$ storing all valid cuts of $S$.}
    \label{fig:subbunches}
\end{figure}

For the multi-graph $G$ in Figure \ref{fig:subbunches}($i$), Figure \ref{fig:subbunches}($ii$) demonstrates the $t$-cactus ${\cal H}$ satisfying properties stated in Lemma \ref{lem: structural-properties-of--cactus-H}. 
Although each crossing valid cut of $S$ is uniquely defined by a pair of non-adjacent edges of a cycle, the same is not necessarily true for a laminar valid cut as stated in the following lemma whose proof is immediate from Property ${\cal P}_2$.
\begin{lemma}
The valid cut of $S$ defined by a pair of adjacent edges of a cycle is the same as the valid cut defined by the tree edge incident on the node common to these edges.  
\label{lem: tree-cut-defines-same-S-partition-as-cut-defined-by-2-adjacent-edges-of-cycle}
\end{lemma}
\begin{note}
It follows from Lemma \ref{lem: tree-cut-defines-same-S-partition-as-cut-defined-by-2-adjacent-edges-of-cycle} that, henceforth, we need to consider only those minimal cuts of ${\cal H}$ that are defined by a tree edge or defined by a pair of non-adjacent edges of a cycle.
\label{note: only-tree-edge-cuts-and-transversal-cuts-of-a-cycle}
\end{note}
 It also follows from Property ${\cal P}_2$ that a tree edge in ${\cal H}$, say $e=(\nu,\nu')$, can be incident on at most 2 cycles. So it follows from Lemma \ref{lem: tree-cut-defines-same-S-partition-as-cut-defined-by-2-adjacent-edges-of-cycle} that the following two are the only cases of multiple cuts in ${\cal H}$ representing the same valid cut of $S$ -- (1)  A tree edge incident on exactly one cycle,~(2) A tree edge incident on exactly 2 cycles. 
Refer to Figure \ref{fig:subbunches}($ii$). Two red cuts in ${\cal H}$ represent the same valid cut of $S$, and so do the three cyan cuts in ${\cal H}$. We can thus state the following lemma based on the above discussion.
\begin{lemma}
For each valid cut of $S$, there are at least 1 and at most 3 minimal cuts in ${\cal H}$ that represent it.    
\label{lem: atleast-1-and-at-most-3-minimal-cuts}
\end{lemma}

Many cacti exist representing the set of all global mincuts or all valid cuts of $S$. 
Nagamochi and Kameda \cite{NK94} defined a canonical representation of a cactus, and proved that there is a unique cactus with this representation that stores all global mincuts of $G$. Recently
He, Huang, and Saranurak \cite{DBLP:conf/soda/HeHS24}
designed an algorithm for computing a canonical cactus that stores all valid cuts of $S$. Their algorithm performs  
${\cal O}({\mbox{polylog}} |S|)$ calls to an algorithm for $(s,t)$-maxflow. 
This cactus can be easily transformed into a $t$-cactus in ${\cal O}(m)$ time.
\subsection{Notations for valid cuts of \texorpdfstring{$S$}{S}}
\label{sec:notations-for-valid-cuts-of-S}
${\cal H}$ stores all valid cuts of $S$. 
In the following 2 sections, we shall analyse the $S$-mincuts associated with these valid cuts. So we introduce notations that will help in compactly representing the subset of $S$ defining any such valid cut. Let $e=(\nu,\mu)$ be a tree edge in ${\cal H}$. Removal of this edge splits ${\cal H}$ into two subcacti - one containing $\nu$ and another containing $\mu$. 
\begin{itemize} 
\item 
$S(\nu,e)$:~ the subset of $S$ mapped to the units of the subcactus containing $\nu$. 
\end{itemize}
Observe that $S(\mu,e)=S\backslash S(\nu,e)$. See Figure \ref{fig:S(nu,e)-and-extension}($i$). The valid cut defined by $S(\nu,e)$ is the opposite of the valid cut defined by $S(\mu,e)$. Each of them is defined by edge $e$. 

Let $O = \langle \nu_0,e_0,\nu_1,\ldots,\nu_{k-1},e_{k-1},\nu_{k}(=\nu_0)\rangle$ be a cycle in ${\cal H}$.
Removal of a pair of adjacent edges in the cycle, say $e_{i-1}$ and $e_i$, splits ${\cal H}$ into two subcacti. We use $S(\nu_i)$ to denote the subset of $S$ mapped to the units of the subcactus containing node $\nu_i$. See Figure \ref{fig:S(nu,e)-and-extension}($ii$). The following notation will help in compactly representing the subset of $S$ that defines the valid cut represented by a pair of edges in $O$.
\[ 
S(\nu_{p}, \nu_{q}) = 
       S(\nu_{p}) \cup  S(\nu_{p+1}) \cup \ldots \cup S(\nu_{q})~~~~~~~~~~~\mbox{(here $+$ is addition {\em mod} $k$)}
\]
Observe that  $S(\nu_{p},\nu_{q})=S\backslash S(\nu_{q+1},\nu_{p-1})$ and $S(\nu_p,\nu_p)=S(\nu_p)$. 
The pair of edges $e_i$ and $e_j$ define two valid cuts -- one defined by $S(\nu_{i+1},\nu_j)$ and another, which is its opposite, defined by $S(\nu_{j+1},\nu_i)$. Figure \ref{fig:S(nu,e)-and-extension}($iii$) illustrates these valid cuts of $S$.
\begin{figure}[H]
 \centering  \includegraphics[width=400pt]{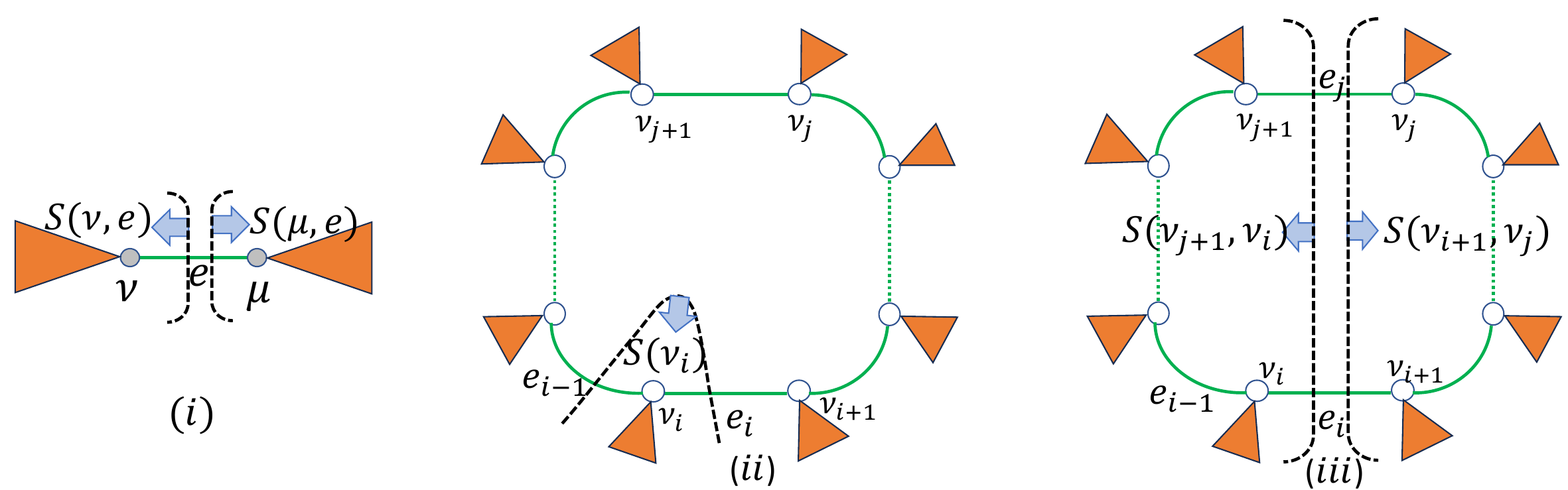} 
  \caption{($i$) $S(\nu,e)$,~($ii$) $S(\nu_i)$,~($iii$) $S(\nu_{i+1},\nu_j)$ and $S(\nu_{j+1},\nu_i)$} 
    \label{fig:S(nu,e)-and-extension}
\end{figure}


\section{Compact structure storing all \texorpdfstring{$S$}{S}-Mincuts}
\label{sec:CompactStructure_for_S-mincuts}
 Let $x$ be any arbitrary vertex from $S$. Observe that any $S$-mincut is also a $(x,y)$-mincut for some suitably chosen vertex $y\in S\backslash\{x\}$, and hence appears as a transversal cut in the strip ${\overrightarrow{\cal D}}_{x,y}$. Therefore, a set of $O(|S|)$ strips, each with source vertex $x$, is sufficient to store and characterize all $S$-mincuts. We store each strip implicitly by keeping only the mapping from $V$ to vertices of the strip. This mapping can be stored in ${\cal O}(n)$ space only. It follows from Lemma \ref{lem: G-from-Gu} that the corresponding strip can be computed in just ${\cal O}(m)$ time using the adjacency lists of $G$ and this mapping. So we can state the following lemma. 
\begin{lemma}
There is a data structure of ${\cal O}(|S|n)$ space that stores and characterizes all $S$-mincuts. 
\label{lem: simple-sm-structure}
\end{lemma}

We now begin the journey of designing a more compact structure.
Let ${\cal F}$ denote the quotient graph of $G$ induced by all $S$-mincuts. We call the vertices of graph ${\cal F}$ as the units of ${\cal F}$; let $\Phi$ be the mapping from the set of vertices of $G$ to the set of units of ${\cal F}$. For any $x,y\in V$, $\Phi(x)=\Phi(y)$  if and only if there is no $S$-mincut separating $x$ and $y$. It is a simple exercise based on the discussion preceding Lemma \ref{lem: simple-sm-structure} that mapping $\Phi$, and so the quotient graph ${\cal F}$, can be computed in ${\cal O}(|S|)$ max-flow computations only.
A unit $\omega$ of ${\cal F}$ is called a $S$-unit if there is at least one vertex from set $S$ mapped to it, that is, $\Phi^{-1}(\omega)\cap S\not=\emptyset$. Otherwise, we call $\omega$ a non-$S$-unit.

In the pursuit of a compact structure storing all $S$-mincuts, it seems quite natural to explore the relationship between the units of ${\cal F}$ and the nodes of ${\cal H}$. This is because ${\mathcal H}$ compactly stores all valid cuts of $S$. 
Let $u$ and $v$ be any two vertices from $S$. Observe that $u$ and $v$ belong to the same node of ${\cal H}$ if and only if they belong to the same unit in ${\cal F}$. So we can define a mapping, denoted by $\pi$ henceforth, from the set of $S$-units of ${\cal F}$ to the set of nodes of ${\cal H}$ as follows -- each $S$-unit is mapped to the node in ${\cal H}$ that stores all those vertices from $S$ that are mapped to the unit. 
The most intriguing question is how to map the non-$S$-units of ${\cal F}$. To find an answer to this question, observe that Theorem \ref{thm: skeleton-theorem} suggests an alternate way to compactly represent all $S$-mincuts  -- For each minimal cut of ${\cal H}$, construct the strip, as follows, that stores all $S$-mincuts associated with the minimal cut.

\begin{definition}[Strip associated with a minimal cut of ${\cal H}$]
Let a minimal cut of ${\cal H}$ split it into two subcacti ${\cal H}_1$ and ${\cal H}_2$. Let $S_1$ and $S_2$ be the sets of $S$-units belonging to ${\cal H}_1$ and ${\cal H}_2$ respectively. 
%
%
The strip associated with the minimal cut is
the strip ${\cal D}_{S_1,S_2}$ obtained after contracting $S_1$ and $S_2$ into the source vertex and the sink vertex respectively.
\label{def: strip associated with minima cut}
\end{definition}

Observe that the set of strips associated with all minimal cuts of ${\cal H}$ stores and characterizes all $S$-mincuts. In particular, each $S$-mincut will appear as a transversal cut in the strip corresponding to the valid cut it defines, which in turn, is represented by at least 1 and at most 3 minimal cuts of ${\mathcal H}$ as stated in Lemma \ref{lem: atleast-1-and-at-most-3-minimal-cuts}. Each $S$-unit will belong to the terminal of the strip depending upon which side of the cut it lies. 
Each non-$S$-unit will either be merged in one of the terminals of the strip or will appear in a non-terminal of the strip.
This motivates us to introduce the following terminologies that will help us understand the role of the non-$S$-units in ${\cal H}$.

\begin{definition}[a non-$S$-unit distinguished by a cut]
Let $\omega$ be any non-$S$-unit in ${\mathcal F}$. $\omega$ is said to be {\em distinguished} by a minimal cut of ${\cal H}$ if $\omega$ does not belong to any terminal of the strip associated with the minimal cut. 
\label{def: unit-distinguished-by-a-cut}
\end{definition}

Based on the above definition, we can classify any non-$S$-unit as follows.
\begin{definition}[terminal unit or stretched unit]
Let $\omega$ be any non-$S$-unit in ${\mathcal F}$. $\omega$ is called a {\em terminal} unit if no minimal cut of ${\mathcal H}$ distinguishes it; otherwise we call $\omega$ a {\em stretched} unit. 
\label{def: terminal or stretched unit}
\end{definition}

Let $\omega$ be any non-$S$-unit in ${\cal F}$. Consider 
the strip ${\cal D}_{S_1,S_2}$ associated with any minimal cut in ${\cal H}$.
Observe that the source and the sink in this strip define the cuts $N(S_1,\overline{S_1})$ and $N(\overline{S_1},S_1)$ respectively. This observation and Definitions \ref{def: unit-distinguished-by-a-cut} and \ref{def: terminal or stretched unit} lead to the following lemma that depicts how any terminal unit and a stretched unit appear in the strip associated with a minimal cut.

\begin{lemma}
Let $(S_1,\overline{S_1})$ be the valid cut of $S$ associated with a minimal cut in ${\cal H}$, and let $\omega \in {\cal F}$.
\begin{enumerate}
  \item[Case] `$\omega$ is a terminal unit':~  
  $\omega$ lies inside $N(S_1,\overline{S_1})$ or $\omega$ lies inside $N(\overline{S_1},S_1)$.
  \item[Case] `$\omega$ is a stretched unit':~ $\omega$ 
  is distinguished by the minimal cut if and only if $\omega$ lies outside each of $N(S_1,\overline{S_1})$ and $N(\overline{S_1},S_1)$, equivalently $\omega$ lies inside each of $F(S_1,\overline{S_1})$ and $F(\overline{S_1},S_1)$. 
\end{enumerate}
\label{lem: alternate-def-distinguished-by-a-minimal-cut}
\end{lemma} 

\begin{note} We may extend the domain of terminal units to include all $S$-units as well. This is because each $S$-unit belongs to one of the terminals of a strip, and hence is not distinguished by any minimal cut of ${\cal H}$. So we have 2 kinds of classifications of the units of ${\cal F}$:~ ($i$)~$S$-units and non-$S$-units, ($ii$)~terminal units and stretched units.
\label{note:steiner_unit=terminal_unit}
\end{note}
Let us consider any stretched unit. While it may be distinguished by multiple minimal cuts of ${\cal H}$, certain combinations of minimal cuts are forbidden
as shown in the following lemma.
\begin{lemma} Let $\omega \in {\cal F}$ be any stretched unit.
\begin{enumerate}
\item 
    $\omega$ can be distinguished by at most 2 laminar cuts of a cycle in ${\cal H}$.
\item 
    $\omega$ cannot be distinguished by any pair of crossing cuts in ${\cal H}$.
\end{enumerate}
\label{lem: forbidden-combination-of-cuts}
\end{lemma}
\begin{proof}
      
For assertion (1), let $O = \langle \nu_0,e_0,\nu_1\ldots \nu_{k-1},e_{k-1},\nu_{k}(=\nu_0)\rangle$ be a cycle in ${\cal H}$. Any laminar cut in the cycle is defined by a pair of adjacent edges. Suppose there are three cuts defined by pairs of adjacent edges $(e_{i-1},e_i)$, $(e_{j-1},e_j)$, and $(e_{\ell-1},e_\ell)$ that distinguish $\omega$. Observe that $S_{\nu_i}, S_{\nu_j}$, and $S_{\nu_{\ell}}$ are pairwise disjoint. Let $C_1,C_2$, and $C_3$ be the subsets of vertices that define $F(S_{\nu_i},\overline{S_{\nu_i}})$, $F(S_{\nu_j},\overline{S_{\nu_j}})$, and  $F(S_{\nu_\ell},\overline{S_{\nu_\ell}})$ respectively. It follows from Lemma \ref{lem: alternate-def-distinguished-by-a-minimal-cut} that $\omega\in C_1\cap C_2\cap C_3$. However, notice that the subsets of $S$ present in $C_1,C_2$, and $C_3$ are pairwise disjoint. So Lemma \ref{lem: corollary of four point lemma} implies that $C_1\cap C_2\cap C_3=\emptyset$. This contradicts $\omega\in C_1\cap C_2\cap C_3$.

For assertion $(2)$, suppose there exists a pair of crossing cuts in ${\cal H}$ that distinguish $\omega$. Let $(S_1,\overline{S_1})$ and $(S_2,\overline{S_2})$ be the valid cuts corresponding to these cuts. Let $(A,\overline{A})=F(S_1,\overline{S_1})$, and $(A',\overline{A'}) = F(\overline{S_1},S_1)$. Lemma \ref{lem: alternate-def-distinguished-by-a-minimal-cut} implies that  $\omega \in A$ and $\omega \in A'$. 
Similarly, let $(B,\overline{B})=F(S_2,\overline{S_2})$ and $(B',\overline{B'})=F(\overline{S_2},S_2)$. Again Lemma \ref{lem: alternate-def-distinguished-by-a-minimal-cut} implies that $\omega \in B$ and $\omega \in B'$.
We define $C_1,C_2$, and $C_3$ as follows.     
\[  C_1 = A'\cap B,~~~~~~~~~~    C_2 = A\cap B,~~~~~~~~~~ C_3 = A\cap B'.\]   

Observe that $\omega$ is present in each of $C_1,C_2,$ and $C_3$. 
So $\omega\in C_1\cap C_2\cap C_3$.  
Moreover, each of them defines a $S$-cut since the cuts in ${\cal H}$ corresponding to $(S_1,\overline{S_1})$ and $(S_2,\overline{S_2})$ form a crossing pair of cuts. So it follows from the submodularity of cuts (Lemma \ref{lem: cor-submodularity-of-cuts})  that each of them defines a $S$-mincut. The subsets of $S$  present in each of $C_1,C_2,C_3$ are $S_2\setminus S_1, S_1 \cap S_2, S_1\setminus S_2$ respectively, and it is easy to verify that they are pairwise disjoint. So Lemma \ref{lem: corollary of four point lemma} implies that $C_1\cap C_2\cap C_3=\emptyset$. This contradicts $\omega\in C_1\cap C_2\cap C_3$. 
\end{proof}

\subsection{Mapping a terminal non-$S$-unit in \texorpdfstring{${\cal H}$}{skeleton}}
Recall that each $S$-unit of ${\cal F}$ is mapped to a unique node in ${\cal H}$. Interestingly, we can also uniquely map each terminal non-$S$-unit to an empty node in ${\cal H}$  as shown by the following theorem. 
\begin{theorem}
    For any terminal non-$S$-unit $\omega$, there exists a unique tree node $\nu$ in ${\cal H}$ with the following property. Let ${\mathcal H}_1$ and ${\mathcal H}_2$ be the two subcacti defined by any minimal cut of ${\mathcal H}$, and let $(S_1,\overline{S_1})$ be the corresponding valid cut of $S$ such that the vertices of $S_1$ are mapped to the nodes of ${\mathcal H}_1$, 
\begin{itemize}
    \item 
If $\nu$ belongs to ${\mathcal H}_1$, $\omega$ belongs to the source node in the strip corresponding to $(S_1,\overline{S_1})$.
\item If $\nu$ belongs to ${\mathcal H}_2$, $\omega$ belongs to the sink node in the strip corresponding to $(S_1,\overline{S_1})$.
\end{itemize}
\label{thm: mapping-a-terminal-unit}
\end{theorem}



We now provide the overview of an algorithmic proof of Theorem \ref{thm: mapping-a-terminal-unit}. 
%
%
Definition \ref{def: terminal or stretched unit} of a terminal unit suggests that 
we can assign a direction to each tree edge depending upon the side of the cut on which $\omega$ is lying. For each tree node, we show that there will be at most one outgoing edge. 
For each cycle, we show that there exists a unique tree edge incident on it which will be directed {\em outward} and the rest will be directed {\em inward} only. So we can effectively compress each cycle into a single node and this node will have exactly one outgoing edge. Traversing this directed tree starting from any leaf node, we are bound to reach a unique internal node with no outgoing edge. We show that this node will be the node $\nu$ for $\omega$ in Theorem \ref{thm: mapping-a-terminal-unit}. 

%
With the overview given above, we begin with assigning {\em direction} to each tree edge in ${\mathcal H}$ as follows. 
Let $e=(\nu,\mu)$ be any tree edge, and let $S'=S(\nu,e)$. 
It follows from Lemma \ref{lem: alternate-def-distinguished-by-a-minimal-cut} that either $\omega$ lies inside $N(S',\overline{S'})$ or $\omega$ lies inside $N(\overline{S'}, S')$.  
If $\omega$ lies inside $N(S',\overline{S'})$, we direct $e$ from $\mu$ to $\nu$; otherwise we direct $e$ from $\nu$ to $\mu$. 
The following lemma states that the direction of an edge incident on a leaf node in ${\cal H}$ will always be directed away from the leaf node.
\begin{lemma}
If $\mu$ is a leaf node in ${\cal H}$, and $\nu$ is its neighbor, $(\mu,\nu)$ is directed from $\mu$ to $\nu$.
\end{lemma}
\begin{proof}
Let $e=(\mu,\nu)$, and let $S_1=S(\mu,e)$. It is given that $\mu$ is a leaf node in ${\mathcal H}$. So Property ${\cal P}_1$ of ${\cal H}$, as stated in Lemma \ref{lem: structural-properties-of--cactus-H}, implies that the $S$-unit mapped to $\mu$ defines an indivisible valid cut of $S$. So it follows from Lemma \ref{lemma: loose-tight-mincut-subset-property}(2) that there does not exist any $S$-mincut that splits $N(S_1,\overline{S_1})$. 
Note that there exists a $S$-mincut that separates $\mu$ and $\omega$ since both $\mu$ and $\omega$ are distinct units in ${\mathcal F}$. 
Therefore, $\omega$ must lie outside $N(S_1,\overline{S_1})$. This fact in conjunction with Lemma \ref{lem: alternate-def-distinguished-by-a-minimal-cut} implies that $\omega$ must lie inside $N(\overline{S_1},S_1)$. Hence $e$ is directed from $\mu$ to $\nu$.
\end{proof}

We now state a lemma that will be used to determine the direction of any other tree edge in ${\cal H}$.

\begin{lemma}
 Let $(S_1,\overline{S_1})$ be a valid cut of $S$, and let 
 $\omega$ lie inside $N(S_1,\overline{S_1})$. 
 Let $e=(\nu,\mu)$ be any tree edge in ${\mathcal H}$, and $S_2=S(\nu,e)$. If $S_1 \subsetneq S_2$,
 $e$ is directed from $\mu$ to $\nu$.
\label{lem: dominance-defines-direction}
\end{lemma}
\begin{proof}
It follows from Lemma \ref{lemma: loose-tight-mincut-subset-property}(1) that $N(S_2,\overline{S_2})$ dominates
$N(S_1,\overline{S_1})$ since $S_1\subsetneq S_2$. Therefore, $\omega$ lies inside $N(S_2,\overline{S_2})$ since it is given that $\omega$ lies inside $N(S_1,\overline{S_1})$. So $e$ is directed from $\mu$ to $\nu$.
\end{proof}

The following lemma states that any tree node in ${\cal H}$ cannot have multiple outgoing edges in ${\mathcal H}$.

\begin{lemma}
There will be at most one outgoing edge for any tree node in ${\mathcal H}$.
\label{lem: internal node in skeleton for terminal units}
\end{lemma}
\begin{proof}
Let $\mu$ be any tree node in ${\mathcal H}$. Let $e_1=(\mu,\nu_1)$ be a tree edge incident on $\mu$ and it is directed outward from $\mu$. Let $e_2=(\mu,\nu_2)$ be any other edge incident on $\mu$. Let $S_1=S(\nu_1,e_1)$, and $S_2=S(\mu,e_2)$. Given that 
$e_1$ is directed from $\mu$ to $\nu_1$, $\omega$ belongs to $N(S_1,\overline{S_1})$. 
$\mu$ is a tree node, so it follows from Property ${\cal P}_3$ of ${\cal H}$, as stated in Lemma \ref{lem: structural-properties-of--cactus-H}, that $S_1\subsetneq S_2$. Hence it follows from Lemma \ref{lem: dominance-defines-direction} that $e_2$ must be directed from $\nu_2$ to $\mu$.
%
\end{proof}

Let $O = \langle \nu_0,e_0,\nu_1,\ldots,\nu_{k-1},e_{k-1},\nu_{k}(=\nu_0)\rangle$ be any cycle in ${\mathcal H}$. 
It follows from property ${\cal P}_2$ of ${\cal H}$, as stated in Lemma \ref{lem: structural-properties-of--cactus-H}), that each node of $O$ has exactly one tree edge incident on it. We now analyze how these tree edges are assigned directions. We recommend the reader to recall the notations for valid cuts of $S$ described in Section \ref{sec:notations-for-valid-cuts-of-S}. For the sake of brevity, we shall use $S_i$ to denote $S(\nu_i)$ and use $S_{i,j}$ to denote $S(\nu_i,\nu_{j})$ for any $0\le i,j \le k-1$. 
%
%
\begin{lemma}
There is a unique $0\le q\le k-1$ such that $\omega$ lies inside $N(S_{q},\overline{S_q})$.
\label{lem: the-cut-defined-by-adjacent-edges}
\end{lemma}
\begin{proof} 
Let us begin with any crossing cut in the cycle $O$. Without loss of generality, assume that it is defined by $e_0$ and $e_\ell$ with $1<\ell<k-1$. It follows from Lemma
\ref{lem: alternate-def-distinguished-by-a-minimal-cut} that $\omega$ lies inside exactly one of $N(S_{1,\ell},S_{\ell+1,0})$ and  $N(S_{\ell+1,0},S_{1,\ell})$. Without loss of generality, suppose $\omega$ lies inside $N(S_{1,\ell},S_{\ell+1,0})$. So $\omega$ does not lie inside $N(S_{\ell+1,0},S_{1,\ell})$. For each $\ell<p\le k$, observe that $S_p \subsetneq S_{\ell+1,0}$; so using Lemma \ref{lemma: loose-tight-mincut-subset-property}(1), $\omega$ does not lie inside $N(S_p,\overline{S_p})$ as well. 
We shall now gradually {\em widen} the range of $p$.

Let $(A',\overline{A'}) = N(S_{\ell+1,0},S_{1,\ell})$. So  $\omega\notin A'$.
Pick a cut in ${\mathcal H}$ that crosses the cut defined by $e_0$ and $e_\ell$. Let it be defined by $e_i$ and $e_j$ with $1\le i<\ell$ and $\ell<j\le k-1$. Let $(B,\overline{B}) = N(S_{j+1,i},S_{i+1,j})$ and 
$(B',\overline{B'}) = N(S_{i+1,j},S_{j+1,i})$. Lemma \ref{lem: alternate-def-distinguished-by-a-minimal-cut} implies that $\omega$ lies inside exactly one of $(B,\overline{B})$ and $(B',\overline{B'})$. 
Without loss of generality, assume $\omega$ lies inside  $(B,\overline{B})$; so $\omega\notin B'$. Thus $\omega\notin A'\cup B'$. The submodularity of cuts (Lemma \ref{lem: cor-submodularity-of-cuts}) implies that $A'\cup B'$ defines a $S$-mincut. This $S$-mincut belongs to the bunch defined by $e_0$ and $e_i$. So, $\omega\notin A'\cup B'$ implies that $\omega$ does not lie inside $N(S_{i+1,0},S_{1,i})$. 
For each $i<p\le k$, observe that $S_p \subsetneq S_{i+1,0}$; so using Lemma \ref{lemma: loose-tight-mincut-subset-property}(1), $\omega$ does not lie inside $N(S_p,\overline{S_p})$ as well. 
%
%
So the range of $p$ has widened. Moreover, given that $\omega$ does not lie inside $N(S_{i+1,0},S_{1,i})$, Lemma \ref{lem: alternate-def-distinguished-by-a-minimal-cut} implies that $\omega$ lies inside $N(S_{1,i},S_{i+1,0})$. If $i=1$, this would imply that $\omega\in N(S_1,\overline{S_{1}})$; thus $q=1$,  and we are done. Otherwise, $e_0$ and $e_i$ define a crossing cut; so, we repeat the above procedure. It is easy to observe that after a finite number of repetitions, we are bound to find the unique node $\nu_q$ in $O$ such that $\omega$ lies inside 
$N(S_q,\overline{S_q})$ and outside $N(S_p,\overline{S_p})$ for each $p\not=q$.
\end{proof}

\begin{lemma}
There is exactly one node of the cycle $O$ such that the tree edge incident on it will be directed away from it.
\label{lem: cycle node}
\end{lemma}
\begin{proof} Using Lemma \ref{lem: the-cut-defined-by-adjacent-edges}, let $\nu_q$ with $0\le q\le k-1$ be the unique node of the cycle such that $\omega$ lies inside $N(S_q,\overline{S_q})$.
Let $e=(\nu_q,\mu)$ be the tree edge incident on $\nu_q$, and let $S'=S(\mu,e)$.
It follows from Lemma \ref{lem: tree-cut-defines-same-S-partition-as-cut-defined-by-2-adjacent-edges-of-cycle} that the bunch associated with the two cycle-edges $e_{q-1}$ and $e_q$ incident on $\nu_q$ is the same as the bunch associated with the tree edge $e$. Therefore, $S'$ is the same as $S_q$. 
Hence $\omega$ lies inside $N(S',\overline{S'})$. So $e$ will be directed from $\nu_q$ to $\mu$.

Now consider any node $\nu_j$ in the cycle with $j\not=q$. Let $e=(\nu_j,\mu')$ be the tree edge incident on $\nu_j$. Let $S''=S(\nu_j,e)$. Observe that $S'\subsetneq S''$. So it follows from Lemma \ref{lem: dominance-defines-direction} that $e$ must be directed from $\mu'$ to $\nu_j$. 
This completes the proof.
\end{proof}

It follows from Lemma \ref{lem: cycle node} that each cycle in ${\mathcal H}$ has precisely one tree edge incident on it which is directed away from it. So we may as well compress each cycle of ${\mathcal H}$ into a single node. Since ${\cal H}$ is a $t$-cactus, what we get will be a tree with the following properties. 
\begin{enumerate} 
\item The edge from each leaf node is directed away from the leaf node.
\item Each node representing a compressed cycle has exactly one outgoing edge.
\item Every other internal node has at most one outgoing edge. 
\end{enumerate}

In this tree, if we start traversing from any node and keep following the outgoing edge of each node, we are bound to stop eventually at a node with no outgoing edge. This is because there is no cycle in the tree. Due to Properties (1) and (2) stated above, this node can neither be a compressed cycle node nor a leaf node. So, there exists an internal tree node in $\cal H$ with no outgoing edge. 
Moreover, there will be exactly one such node in ${\cal H}$ as follows. Suppose there exist two nodes, say $\nu$ and $\mu$, with no outgoing edge. Consider the path, ignoring directions, in the tree between $\nu$ and $\mu$. 
Each intermediate node on this path is either an internal tree node or a compressed cycle node. 
Observe that the first edge and the last edge of this path are in opposite directions. So there must exist a node on this path with two outgoing edges -- a contradiction to Lemma \ref{lem: internal node in skeleton for terminal units} if the node is a tree node or contradiction to Property (2) if the node is a compressed cycle node. Hence there exists a unique tree node $\nu$ in ${\mathcal H}$ with no outgoing edge. We can thus {\em project} $\omega$ to $\nu$. Observe that each pair of units in ${\cal F}$ are separated by at least one $S$-mincut. Therefore, if $\omega$ is projected to $\nu$, there cannot be any other terminal unit projected to $\nu$. 
This completes the proof of Theorem \ref{thm: mapping-a-terminal-unit}.

\subsection{Mapping a stretched unit in \texorpdfstring{${\cal H}$}{skeleton}}
We first state formally the two basic properties satisfied by a stretched unit. The proof for these properties given by Dinitz and Vainshtein in their article \cite{DBLP:journals/siamcomp/DinitzV00} works irrespective of whether the capacity of a $S$-mincut is odd or even. While their proof of the 1st property ({\textsc{Distinctness}}) is based on the submodularity of cuts, and we reproduce the proof below; their proof of the 2nd property ({\textsc{unique-inherent-partition}}) is based on the theory of locally orientable graphs. Here, we give an alternate and much shorter proof of the 2nd property which is based on the submodularity of cuts.

\begin{figure}[ht]
 \centering  \includegraphics[width=500pt]{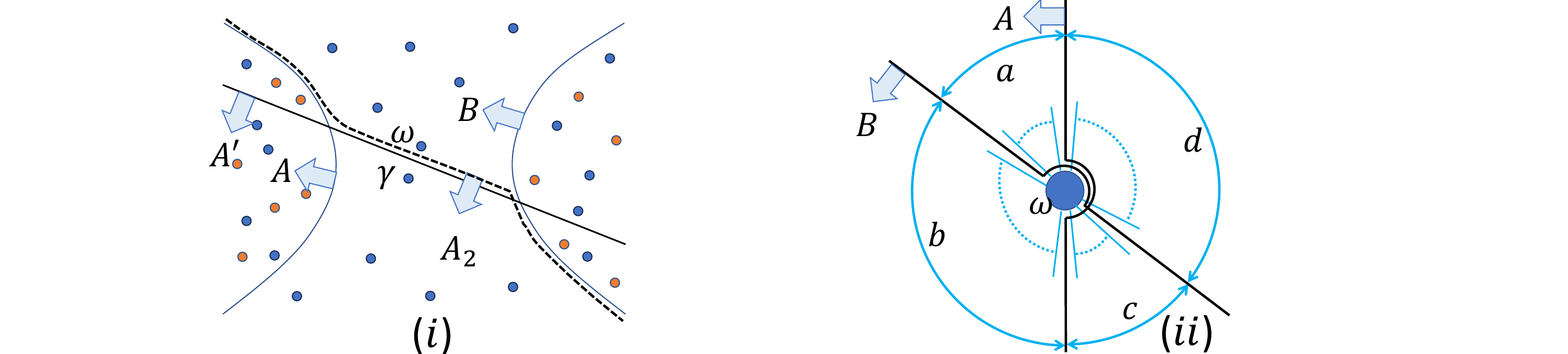} 
  \caption{($i$) Cut defined by $A_2$ separates $\omega$ and $\gamma$ and belongs to the same bunch as that of $A$ and $B$~ 
  ($ii$) the inherent partitions of $E(\omega)$ in the two strips.}
    \label{fig:2-properties}
\end{figure}

\begin{theorem}[{\textsc{Distinctness}}]
(Lemma 5.2 in \cite{DBLP:journals/siamcomp/DinitzV00})
If, in the strip corresponding to a valid cut of $S$, a stretched unit $\omega$ does not belong to any terminal, $\omega$ appears as a distinct non-terminal in the strip. 
\label{thm:distinct-nodes-in-strip}
\end{theorem}
\begin{proof}
Consider any minimal cut of ${\cal H}$ that distinguishes $\omega$. Let $(S_1,\overline{S_1})$ be the valid cut of $S$ corresponding to this minimal cut. Let 
$\gamma$ be any other stretched unit distinguished by the same minimal cut (if $\gamma$ does not exist, the lemma holds trivially). 
%
%
Since $\omega$ and $\gamma$ are distinct units in ${\cal F}$, there must exist a 
$S$-mincut, say $(A',\overline{A'})$, that separates $\gamma$ and $\omega$. Without loss of generality, assume $\gamma\in A'$ and $\omega\notin A'$. If this $S$-mincut belongs to the bunch associated with $(S_1,\overline{S_1})$, we are done. Otherwise, we shall use this $S$-mincut to construct another $S$-mincut that belongs to the bunch associated with $(S_1,\overline{S_1})$ and separates $\gamma$ and $\omega$ as follows. This would establish the lemma.

Since $A'$ divides $S$, therefore, 
without loss of generality, we can assume that $A' \cap S_1 \not=\emptyset$ and $\overline{A'}\cap \overline{S_1}\not=\emptyset$. 
Let $(A,\overline{A})=N(S_1,\overline{S_1})$ and $(B,\overline{B})=F(S_1,\overline{S_1})$.
Figure \ref{fig:2-properties}($i$) illustrates the cuts defined by $A,B$, and $A'$. 
Since $\omega$ and $\gamma$ don't belong to the terminals of the strip associated with $(S_1,\overline{S_1})$, so $\gamma,\omega \notin A$ and $\gamma,\omega \in B$. 
The cuts defined by $A'\cap B$ as well as $A'\cup B$ are $S$-cuts.
Hence, using the submodularity of cuts (Lemma \ref{lem: cor-submodularity-of-cuts}), $A_1 = A'\cap B$ defines a $S$-mincut. Observe that $\gamma \in A_1$ and $\omega\notin A_1$. The cuts defined by $A\cap A_1$ and $A\cup A_1$ are also $S$-cuts. So using the submodularity of cuts  (Lemma \ref{lem: cor-submodularity-of-cuts}) again, we can infer that $A_2=A\cup A_1=A\cup (A'\cap B)$ defines a $S$-mincut. Figure \ref{fig:2-properties}($i$) illustrates the cut defined by $A_2$.  
Observe that $\gamma\in A_2$ and $\omega \notin A_2$. Hence, $(A_2,\overline{A_2})$ separates $\gamma$ and $\omega$. 
It follows from the construction that $A\subsetneq A_2\subsetneq B$. So $(A_2,\overline{A_2})$ belongs to the bunch associated with $(S_1,\overline{S_1})$. 
\end{proof}

Theorem \ref{thm:distinct-nodes-in-strip} suggests a neat way to represent a stretched unit $\omega$ in ${\cal H}$ as follows. Map $\omega$ to each minimal cut of ${\cal H}$ that distinguishes it. We now state the 2nd property satisfied by $\omega$.

\begin{theorem}[\textsc{unique-inherent-partition}](Lemma 5.3 in \cite{DBLP:journals/siamcomp/DinitzV00})
 Let $\omega\in {\cal F}$ be distinguished by two distinct bunches. In the strips corresponding to these bunches, the inherent partitions of $\omega$ coincide.  
\label{thm:inherent-partition}
\end{theorem}

\begin{proof}
  It follows from Lemma \ref{lem: forbidden-combination-of-cuts}(2)
  that the cuts defining these 2 bunches must be laminar cuts in ${\cal H}$. So there exist subsets $S_1,S_2\subset S$ that define these bunches such that $S_1\subset S_2$. 
  Let strips ${\cal W}_1$ and ${\cal W}_2$ respectively be the strips corresponding to the valid cut $(S_1,\overline{S_1})$ and $(S_2,\overline{S_2})$.
  It follows from Theorem \ref{thm:distinct-nodes-in-strip} that $\omega$ appears as a distinct non-terminal in each of these 2 strips.  
  Let $(A,\overline{A})$ be the $S$-mincut defined by ${\cal R}_s(\omega)$ in strip ${\cal W}_1$, and $(B,\overline{B})$ be the $S$-mincut defined by ${\cal R}_s(\omega)$ in strip ${\cal W}_2$. 
  It follows from Lemma \ref{lem:R_s(x)_is_(s,t)-mincut} that $(A,\overline{A})$ and $(B,\overline{B})$ define $(s,t)$-mincuts in the respective strips, and hence are $S$-mincuts. Suppose the 2-sides of the inherent partition of $\omega$ in ${\cal W}_1$ do not coincide with the 2-sides of the inherent partition of $\omega$ in ${\cal W}_2$. This implies that there exists a partition of $E(\omega)$ (the set of all edges incident on $\omega$) into 4 sets of cardinality $a,b,c,d>0$ such that (1) the number of edges contributed by $\omega$ to the $S$-mincut $(A,\overline{A})$ is $d+c$, and (2) the number of edges contributed by $\omega$ to the $S$-mincut $(B,\overline{B})$ is $a+d$. Refer to Figure \ref{fig:2-properties}($ii$). 
  
  Recall that there is a balanced partition of all edges incident on $\omega$ in each of the two strips. Therefore,  $a+b=c+d=a+d=b+c$. So $a+d-c=b$. Since $c>0$, the following inequality holds. 
  \begin{equation}        
  a+d+c>b       
  \label{Eq:a+d+c>b}
  \end{equation}
    It follows from the submodularity of cuts (Lemma \ref{lem: cor-submodularity-of-cuts}) that $A\cap B$ and $A\cup B$ define $S$-mincuts. Notice that $\omega\in A\cap B$, and the number of edges contributed by $\omega$ to the $S$-mincut defined by $A\cap B$ is $a+d+c$. Since $\omega$ is a non-$S$-unit, moving only $\omega$ to the other side of this cut gives us another $S$-cut. However, the edges contributed by $\omega$ in this cut is $b$ which is strictly less than $a+d+c$ (Inequality \ref{Eq:a+d+c>b}). So this
  $S$-cut has capacity strictly less than that of the $S$-mincut defined by $A \cap B$ -- a contradiction. 
\end{proof}  

Theorem \ref{thm:distinct-nodes-in-strip}  and Theorem \ref{thm:inherent-partition} together reveal a close relation between ${\cal F}$ and the strips associated with all minimal cuts of ${\cal H}$ as follows. 
Let $(S_1,\overline{S_1})$ be the valid cut of $S$ associated with a given minimal cut of ${\cal H}$. Let $G'$ be the quotient graph of ${\cal F}$ obtained by contracting $N(S_1,\overline{S_1})$ into a node and contracting $N(\overline{S_1},S_1)$ into another node. Theorem \ref{thm:distinct-nodes-in-strip} implies that every stretched unit, say $\omega$, that survives in $G'$ will appear as a distinct non-terminal in the strip associated with $(S_1,\overline{S_1})$. Theorem \ref{thm:inherent-partition} shows that 
the inherent partition of $\omega$ in this strip will be the same (upto reversal) as in any other strip in which $\omega$ appears as a non-terminal. So, $(1)$ the strip associated with every minimal cut of ${\cal H}$ is implicitly present in ${\cal F}$, and $(2)$ every stretched unit of ${\cal F}$ appears identically (same inherent partition) in all the strips that distinguish it. It is because of these properties that ${\cal F}$ can be viewed as a {\em multi-terminal} DAG. 


Suppose we store, for each stretched unit $\omega$, all those minimal cuts in ${\cal H}$ that distinguish it. This would lead to a data structure for storing and characterizing all $S$-mincuts occupying $O(n|S|^2)$ space. This is because there can be $\Theta(n)$ stretched units in ${\cal F}$ and at most $|S|\choose 2$ minimal cuts in ${\cal H}$. But, this bound can be potentially even worse than the ${\cal O}(n|S|)$ size of the obvious data structure as stated in Lemma \ref{lem: simple-sm-structure}. Interestingly and very amazingly, the set of minimal cuts that distinguish any stretched unit has an elegant structure as stated in the following theorem. 

\begin{theorem}
For each stretched unit $\omega$, there exists a proper path $P(\omega)$ in skeleton ${\cal H}$ with the following properties satisfied by each structural edge $e\in P(\omega)$.
 \begin{enumerate}
 \item 
 if $e$ is a tree edge, $\omega$ is distinguished by the minimal cut defined by $e$. 
 \item 
 if $e$ is a cycle edge, then $\omega$ is distinguished by each minimal cut defined by $e$ and any other edge of the cycle. 
\end{enumerate}
Furthermore, the minimal cuts stated in (1) and (2) are the only minimal cuts that distinguish $\omega$. Thus $P(\omega)$ captures all minimal cuts that distinguish $\omega$. In other words, $\omega$ is {\em projected} onto each edge of the path $P(\omega)$. 
\label{thm: projection-of-stretched-unit-to-proper-path}
\end{theorem}

Recall that $\pi$ was defined as an injective mapping from the set of $S$-units of ${\cal F}$ to the set of nodes in ${\cal H}$. Theorem \ref{thm: mapping-a-terminal-unit} and Theorem \ref{thm: projection-of-stretched-unit-to-proper-path} help us extend this mapping to all units of ${\cal F}$ as follows (Theorem 4.4 in \cite{DBLP:conf/soda/DinitzV95}). 
\begin{center}
\fbox{\parbox{6.5in}{    
\begin{itemize}
    \item 
    If $\omega$ is a terminal unit, $\pi(\omega)$ is the node of ${\cal H}$ to which $\omega$ is mapped. 
    \item 
    If $\omega$ is a stretched unit, $\pi(\omega)$ stores the endpoints of path $P(\omega)$ to which $\omega$ is projected. 
\end{itemize}
}}
\end{center}

 Henceforth, $\pi$ will be referred to as the projection mapping. The following section is devoted to the detailed proof of Theorem \ref{thm: projection-of-stretched-unit-to-proper-path}.


\section{Projection of a stretched unit to a proper path}
\label{sec:projection-of-a-stretched-unit}
 We now analyze the set of minimal cuts of ${\cal H}$ that distinguish a stretched unit. 
 %
 %
 Observe that a tree edge or a pair of edges from a cycle in ${\cal H}$ define two minimal cuts that are opposite to each other. Without causing any ambiguity, henceforth, a minimal cut defined by a tree edge or a pair of edges of a cycle will refer to any one of these two minimal cuts, unless stated explicitly, possibly through a figure. 

 We shall now explore the structure underlying all the minimal cuts that distinguish a stretched unit $\omega \in {\cal F}$. To this end, we first establish interesting properties of certain minimal cuts in ${\cal H}$.
 %
We begin with the following lemma.

\begin{lemma}
    Let $C_1,C_2,C_3$ be any three minimal cuts in skeleton ${\cal H}$, and let $(S_1,\overline{S_1}), (S_2,\overline{S_2}), (S_3,\overline{S_3})$ be the corresponding valid cuts of $S$ defined by them. Let $\omega$ be a stretched unit distinguished by $C_1$ as well as $C_3$. If $S_1\subsetneq S_2 \subsetneq S_3$, $\omega$ is distinguished by $C_2$ as well.
    \label{lem:S1-S_2-S_3}
\end{lemma}
\begin{proof}
    $\omega$ is distinguished by $C_3$, so $\omega$ lies outside $N(S_3,\overline{S_3})$. It is given that $S_2 \subsetneq S_3$. So, it follows from Lemma \ref{lemma: loose-tight-mincut-subset-property}(1) that $\omega$ lies outside $N(S_2,\overline{S_2})$ as well. $\omega$ is distinguished by $C_1$ as well, so $\omega$ lies outside $N(\overline{S_1},S_1)$. Observe that $\overline{S_2} \subsetneq \overline{S_1}$ since $S_1 \subsetneq S_2$. So, it follows from Lemma \ref{lemma: loose-tight-mincut-subset-property}(1) that $\omega$ lies outside $N(\overline{S_2},S_2)$. Hence, Lemma \ref{lem: alternate-def-distinguished-by-a-minimal-cut} implies that $\omega$ is distinguished by $C_2$ as well.
\end{proof}

The following lemma states an interesting property for the minimal cuts of a cycle in ${\cal H}$.

\begin{lemma}
Let $O = \langle \nu_0,e_0,\nu_1\ldots \nu_{k-1},e_{k-1},\nu_{k}(=\nu_0)\rangle$ be a cycle in skeleton ${\cal H}$. Suppose a stretched unit $\omega$ is distinguished by the minimal cut defined by $e_0$ and $e_1$ as well as the minimal cut defined by $e_0$ and $e_{k-1}$. $\omega$ must be distinguished by the cut defined by $e_0$ and $e_{i}$ for each $1<i<k-1$. 
    
\label{lem: distinguish-contiguously-in-a-cycle}
\end{lemma}
\begin{proof}
Let us pick any $i$ such that $1<i<k-1$. Let us define 3 cuts in ${\cal H}$ as follows. $C_1$ is the minimal cut defined by $e_0$ and $e_{k-1}$, $C_2$ is the minimal cut  defined by $e_0$ and $e_i$, and  $C_3$ is the minimal cut defined by $e_0$ and $e_1$. Refer to Figure \ref{fig:Cuts-of-a-cycle}. 
\begin{figure}[ht]
  \centering  \includegraphics[width=0.29\textwidth]{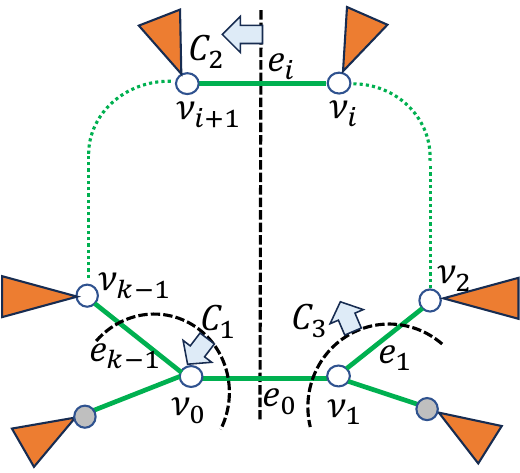} 
  \caption{$S_1\subsetneq S_2 \subsetneq S_3$}
    \label{fig:Cuts-of-a-cycle}
\end{figure}
Let $(S_1,\overline{S_1}), (S_2,\overline{S_2})$, and $(S_3,\overline{S_3})$ be the valid cuts corresponding to $C_1,C_2$, and $C_3$ respectively. Observe that $S_1\subsetneq S_2\subsetneq S_3$.
So Lemma \ref{lem:S1-S_2-S_3} implies that $\omega$ is distinguished by $C_2$ as well. 
\end{proof}

The following lemma states a property of a proper path beginning and ending with a tree edge.
\begin{lemma}
    Let $P = \langle \nu_1,e_1,\nu_2\ldots \nu_{k},e_k,\nu_{k+1}\rangle$ be any proper path of length at least 3 in skeleton ${\cal H}$ with $e_1$ and $e_k$ as tree edges. If $\omega$ is distinguished by the minimal cut defined by $e_1$ and the minimal cut defined by $e_k$, $\omega$ is distinguished by the minimal cut defined by $e_{i+1}$ for each $1\le i \le k-2$.
\label{lem: chain-of-subsets-on-proper-path}
\end{lemma}
\begin{proof}

We first show that $\omega$ is distinguished by the minimal cut defined by each tree edge of $P$. Let $e_{i+1}$ be any tree edge with $1\le i\le k-2$. We define $S_{i+1}=S(\nu_{i+1},e_{i+1})$.
If $e_i$ is a tree edge, there are two possibilities as shown in Figure \ref{fig:Cuts-of-a-proper-path}($i$) and Figure \ref{fig:Cuts-of-a-proper-path}$(ii)$. It is easy to observe that $S_1\subseteq S_i\subsetneq S_{i+1}$ in both the cases. If $e_i$
is a cycle edge, $e_{i-1}$ must be a tree edge since $P$ is a proper path. Moreover, $i\ge 2$ since $e_1$ is a tree edge. Refer to Figure \ref{fig:Cuts-of-a-proper-path}($iii$). 
So $S_1\subseteq S_{i-1}\subsetneq S_{i+1}$. Hence $S_1\subsetneq S_{i+1}$  for each $1\le i\le k-2$. Using the same reasoning, we can show that 
$S_{i+1}\subsetneq S_k$ for each $1\le i\le k-2$. Hence $S_1\subsetneq S_{i+1}\subsetneq S_k$ for each $1\le i\le k-2$. 
Since $\omega$ is distinguished by the minimal cut defined by $e_1$ and the minimal cut defined by $e_k$, Lemma \ref{lem:S1-S_2-S_3} implies that $\omega$ is distinguished by the minimal cut defined by $e_{i+1}$ as well. 
\begin{figure}[ht]
 \centering  \includegraphics[width=460pt]{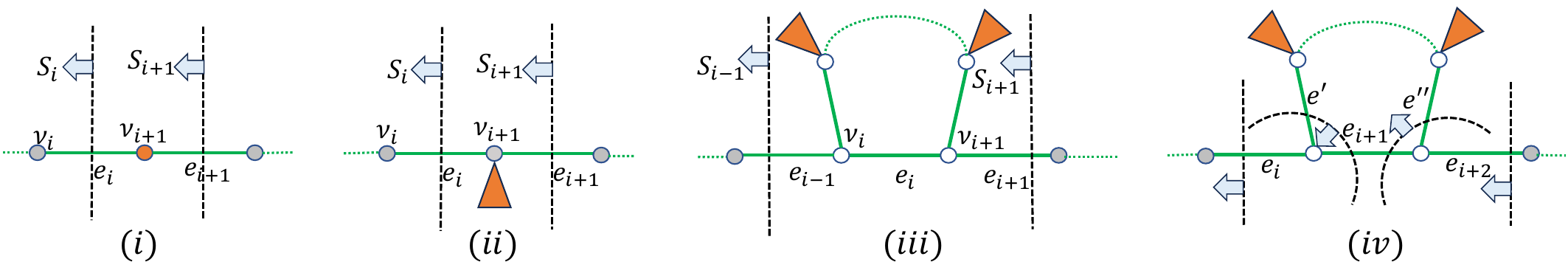} 
  \caption{($i$)  
  $\nu_{i+1}$, having degree 2, must store a $S$-unit, ($ii$) $\nu_{i+1}$ has degree $>2$, ($iii$) $S_{i-1} \subsetneq S_{i+1}$, ($iv$) Cut defined by $e'$ and
  $e_{i+1}$ represents the same bunch as that of the minimal cut defined by $e_{i}$.}
    \label{fig:Cuts-of-a-proper-path}
\end{figure}

Let $e_{i+1}$ be any cycle edge on $P$ with $1\le i\le k-2$. Observe that $e_i$ and $e_{i+2}$ must be tree edges since $P$ is a proper path. Refer to Figure \ref{fig:Cuts-of-a-proper-path}$(iv)$. We have already shown that $\omega$ is distinguished by the cut defined by $e_i$ and the cut defined by $e_{i+2}$. Let $e'$ and $e''$ be the two edges belonging to the cycle and neighboring to $e_{i+1}$.
 Lemma \ref{lem: tree-cut-defines-same-S-partition-as-cut-defined-by-2-adjacent-edges-of-cycle} implies that
the minimal cut defined by $e_i$ represents the same bunch as that of the minimal cut defined by $e_{i+1}$ and $e'$. So $\omega$
is distinguished by the minimal cut defined by $e_{i+1}$ and $e'$.
Using the same reasoning, $\omega$ is distinguished by the minimal cut defined by $e_{i+1}$ and $e''$. 
So Lemma \ref{lem: distinguish-contiguously-in-a-cycle} implies that $\omega$ is distinguished by each minimal cut defined by $e_{i+1}$ in the cycle.
\end{proof} 
If a minimal cut defined by a tree edge of ${\cal H}$ distinguishes $\omega$, we can project $\omega$ to the tree edge. The following section analyses the minimal cuts of a cycle distinguishing $\omega$.
%
%
%


\subsection{Cuts of a cycle distinguishing a stretched unit}
It follows from Note \ref{note: only-tree-edge-cuts-and-transversal-cuts-of-a-cycle} that we need to deal only with the case where a stretched unit $\omega$ is distinguished by a minimal cut defined by a crossing cut of a cycle. 
The following theorem provides an interesting insight into such cuts. 
\begin{theorem}
If a stretched unit $\omega$ is distinguished by any crossing cut of a cycle, there exists a unique edge $e$ of the cycle such that $\omega$ is distinguished by all those and exactly those minimal cuts of the cycle that are defined by $e$.
\label{thm:all-cuts-defined-by-e}
\end{theorem}

We shall now establish the proof of Theorem \ref{thm:all-cuts-defined-by-e}. Let $O=\langle \nu_0,e_0,\nu_1\ldots \nu_{k-1},e_{k-1},\nu_{k}\rangle$ with $\nu_{k} = \nu_0$ be a cycle in skeleton ${\cal H}$. Observe that $k\ge 4$ due to Property ${\cal P}_2$ of skeleton ${\cal H}$, as stated in Lemma \ref{lem: structural-properties-of--cactus-H}. For the sake of brevity, we shall use $S_{i,j}$ to denote $S(\nu_i,\nu_{j})$ for any $0\le i,j \le k-1$. 

\begin{figure}[ht]
 \centering  \includegraphics[width=480pt]{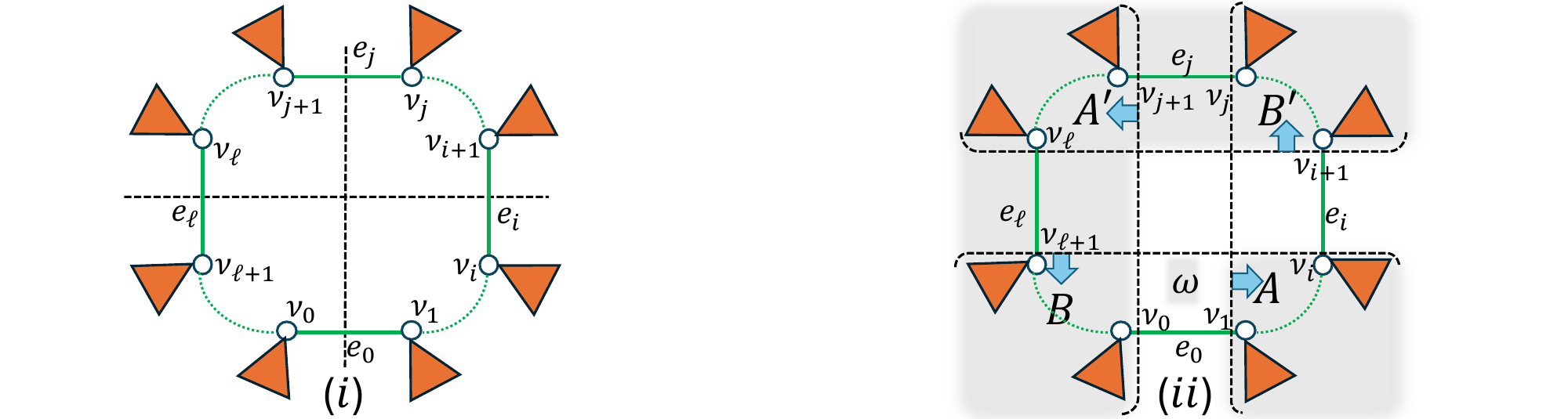} 
 \caption{($i$) Two crossing cuts in the cycle.
 ($ii$) $A\cap B$ and $A'\cup B'$ are shown shaded.}
 \label{Fig:Cuts-of-cycle-defined-by-2-edges}
\end{figure}

Suppose $\omega$ is distinguished by a crossing cut of the cycle $O$. Without loss of generality, suppose this minimal cut is defined by edges $e_0$ and $e_j$. Observe that $1 < j < k-1$ since the minimal cut is a crossing cut. Let us choose any pair of edges $e_i$ and $e_\ell$ with $1\le i<j$ and $j<\ell\le k-1$. Observe that the minimal cut defined by $e_i$ and $e_\ell$ crosses the minimal cut defined by $e_0$ and $e_j$. Refer to Figure \ref{Fig:Cuts-of-cycle-defined-by-2-edges}$(i)$. 
So Lemma \ref{lem: forbidden-combination-of-cuts}(2) implies that $\omega$ is not distinguished by the minimal cut defined by $e_i$ and $e_\ell$. So it follows from Lemma \ref{lem: alternate-def-distinguished-by-a-minimal-cut} that either $\omega$ lies inside
$N(S_{i+1,\ell},S_{\ell+1,i})$ or inside $N({S_{\ell+1,i}},S_{i+1,\ell})$. 
The following lemma will play a crucial role in proving Theorem \ref{thm:all-cuts-defined-by-e}. 
\begin{lemma} The following assertions hold.
\begin{enumerate}
    \item 
    If $\omega$ lies inside $N(S_{\ell+1,i},S_{i+1,\ell})$, then $\omega$ is distinguished by
    the minimal cut defined by $e_0$ and $e_i$, and
    the minimal cut defined by $e_0$ and $e_\ell$. 
    \item If $\omega$ lies inside  $N(S_{i+1,\ell},S_{\ell+1,i})$, then $\omega$ is distinguished by
    the minimal cut defined by $e_j$ and $e_i$, and
    the minimal cut defined by $e_j$ and $e_\ell$.
    \end{enumerate}
\label{lem: distinguished-by-two-edges}
\end{lemma}
\begin{proof}
We establish the proof for (1). The proof for 
case (2) is along similar lines. 
Let $(A,\overline{A}) = N(S_{1,j},S_{j+1,0})$ and $(A',\overline{A'
}) = N(S_{j+1,0},S_{1,j})$.
Observe that $\omega \notin A$ and $\omega \notin A'$ since $\omega$ is distinguished by the cut defined by $e_0$ and $e_j$.
Let $(B,\overline{B})= N(S_{\ell+1,i},S_{i+1,\ell})$ and $(B',\overline{B'})=N(S_{i+1,\ell},S_{\ell+1,i})$. $\omega\notin B'$ since $\omega\in B$. Refer to Figure \ref{Fig:Cuts-of-cycle-defined-by-2-edges}$(ii)$ that illustrates $A,A',B$, and $B'$. 

Observe that $A\cap B$ and $A\cup B$ define $S$-cuts, and so do $A'\cup B'$ and $A'\cap B'$. So the submodularity of cuts (Lemma \ref{lem: cor-submodularity-of-cuts}) implies that each of them defines a $S$-mincut. The cut defined by 
$A\cap B$ and the cut defined by $A'\cup B'$ belong to the bunch corresponding to the minimal cut of ${\cal H}$ defined by $e_0$ and $e_i$. 
Observe that $\omega\notin A\cap B$ since $\omega \notin A$. So $\omega$ lies outside $N(S_{1,i},S_{i+1,0})$. Also $\omega\notin A'\cup B'$ since $\omega \notin A'$ and $\omega\notin B'$. So $\omega$ lies outside $N(S_{i+1,0},S_{1,i})$. Therefore, Lemma \ref{lem: alternate-def-distinguished-by-a-minimal-cut} implies that $\omega$ is distinguished by the minimal cut defined by $e_0$ and $e_i$. 
To show that $\omega$ is distinguished by the cut defined by $e_0$ and $e_\ell$, we use similar arguments, and show that $\omega \notin A'\cap B$ and $\omega\notin A\cup B'$. 
\end{proof}

\begin{figure}[ht]
 \centering  \includegraphics[width=380pt]{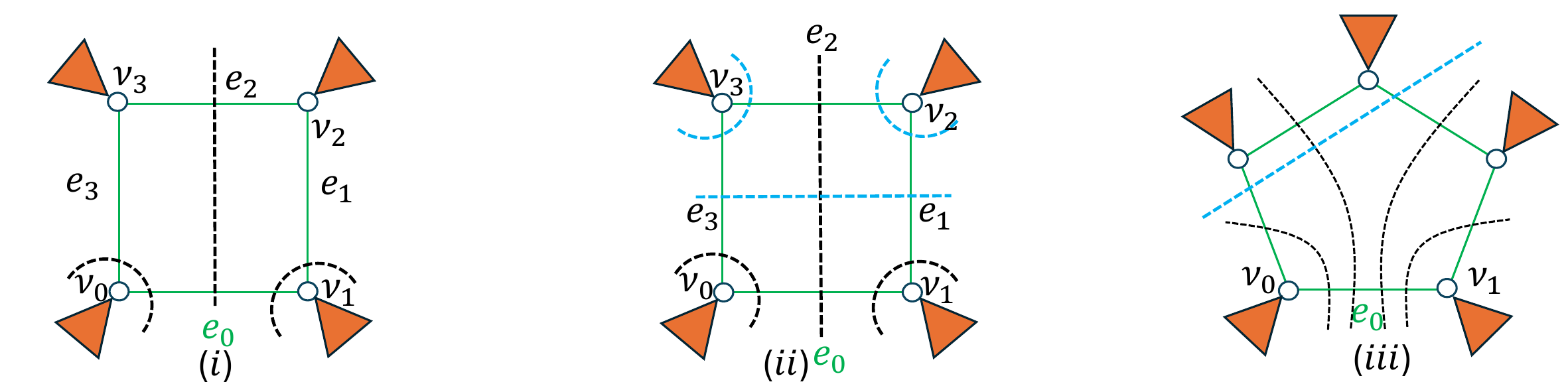} 
 \caption{($i$) the 3 cuts defined by $e_0$, 
 ($ii$) the cut defined by $e_0$ and $e_1$ and the cut defined by $e_0$ and $e_3$ are laminar cuts, ($iii$) the cut in cyan color must cross at least one of the cuts defined by $e_0$.}
 \label{Fig:No-2-edges-in-cycle-distinguish-omega}
\end{figure}

If the cycle $O$ consists of exactly 4 edges,
we can establish Theorem \ref{thm:all-cuts-defined-by-e} easily as follows. Suppose $\omega$ is distinguished by the cut defined by $e_0$ and $e_2$. 
Without loss of generality, assume that $\omega$ lies inside $N(S_{0,1},S_{2,3})$. So, applying  
Lemma \ref{lem: distinguished-by-two-edges}(1) with $i=1$ and $\ell=3$ implies that $\omega$ is distinguished by the 3 minimal cuts defined by $e_0$ as shown in Figure \ref{Fig:No-2-edges-in-cycle-distinguish-omega}($i$). We shall now show that none of the remaining 3 minimal cuts of the cycle can distinguish $\omega$. 
The minimal cut defined by $e_1$ and $e_3$ crosses the minimal cut defined by $e_0$ and $e_2$ which already distinguishes $\omega$, so the former cannot distinguish $\omega$ due to Lemma \ref{lem: forbidden-combination-of-cuts}(2).
The remaining two minimal cuts ---
the minimal cut defined by $e_2$ and $e_3$ as well as the minimal cut defined by $e_1$ and $e_2$ --- are laminar cuts. None of them can distinguish $\omega$ due to Lemma \ref{lem: forbidden-combination-of-cuts}(1) since there are already two laminar cuts defined by $e_0$
that distinguish $\omega$ (see Figure \ref{Fig:No-2-edges-in-cycle-distinguish-omega}($ii$)).  

So, henceforth, assume that the cycle $O$ has length at least 5. For this case, we establish the following lemma.
\begin{lemma}
    $\omega$ lies inside $N(S_{0,1}, S_{2,k-1})$ or inside $N(S_{j,j+1}, S_{j+2,j-1})$.
    \label{lem:omega-in-N(S_1,2)}
\end{lemma}
\begin{proof}
Recall that the minimal cut defined by $e_0$ and $e_j$ is a crossing cut, and the cycle has a length at least 5. So either $e_1\not=e_{j-1}$ or $e_{j+1}\not=e_{k-1}$. Without loss of generality, assume that $e_1\not=e_{j-1}$. To prove the lemma, we shall rule out the possibility of $\omega$ lying inside neither $N(S_{0,1},S_{2,k-1})$ nor $N(S_{j,j+1}, S_{j+2,j-1})$ as follows.
If $\omega$ does not lie inside $N(S_{0,1}, S_{2,k-1})$, $\omega$ must lie inside
$N(S_{2,k-1},S_{0,1})$ due to Lemma \ref{lem: alternate-def-distinguished-by-a-minimal-cut} since $\omega$ is not distinguished by the minimal cut defined by $e_1$ and $e_{k-1}$. So applying Lemma \ref{lem: distinguished-by-two-edges}(2) with $i=1,\ell=k-1$ implies that the cut defined by $e_j$ and $e_1$ distinguishes $\omega$. 
If $\omega$ does not lie inside   $N(S_{j,j+1}, S_{j+2,j-1})$, it must lie inside $N(S_{j+2,j-1},S_{j,j+1})$. So applying Lemma \ref{lem: distinguished-by-two-edges} (1) with $i=j-1,\ell=j+1$ implies that the cut defined by $e_0$ and 
$e_{j-1}$ distinguishes $\omega$. However, Lemma \ref{lem: forbidden-combination-of-cuts}(2)  implies that it is not possible. This is because the cut defined by $e_0$ and $e_{j-1}$ crosses the cut defined by $e_1$ and $e_j$, as $e_1\not=e_{j-1}$.
\end{proof}

Using Lemma \ref{lem:omega-in-N(S_1,2)}, we can assume, without loss of generality, that $\omega \in N(S_{0,1},S_{2,k-1})$. So applying Lemma \ref{lem: distinguished-by-two-edges}(1) with $\ell=k-1$ and $i=1$ implies that $\omega$ is distinguished by the cut defined by $e_0$ and $e_1$ as well as the cut defined by $e_0$ and $e_{k-1}$. It thus follows from Lemma \ref{lem: distinguish-contiguously-in-a-cycle} that $\omega$ is distinguished by each of the $k-1$ minimal cuts of ${\cal H}$ defined by edge $e_0$.
Can there be any other cut in the cycle that also distinguishes $\omega$ ? The answer to this question is in negation, as follows, thus completing the proof of Theorem \ref{thm:all-cuts-defined-by-e}. Observe that $\omega$ is already distinguished by 2 laminar cuts -- one defined by $e_0$ and $e_1$, and another defined by $e_0$ and $e_{k-1}$. So Lemma \ref{lem: forbidden-combination-of-cuts}(1) implies that no other laminar cut of the cycle can distinguish $\omega$. Let us consider any crossing cut in the cycle other than those defined by $e_0$. Such a cut will surely cross at least one of the minimal cuts defined by $e_0$ as shown in Figure \ref{Fig:No-2-edges-in-cycle-distinguish-omega}($iii$). Hence, Lemma \ref{lem: forbidden-combination-of-cuts}(2) implies that this cut can not distinguish $\omega$. 

\subsection{The projection of a stretched unit} 
Getting inspired by Theorem \ref{thm:all-cuts-defined-by-e}, we can define a mapping for a stretched unit $\omega$ to any structural edge $e$ of ${\cal H}$ as follows.
\begin{itemize}
\item If $e$ is tree edge, we project $\omega$ to $e$ if $\omega$ is distinguished by 
the minimal cut defined by $e$. 
\item 
If $e$ is a cycle edge, we project $\omega$ to $e$ if it is distinguished by each minimal cut of the cycle defined by $e$.
\end{itemize}

With the projection defined above, the following lemma is an immediate corollary of Theorem \ref{thm:all-cuts-defined-by-e}.
\begin{lemma}
In any cycle in ${\mathcal H}$, there exists at most one structural edge to which $\omega$ can be projected.
\label{lem: atmost-one-edge-of-cycle}
\end{lemma}

Let $e$ be a cycle edge in ${\cal H}$.
Recall that ${\cal H}$ is a $t$-cactus, therefore, each endpoint of $e$ has exactly one tree edge incident on it. 
So, based on the projection defined above and 
Lemma \ref{lem: tree-cut-defines-same-S-partition-as-cut-defined-by-2-adjacent-edges-of-cycle}, the following lemma is immediate.
 
\begin{lemma}
If $\omega$ is projected to a cycle edge $e$ in ${\cal H}$, $\omega$ is projected to each tree edge incident on its endpoints. 
\label{lem: cycle-edge-to-two-tree-edges}
\end{lemma}

The following is an immediate corollary of Lemma \ref{lem: cycle-edge-to-two-tree-edges}. 
\begin{corollary}
    If $\omega$ is a stretched unit, it is projected to at least one tree edge.
    \label{cor: at-least-one-tree-edge-in-pi(w)}
\end{corollary}

It follows from the projection defined above that a stretched unit will be projected to $O(|S|)$ edges in ${\cal H}$. So we store the entire projection mapping in $O(n|S|)$ space. 
Amazingly, this bound can be further improved to $O(n)$. The brilliant insight is that the edges to which any stretched unit is projected form a proper path in ${\cal H}$. So we just need to store the two endpoints of the path, which takes only $O(1)$ space. The rest of this section is devoted to establish this result.

Consider any stretched unit $\omega$. Suppose it is projected to any two tree edges in ${\cal H}$. The following lemma reveals an important property about any cycle in ${\cal H}$ lying on the path between the two.

\begin{lemma}
  Let $e$ and $e'$ be any pair of tree edges in ${\cal H}$ to which a stretched unit $\omega$ is projected. 
  Let $O$ be any cycle on the path between $e$ and $e'$ in ${\cal H}$. 
  Let $\mu$ and $\mu'$ be respectively the two nodes of cycle $O$ through which any path from $e$ to $e'$ in ${\cal H}$ enters and leaves the cycle. $\mu$ and $\mu'$ must be joined by a structural edge in ${\cal H}$.
  \label{lem:only-one-edge-shared-with-cycle}
\end{lemma}
\begin{proof}
    Let $C_1$ and $C_3$ be the minimal cuts in ${\cal H}$ defined by $e$ and $e'$ respectively, as shown in Figure \ref{Fig:Only-one-edge-of-cycle-shared}($i$). 
   It is given that cycle $O$ lies on the path between $e$ and $e'$. So, there exist 2 paths, say $P_1$ and $P_2$, that originate from $\mu$ and terminate at $\mu'$ in cycle $O$. Let us suppose $\mu$ and $\mu'$ are not joined by a structural edge. So $P_1$ and $P_2$ must consist of at least 2 edges each. Refer to Figure \ref{Fig:Only-one-edge-of-cycle-shared}($i$).
\begin{figure}[ht]
 \centering  \includegraphics[width=480pt]{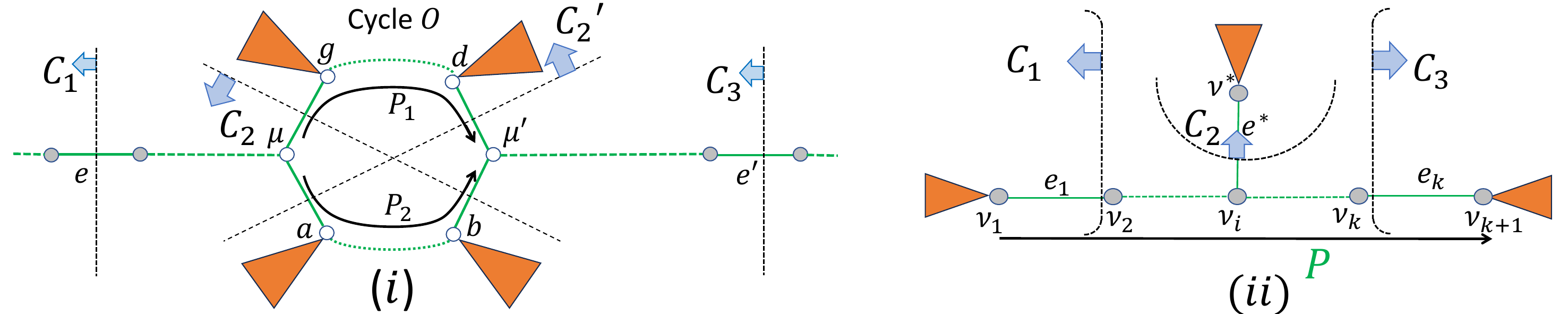} 
 \caption{($i$) Subcacti hanging from $a,b,g,d$ have $S$-vertices. It is possible that $a=b$ $\&$ $g=d$.~($ii$) The subsets of $S$ present in $C_1$, $C_2$, and $C_3$ are pairwise disjoint.}
 \label{Fig:Only-one-edge-of-cycle-shared}
\end{figure}
   Let $C_2$ be the minimal cut in ${\cal H}$ defined by the first edge of $P_1$ and the last edge of $P_2$. Similarly, let $C_2'$ be the minimal cut in ${\cal H}$ defined by the first edge of $P_2$ and the last edge of $P_1$. Let $(S_1,\overline{S_1}),(S_2,\overline{S_2}), (S_2',\overline{S_2'})$, and $(S_3,\overline{S_3})$ be the valid cuts of $S$ defined by $C_1,C_2,C_2',$ and $C_3$ respectively. Observe that $S_1\subsetneq S_2\subsetneq S_3$ and $S_1\subsetneq S_2'\subsetneq S_3$. So, Lemma \ref{lem:S1-S_2-S_3} implies that $\omega$ is distinguished by $C_2$ as well as $C_2'$. This would contradict Lemma \ref{lem: forbidden-combination-of-cuts}(2) since $C_2$ and $C_2'$ form a crossing pair of cuts in ${\cal H}$.
\end{proof}


\begin{theorem}
Let $\omega$ be a stretched unit present in ${\cal F}$. The set of structural edges of ${\cal H}$ to which $\omega$ is projected constitutes a proper path, and its first edge as well as the last edge is a tree edge.
\label{thm: projection-of-nonSteiner-unit-to-a-proper-path}
\end{theorem}
\begin{proof}
Let ${\cal E}(\omega)$ be the set of structural edges of ${\mathcal H}$ to which $\omega$ is projected. 
%
If ${\cal E}(\omega)$ consists of only one structural edge, Corollary \ref{cor: at-least-one-tree-edge-in-pi(w)} implies that it must be a tree edge, and so the theorem holds vacuously. Otherwise, let $e$ and $e'$ be any two structural edges in ${\cal E}(\omega)$ separated by the maximum distance in ${\cal H}$. $e$ (likewise $e'$) cannot be a cycle edge as follows. If $e$ is a cycle edge, it follows from Lemma \ref{lem: cycle-edge-to-two-tree-edges} that there are two tree edges incident on the endpoints of $e$ that also belong to ${\cal E}(\omega)$. The distance of one of them from $e'$ has to be greater than the distance of $e$ from $e'$, which is not possible by construction. So $e$ and $e'$ must be tree edges. 
Lemma \ref{lem:only-one-edge-shared-with-cycle} implies that there exists a proper path in ${\cal H}$ with $e$ and $e'$ as its first and last edges respectively. Let $P = \langle \nu_1,e_1(=e),\nu_2 \ldots,e_k(=e'),\nu_{k+1}\rangle$ be this proper path. Lemma \ref{lem: chain-of-subsets-on-proper-path} implies that each edge of $P$ also belongs to ${\cal E}(\omega)$. To establish the theorem, we shall prove that ${\cal E}(\omega)$ consists of only the edges of path $P$ as follows. 

The proof is by contradiction. Let us suppose ${\cal E}(\omega)\not=P$. So let $e'' $
be any edge from ${\cal E}(\omega)\setminus P$. Using Lemma \ref{lem: cycle-edge-to-two-tree-edges}, we can assume without loss of generality that $e''$ is a tree edge. It follows from Lemma \ref{lem:only-one-edge-shared-with-cycle} that there exists a proper path in ${\cal H}$ with $e$ and $e''$ as its first and last edges. 
Moreover, Lemma \ref{lem: chain-of-subsets-on-proper-path} implies that all edges of this path belong to 
${\cal E}(\omega)$. To complete the proof by contradiction, we shall show that there exists an edge of this path that cannot belong to ${\cal E}(\omega)$ as follows. Traversing this path from $e$ towards $e''$, let $e^*=(\nu_i,\nu^*)$ be the first edge that does not belong to $P$. Observe that $1<i<k$, otherwise it would refute that $e$ and $e'$ are the farthest pair of edges in ${\cal E}(\omega)$. 
We shall now show that $e^*\notin {\cal E}(\omega)$. 

Let $S_1=S(\nu_1,e_1)$ and $S_3=S(\nu_{k+1},e_k)$. 
Let $(C_1,\overline{C_1})=F(S_1,\overline{S_1})$ and $(C_3,\overline{C_3})=F(S_3,\overline{S_3})$. Refer to Figure \ref{Fig:Only-one-edge-of-cycle-shared}($ii$).
Observe that $\omega\in C_1$ and $\omega\in C_3$ due to Lemma \ref{lem: alternate-def-distinguished-by-a-minimal-cut}. If $e^*$ belongs to a cycle, it follows from Property ${\cal P}_2$ of ${\mathcal H}$ (refer to Lemma \ref{lem: structural-properties-of--cactus-H}) that $\nu_i$ has to have degree 3. So the cycle must pass through one of $e_{i-1}$ and $e_i$. 
But this possibility is ruled out since at most one edge of any cycle in ${\cal H}$ may belong to ${\cal E}(\omega)$ as stated in Lemma \ref{lem: atmost-one-edge-of-cycle}. 
Hence, $e^*$ has to be a tree edge as shown in Figure \ref{Fig:Only-one-edge-of-cycle-shared}($ii$). 
Define $S_2 = S(\nu^*,e^*)$ and $(C_2,\overline{C_2})=F(S_2,\overline{S_2})$. 
Observe that $S_1,S_2,$ and $S_3$ are pairwise disjoint. 
So, Lemma \ref{lem: corollary of four point lemma} implies that $C_1 \cap C_2 \cap C_3=\emptyset$. But $\omega\in C_1\cap C_3$ as mentioned above. So $\omega\notin C_2$, hence
$e^* \notin {\cal E}(\omega)$.
\end{proof}

\begin{figure}[ht]
 \centering  \includegraphics[width=480pt]{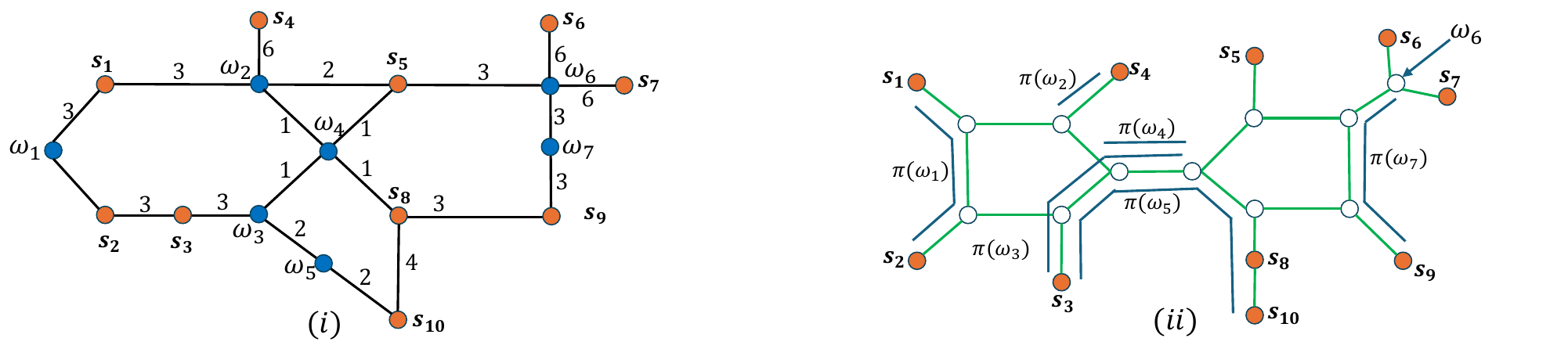} 
  \caption{($i$) Flesh ${\cal F}$, ($ii$) Skeleton ${\cal H}$ with projection mapping $\pi$ for all non-$S$-units.}
    \label{fig:projection-mapping}
\end{figure}

For the flesh graph ${\cal F}$ shown in Figure \ref{fig:projection-mapping}($i$), its skeleton ${\cal H}$ is shown in Figure \ref{fig:projection-mapping}($ii$). The  projection mapping of all non-$S$-units of ${\cal F}$ is also shown in this figure. Note that $\omega_6$ is a terminal unit, hence $\pi(\omega_6)$ is a node. $\pi(\omega_4)$ is a tree edge joining two cycles.

The tuple consisting of ${\cal F}$, ${\cal H}$, and the projection mapping $\pi$ is called the connectivity carcass by Dinitz and Vainshtein \cite{DBLP:conf/stoc/DinitzV94}. 
It follows from Theorem \ref{thm: projection-of-nonSteiner-unit-to-a-proper-path} that we can easily compute projection mapping of all non-$S$-units if we compute a strip for each tree edge in ${\cal H}$. There are ${\cal O}(|S|)$ tree edges in ${\cal H}$. So it takes ${\cal O}(|S|)$ max-flow computations to compute $\pi$. As discussed earlier, ${\cal F}$ and ${\cal H}$ can be computed in ${\cal O}(|S|)$ max-flow computations, thus leading to the following lemma.

\begin{lemma}
The connectivity carcass for any given set $S\subseteq V$ can be built in ${\cal O}(|S|)$ max-flow computations.
\label{lem: algorithm for computing projection mapping}
\end{lemma}

\section{Connectivity carcass as an efficient data structure}
\label{sec:carcass_as_an_efficient_datastructure}
The connectivity carcass can serve as an efficient data structure to answer 
various fundamental queries on $S$-mincuts as shown in the following sections.

\subsection{Computing the strip for a minimal cut of \texorpdfstring{${\cal H}$}{skeleton}}
Consider any minimal cut of ${\cal H}$. Referring to Note \ref{note: only-tree-edge-cuts-and-transversal-cuts-of-a-cycle}, without loss of generality, we can assume that this minimal cut is defined by a tree edge or a pair of non-adjacent edges of some cycle in ${\cal H}$. Suppose the minimal cut breaks ${\mathcal H}$ into 2 subcacti, say ${\mathcal H}_1$ and ${\mathcal H}_2$. Let $S_1\subsetneq S$ be the set of vertices belonging to ${\mathcal H}_1$. Let ${\cal W}$ be the strip corresponding to the valid cut $(S_1, \overline{S_1})$. We shall now show that ${\cal W}$ can be built in ${\cal O}(m)$ time. 

Consider any unit $\omega\in {\cal F}$. If $\omega$ is a terminal unit, $\pi(\omega)$ is a node in ${\cal H}$.  
If this node belongs to ${\cal H}_1$, $\omega$ belongs to the source node of ${\cal W}$, 
 otherwise $\omega$ belongs to the sink node. Let us discuss the case when $\omega$ is a stretched unit. If $\pi(\omega)$ shares an edge with the minimal cut, $\omega$ appears as a non-terminal in ${\cal W}$. 
Otherwise, $\pi(\omega)$ is either present entirely in  ${\mathcal H}_1$ or present entirely in ${\mathcal H_2}$. In each of these cases, $\omega$ belongs to one of the two terminals of ${\cal W}$, and the following lemma can be used to determine this terminal.
\begin{lemma} 
    If $\pi(\omega)$ is present entirely in ${\cal H}_2$, then $\omega$ is mapped to the sink in ${\cal W}$; otherwise $\omega$ belongs to the source.
    \label{lem:deciding-the-terminal-for-a-stretched-unit}
\end{lemma}
    \begin{proof}
    If $\pi(\omega)$ is present entirely in ${\cal H}_2$, it follows from Corollary \ref{cor: at-least-one-tree-edge-in-pi(w)} that  there exists at least one tree edge, say $e$, in ${\cal H}_2$ such that $e\in \pi(\omega)$. As stated above, the minimal cut corresponding to ${\cal W}$ is defined by a tree edge or a pair of non-adjacent edges of some cycle in ${\cal H}$. 
    In either case, it can be observed that there exists a valid cut $(S_2,\overline{S_2})$ defined by $e$ such that $S_1 \subsetneq S_2$. Observe that $\omega$ lies outside the tight mincut from $S_2$ to $\overline{S_2}$ since $\omega$ is distinguished by the valid cut $(S_2,\overline{S_2})$. So it follows from Lemma \ref{lemma: loose-tight-mincut-subset-property}(1) that $\omega$ lies outside the tight mincut from $S_1$ to $\overline{S_1}$ as well. Therefore, $\omega$ does not belong to the source in ${\cal W}$. So $\omega$ 
    belongs to the sink in ${\cal W}$. 
    \end{proof}

 It follows from the tree data structure for ${\cal H}$,
 as stated in Lemma \ref{lem:skeleton-tree-queries}, that we can determine in ${\cal O}(1)$ time whether $\pi(\omega)$ shares an edge with any given minimal cut of ${\cal H}$. 
So we can state the following theorem.
\begin{theorem}
Let $(S_1,\overline{S_1})$ be a valid cut defined by any minimal cut in ${\mathcal H}$. Given the projection mapping $\pi$, it takes ${\cal O}(1)$ time
to determine whether any unit $\omega\in {\cal F}$ belongs to the source or the sink, or appears as a non-terminal in the strip corresponding to $(S_1,\overline{S_1})$.
\label{thm: strip details from H and pi}
\end{theorem}

Once we know all the vertices of a strip, the strip can be built in $O(m)$ time using our algorithm stated in Lemma \ref{lem: G-from-Gu}.

\subsection{Reporting \texorpdfstring{$(s,t)$}{(s,t)}-strip for any \texorpdfstring{$s,t \in S$}{s,t in V}} \label{appendix:xxx}

Let $s$ and $t$ be any two vertices of $S$ separated by a $S$-mincut. Let $\nu_1$ and $\nu_2$ be the two nodes in ${\cal H}$ storing $s$ and $t$ respectively. Observe that $\nu_1$ and $\nu_2$ must be tree nodes. 
Recall that strip ${\cal D}_{s,t}$ stores all $(s,t)$-mincuts. We shall show that ${\cal D}_{s,t}$ can be constructed in just ${\mathcal O}(m)$ time 
using ${\cal H}$, $\Phi$, and the projection mapping $\pi$ as follows.

We first state the following lemma that captures the necessary and sufficient condition for any valid cut of $S$ to separate $s$ and $t$. Its proof is immediate from the structure of ${\cal H}$ and Theorem \ref{thm: skeleton-theorem}. 

\begin{lemma}[Cut separating two nodes in a $t$-cactus] 
Let $e$ be any structural edge in a $t$-cactus ${\mathcal H}$. There exists a minimal cut defined by $e$ that separates $\nu_1$ and $\nu_2$ in ${\mathcal H}$ if and only if there exists a path between $\nu_1$ and $\nu_2$ in ${\mathcal H}$ that passes through $e$.
\label{lem:cut-in-cactus}
\end{lemma}

We use the insight from Lemma \ref{lem:cut-in-cactus} to build a cactus structure that stores all those valid cuts of $S$ that separate $s$ and $t$. This cactus, denoted by ${\mathcal H}_{s,t}$, is a quotient graph of ${\cal H}$, and potentially much smaller than ${\mathcal H}$. First, we introduce two terminologies.

\begin{definition}[Subcactus ${\cal H}(\nu,e)$]
Given a node $\nu$ and a tree edge $e$ incident on it in ${\cal H}$, ${\cal H}(\nu,e)$ is the subcactus containing $\nu$ obtained after removing the edge $e$. 
\label{def: subcactus}
\end{definition}

\begin{definition}[Subcactus ${\cal H}(e,e')$]
Let $\nu$ be any tree node in ${\cal H}$, and let $e$ and $e'$ be any two structural edges incident on $\nu$. ${\cal H}(e,e')$ is the subcactus containing $\nu$ obtained after removal of edges $e$ and $e'$ from ${\mathcal H}$. This notation is extended to the case when $\nu$ belongs to a cycle, albeit with the following restriction -- $e$ and $e'$ are the two edges incident on $\nu$ in the cycle .
\label{def: Hee'}
\end{definition}

 Recall that $\nu_1$ and $\nu_2$ are tree nodes
 in ${\cal H}$. We transform ${\mathcal H}$ as follows to construct
${\mathcal H}_{s,t}$.
\begin{enumerate}
\item 
On a path between $\nu_1$ and $\nu_2$ in ${\mathcal H}$, let $e_1$ and $e_2$ be the tree edges incident on $\nu_1$ and $\nu_2$ respectively. Compress ${\cal H}(\nu_1,e_1)$ into a single node, and compress ${\cal H}(\nu_2,e_2)$ into a single node. 
\item
Let $\mu \notin \{\nu_1,\nu_2\}$ be any node of ${\mathcal H}$ lying on a path between $\nu_1$ and $\nu_2$. If $\mu$ is the entry or exit node of some cycle, we keep it as it is. Otherwise, let $e$ and $e'$ be the two edges incident on $\mu$ that belong to a path between $\nu_1$ and $\nu_2$.
Compress ${\cal H}(e,e')$ into a single node.
\end{enumerate}
Construction of ${\cal H}_{s,t}$ described above and Lemma \ref{lem:cut-in-cactus} directly lead to the following property of ${\cal H}_{s,t}$.
\begin{lemma}
    ${\cal H}_{s,t}$ stores all those and exactly those valid cuts of $S$ that separate $s$ and $t$.
    \label{lem: H_st}
\end{lemma}

The vertex set of ${\cal D}_{s,t}$ will consist of 
all non-empty nodes of ${\cal H}_{s,t}$ and all those stretched units of ${\cal F}$ whose projection in ${\cal H}$ intersects any path between $\nu_1$ and $\nu_2$. We use Lemma \ref{lem: H_st} to define the mapping $\Phi_{s,t}$ from $V$ to vertices of ${\cal D}_{s,t}$ as follows. Let $u\in V$. There are two cases.
\begin{enumerate}
    \item $\Phi(u)$ is a terminal unit in ${\mathcal F}$:~~ 
    Let $\omega = \Phi(u)$. It follows from the construction of ${\cal H}_{s,t}$ that node $\pi(\omega)$, which is a node in ${\cal H}$, belongs to 
    a non-empty node, say $\nu$, in ${\cal H}_{s,t}$. So we define $\Phi_{s,t}(u) \leftarrow \nu$.     
    \item $\Phi(u)$ is a stretched unit in ${\mathcal F}$:~~
    Let $\omega = \Phi(u)$. 
    We use the data structure from Lemma \ref{lem:skeleton-tree-queries} to determine in ${\cal O}(1)$ time if path $\pi(\omega)$ intersects any path between 
    $\nu_1$ and $\nu_2$ in ${\cal H}$. If there is intersection, $\omega$ is going to appear as a non-terminal in the strip ${\cal D}_{s,t}$, so we define $\Phi_{s,t}(u)\leftarrow \omega$. Otherwise, path $\pi(\omega)$ is present fully inside a subcactus of ${\cal H}$ that was compressed to a node, say $\nu$,
    during the construction of ${\cal H}_{s,t}$. 
    So we define $\Phi_{s,t}(u)\leftarrow \nu$.
\end{enumerate}

The mapping $\Phi_{s,t}$ described above and the algorithm stated in Lemma \ref{lem: G-from-Gu} lead to the following theorem.
\begin{theorem}
    Given $\Phi, {\cal H},$ and $\pi$, we can compute the mapping $\Phi_{s,t}$ from $V$ to the set of vertices of strip ${\cal D}_{s,t}$ in ${\mathcal O}(n)$ time. The strip ${\cal D}_{s,t}$ itself can be built in ${\cal O}(m)$ time.
\end{theorem}

\subsection{Reporting a \texorpdfstring{$S$}{S}-mincut separating any two units in \texorpdfstring{${\cal F}$}{flesh}}
Let $x$ and $y$ be any two units in ${\cal F}$. We describe an ${\cal O}(m)$ time algorithm to report a 
$S$-mincut separating $x$ and $y$.
It follows from the discussion above that the only non-trivial case left is when $x$ and $y$ are stretched units. 
Consider any structural edge $e\in \pi(x)$. Using 
Theorem \ref{thm: strip details from H and pi}, we construct the strip corresponding to any minimal cut defined by $e$. Observe that $x$ is a non-terminal in this strip. If $e\notin \pi(y)$, $y$ must be mapped to the source (or sink) of the strip. So the cut defined by the source (or the sink) of the strip is a $S$-mincut separating $x$ and $y$.
Let us consider the case when $e\in \pi(y)$. So both $x$ and $y$ appear as non-terminals in the strip. 
We compute ${\cal R}_{s}(x)$ and ${\cal R}_{s}(y)$. It follows from Lemma \ref{lem:Properties-of-R_s(x)} (2) that one of these reachability cones serves as $S$-mincut separating $x$ and $y$. This leads to the following theorem.
\begin{theorem}
    Given $\Phi, {\cal H},$ and $\pi$, we can compute the $S$-mincut separating any two given units of ${\cal F}$ in ${\cal O}(m)$ time.
\label{thm: reporting-S-mincut-separating-x-and-y}
\end{theorem}


\section{More insight into the relation between the flesh graph and the skeleton}
\label{sec:projection_of_an_edge}

In this section, we discuss additional relations between ${\cal F}$ and ${\cal H}$. 

\subsection{Flesh graph \texorpdfstring{${\cal F}$}{} from perspective of a cycle in \texorpdfstring{${\cal H}$}{skeleton}}  \label{sec:the-ring-graph-structure}
${\cal F}$ viewed from the perspective of a minimal cut in ${\cal H}$ is a strip.
We now explore the structure of ${\cal F}$ from the perspective of a cycle in ${\cal H}$. 

Let $O=\langle \nu_0,e_0,\nu_1\ldots \nu_{k-1},e_{k-1},\nu_{k}\rangle$ with $\nu_{k} = \nu_0$ be a cycle in skeleton ${\cal H}$. 
Consider $\nu_i$ for any $0\le i \le k-1$. We compress all terminal units present in $S(\nu_i)$ into $\nu_i$. We also compress each stretched unit $\omega\in {\cal F}$ into $\nu_i$ if the endpoints of $\pi(\omega)$ lie within the subcactus ${\cal H}(e_{i-1},e_i)$ (refer to Definition \ref{def: Hee'}). After performing this step for each $0\le i \le k-1$, let ${\cal F}(O)$ be the resulting quotient graph of ${\cal F}$.  
Observe that (1) all those and only those stretched units of ${\cal F}$ remain intact in ${\cal F}(O)$ whose projection mapping shares an edge with the cycle $O$, (2) ${\cal F}(O)$ preserves all $S$-mincuts that are defined by all minimal cuts of $O$. 
We now derive the constraints on the edges present in ${\cal F}(O)$.

\begin{figure}[ht]
 \centering  \includegraphics[width=420pt]{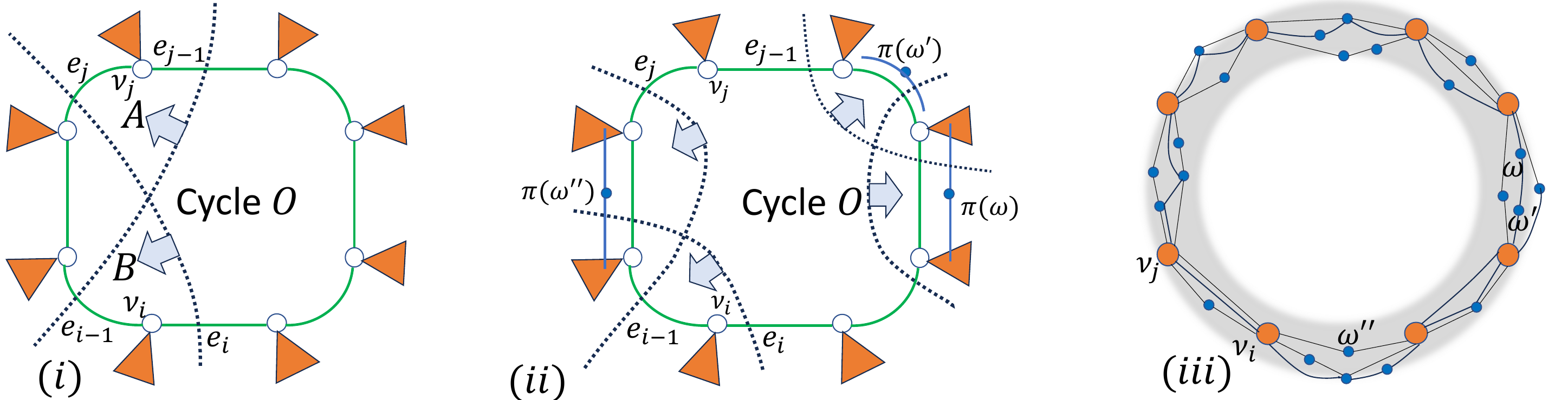} 
  \caption{($i$) Cycle $O$ and a pair of non-adjacent nodes $\nu_i$ and $\nu_j$, ($ii$) stretched unit of ${\cal F}$ projected to the edges of cycle $O$, ($iii$) The quotient graph ${\cal F}(O)$.}
    \label{fig:F-induced-by-cuts-of-O}
\end{figure}

\begin{lemma}
    Let $\nu_i$ and $\nu_j$ be any two nodes of the cycle $O$ after the compression described above to form ${\cal F}(O)$. If $\nu_i$ and $\nu_j$ are not adjacent in ${\cal O}$, there cannot be any edge between $\nu_i$ and $\nu_j$.
\label{lem:no-diagonal-edges}
\end{lemma}
\begin{proof}
%
We define $S$-mincuts $(A,\overline{A})$ and $(B,\overline{B})$ as shown in Figure  \ref{fig:F-induced-by-cuts-of-O}($i$). These cuts satisfy the condition stated in Lemma \ref{lem:no-edge-CminusC'-and-C'-minus-C} since $\nu_i$ and $\nu_j$ are not adjacent in the cycle $O$. So, there cannot be any edge between $\nu_i$ and $\nu_j$. 
\end{proof}
Along similar lines as Lemma \ref{lem:no-diagonal-edges}, we can establish the following constraints on the edges incident on the stretched units of ${\cal F}$ present in ${\cal F}(O)$. Refer to Figure \ref{fig:F-induced-by-cuts-of-O}($ii$) for the illustration helpful for the proof.
\begin{enumerate}
    \item Let $\omega$ and $\omega'$ be any two stretched units in ${\cal F}$. If they are joined by an edge in ${\cal F}(O)$, they must be projected to the same edge in $O$.
    \item Let $\omega''$ be any stretched unit in ${\cal F}$. If $\omega''$ is joined to $\nu_i$ in ${\cal F}(O)$, $\omega''$ is projected to exactly one of the two edges of $O$ that are incident on $\nu_i$.
\end{enumerate}
So ${\cal F}(O)$, the quotient graph of ${\cal F}$ induced by the cycle $O$, looks like a {\em ring} as shown in Figure \ref{fig:F-induced-by-cuts-of-O}($iii$).

\subsection{Projection of an edge}
Let $e$ be an edge in $G$ that appears in ${\cal F}$ as well. Let $x$ and $y$ be the units in ${\cal F}$ to which the endpoints of $e$ belong. Hence $e$ appears as edge $(x,y)$ in ${\cal F}$. Along similar lines of Definition \ref{def: unit-distinguished-by-a-cut}, we introduce the following definition.

\begin{definition}[an edge of ${\cal F}$ distinguished by a cut]
An edge $(x,y)\in {\cal F}$ is said to be distinguished by a minimal cut in ${\cal H}$ if it appears in the strip corresponding to the minimal cut. That is, either $x$ and $y$ belong to different terminals of the strip, or at least one of them appears as a non-terminal in the strip.
\label{def: edge-distinguished-by-a-cut}
\end{definition}
Let $G'$ be the graph obtained by introducing a vertex $v_{xy}$ at the center of edge $e$ in graph $G$. Observe that the $S$-mincut capacity of $G'$ is the same as that of $G$. 
Any cut in $G$ that does not separate $x$ and $y$ is preserved in $G'$ as well. Therefore, any $S$-mincut in $G$ that does not separate $x$ and $y$ is also a $S$-mincut in $G'$. Let us consider the $S$-mincuts in $G$ that separate $x$ and $y$. Let $A$ be any subset of $V$ such that $x\in A$ and $y\in \overline{A}$. The following observations for $A$ are immediate from the construction of $G'$. 
\begin{enumerate} 
    \item 
    $A$ defines a $S$-mincut in $G$ if and only if $A$ defines a $S$-mincut in $G'$.
    \item 
    $A$ defines a $S$-mincut in $G'$ with $v_{xy}\in \overline{A}$ if and only if $A\cup \{v_{xy}\}$ defines a $S$-mincut in $G'$.
\end{enumerate}
These observations lead to the following inferences about the skeleton and flesh graph of $G'$. ${\cal H}$ serves as a skeleton for $G'$ as well. Let ${\cal F}(G)$ and ${\cal F}(G')$ denote the flesh graphs for $G$ and $G'$ respectively. 
${\mathcal F}(G')$ will be identical to ${\mathcal F}(G)$ except the following changes. 
There will be an additional non-$S$-unit $v_{xy}$ in ${\mathcal F}(G')$. The edge $(x,y)$ in ${\mathcal F}(G)$ gets replaced by two edges, say $e_x$ and $e_y$, incident on unit $v_{xy}$ from $x$ and $y$ respectively. $e_x$ and $e_y$ belong to different sides of the inherent partition of $v_{xy}$. We can use these insights to derive the following lemma.

\begin{lemma}
A minimal cut in ${\mathcal H}$ distinguishes an edge, say $(x,y)$, of ${\cal F}(G)$ if and only if it distinguishes unit $v_{xy}$ of ${\cal F}(G')$. 
\label{lem: v_{xy}-and-(x,y)}
\end{lemma}
\begin{proof}
Consider any minimal cut in ${\cal H}$ that distinguishes edge $(x,y)$. The strip in $G$ corresponding to this minimal cut contains edge $(x,y)$. 
Introducing $v_{xy}$ in the middle of this edge transforms it into the strip in $G'$ for the same minimal cut, and $v_{xy}$ appears as a non-terminal in it. So the minimal cut distinguishes $v_{xy}$ as well. We now establish the validity of the assertion in the other direction. Consider any minimal cut in ${\cal H}$ that distinguishes $v_{xy}$ in ${\cal F}(G')$. So, the strip in $G'$ corresponding to this minimal cut keeps $v_{xy}$ as a non-terminal with $e_x$ and $e_y$ lying on the different sides of its inherent partition. Joining the two neighbors of $v_{xy}$ in the strip by an edge followed by the removal of $v_{xy}$ and edges $\{ e_x,e_y\}$ transforms it into the strip in $G$ for the same minimal cut with edge $(x,y)$ appearing in it. So the minimal cut distinguishes edge $(x,y)$ also.
\end{proof}

Lemma \ref{lem: v_{xy}-and-(x,y)} and Theorem \ref{thm: projection-of-nonSteiner-unit-to-a-proper-path} lead to the following theorem.

\begin{theorem}[Projection of an edge]
    Let $(x,y)$ be an edge in ${\cal F}$. There exists a proper path $\pi(x,y)$ in ${\cal H}$ such that edge $(x,y)$ is distinguished by those and exactly those minimal cuts of ${\cal H}$ that are defined by the structural edges of $\pi(x,y)$. The edge $(x,y)$ is said to be projected to path $\pi(x,y)$.
\label{thm: projection-of-an-edge}
\end{theorem}

For an edge $(x,y)\in {\cal F}$, we shall now explore relationship among $\pi(x),\pi(y)$, and $\pi(x,y)$. 
To this end, we first introduce notations for the bunches defined by structural edges of ${\cal H}$.
\begin{itemize}
\item Bunch defined by a tree edge $e$:\\ 
Let $e$ be a tree edge incident on a node $\nu$ in ${\cal H}$. We use ${\mathcal C}(\nu,e)$ to denote the set of all $S$-mincuts $(A,\overline{A})$ satisfying $S\cap A = S(\nu,e)$. 
\item Bunch defined by a pair of edges from a cycle:\\
Let $\langle \nu_0,e_0,\nu_1,\ldots,\nu_{\ell-1},e_{\ell-1},\nu_{\ell}(=\nu_0)\rangle$ be a cycle in skeleton ${\cal H}$.  
The bunch defined by edges $e_{i}$ and $e_j$ is the 
set of all $S$-mincuts $(A,\overline{A})$ satisfying $S\cap A = S(\nu_{i+1},\nu_{j})$. We denote this bunch by ${\mathcal C}(\nu_{i+1},\nu_{j})$.
\end{itemize}

Let $e$ be any structural edge in ${\cal H}$, and $\nu$ be one of its endpoints.
Henceforth, 
we overload the notation ${\mathcal C}(\nu,e)$ in case $e$ belongs to a cycle as follows.  
We pick any other edge, say $e'=(\nu',\nu'')$, from the cycle. Without loss of generality, suppose $\nu'$ and $\nu$ belong to the same side in the cut of ${\cal H}$ defined by $e$ and $e'$. 
${\mathcal C}(\nu,e)$ would refer to ${\mathcal C}(\nu',\nu)$. 
The following lemma shows that there is an {\em implicit} direction to the edges constituting $\pi(x,y)$.
\begin{lemma}[Unidirectionality Property]
Suppose $\pi(x,y) = \langle \nu_1,e_1,\ldots, \nu_k,e_k,\nu_{k+1}\rangle$.
Let $(A,\overline{A})$ be any $S$-mincut from ${\mathcal C}(\nu_i,e_i)$ such that $x\in A$ and $y\in \overline{A}$. If $(B,\overline{B})$ is any $S$-mincut from ${\mathcal C}(\nu_j,e_j)$ for any $1\le j \le k$ that separates $x$ and $y$, then $x\in B$ and $y\in \overline{B}$.
\label{lem: x precedes y for each edge on pi(x,y)}
\end{lemma}
\begin{proof}
 Either $x\in \overline{B}$ and $y\in B$ or $x\in B$ and $y\in \overline{B}$ because $(B,\overline{B})$ separates $x$ and $y$.  Observe that $A\cap B$ and $A\cup B$ define $S$-cuts, and hence are $S$-mincuts using the submodularity of cuts (Lemma \ref{lem: cor-submodularity-of-cuts}). 
 If $x\in \overline{B}$ and $y\in B$,
 it would imply that there is an edge between $A\backslash B$ and $B\backslash A$, which contradicts Lemma \ref{lem:no-edge-CminusC'-and-C'-minus-C}.
Hence $x\in B$ and $y\in \overline{B}$.
\end{proof}

The following lemma can be viewed as a corollary of Lemma \ref{lem: x precedes y for each edge on pi(x,y)}.
\begin{lemma}
    Let $x$ be a stretched unit in ${\cal F}$. The presence of edge $(x,y)$ in ${\cal F}$ imposes the following necessary conditions on $\pi(y)$. 
    \begin{enumerate} 
    \item If $y$ is a terminal unit, $\pi(y)$ cannot be any internal node of the path $\pi(x)$.
    \item If $y$ is a stretched unit, $\pi(x)\cap \pi(y)$ is either $\emptyset$ or a prefix/suffix of $\pi(x)$.
    \end{enumerate}
\label{cor: unidirectional-property}
\end{lemma}

The following lemma illustrates the complete relation among $\pi(x), \pi(y)$, and $\pi(x,y)$.
\begin{lemma}
Let $(x,y)$ be an edge in ${\cal F}$. $\pi(x)$ forms the initial part and $\pi(y)$ forms the final part of $\pi(x,y)$.
\label{lem: pi(x),pi(y),pi(x,y)}
\end{lemma}
\begin{proof}
Suppose $\pi(x)\cap \pi(y)=\emptyset$. Consider any path $P$ in ${\cal H}$ originating from a node in $\pi(x)$ and terminating at a node in $\pi(y)$ such that $P$ does not share any other node with $\pi(x)$ and $\pi(y)$. Suppose $P$ passes through any cycle, say $O$, in ${\cal H}$. Let $\nu$ and $\nu'$ be the nodes through which $P$ enters and leaves the cycle $O$. It follows from Lemma \ref{lem:no-diagonal-edges} that $\nu$ and $\nu'$ must be neighbors in ${\cal H}$. Hence, without loss of generality, we can assume that $P$ shares exactly one edge with each cycle of ${\cal H}$ that it passes through. Hence, $P$ is a proper path in ${\cal H}$. It follows from the construction of $P$ that $\pi(x)$ and $\pi(y)$ lie on the opposite sides of the minimal cut defined by each structural edge belonging to $P$. So it follows from Lemma \ref{lem:deciding-the-terminal-for-a-stretched-unit} that $\pi(x)$ and $\pi(y)$ 
belong to the different terminals of the strip associated with the minimal cut, so edge $(x,y)$ is distinguished by this minimal cut. This fact in conjunction with Theorem \ref{thm: projection-of-an-edge} implies that $P$ is a subpath of $\pi(x,y)$. 
%
%
%
%
Therefore, if $\pi(x)$ and $\pi(y)$ are terminal units in ${\cal H}$, $P=\pi(x,y)$, and we are done. If $x$ is a stretched unit, consider the strip corresponding to any minimal cut defined by a structural edge from $\pi(x)$. It follows from {\textsc{Distinctness Property}} (Theorem \ref{thm:distinct-nodes-in-strip}) that $x$ appears as a non-terminal in this strip. 
It follows from Lemma \ref{lem:deciding-the-terminal-for-a-stretched-unit} that $y$ belongs to a terminal in this strip since $\pi(x)\cap \pi(y)=\emptyset$. 
So $(x,y)$ also appears in this strip, and hence, is distinguished by the minimal cut. Hence, $\pi(x)$ is a subpath of $\pi(x,y)$. Using similar reasoning, $\pi(y)$ is also a subpath of $\pi(x,y)$. Recall that path $P$ shares just one node with $\pi(x)$ and one node with $\pi(y)$. Thus, $\pi(x,y)=\pi(x)::P::\pi(y)$. Hence, $\pi(x)$ forms the initial part and $\pi(y)$ forms the final part of $\pi(x,y)$.

Suppose $\pi(x)\cap \pi(y)\not=\emptyset$. Surely, at least one of $x$ and $y$ has to be a stretched unit. Without loss of generality, assume $x$ is a stretched unit. If $y$ is a terminal unit, it follows from Lemma \ref{cor: unidirectional-property}(1) that $\pi(y)$ must be one of the endpoints of $\pi(x)$. 
If $y$ is a stretched unit, it follows from Lemma  \ref{cor: unidirectional-property}(2) that $\pi(x)\cap \pi(y)$ is a prefix or a suffix of $\pi(x)$. In all these cases, $\pi(x)$ forms the initial part, and $\pi(y)$ forms the final part of $\pi(x,y)$. 
\end{proof}

An immediate inference from Lemma \ref{lem: pi(x),pi(y),pi(x,y)} is that for any edge $(x,y)\in {\cal F}$, $\pi(x,y)$ can be determined completely from $\pi(x)$ and $\pi(y)$. Moreover, Lemma \ref{lem: x precedes y for each edge on pi(x,y)} and Lemma \ref{lem: pi(x),pi(y),pi(x,y)} immediately lead to the following theorem that states that we can infer precisely the sides of $\pi(x,y)$ that constitute $\pi(x)$ and $\pi(y)$ if we are given any $S$-mincut separating $x$ and $y$.
\begin{theorem}
Let $(x,y)$ be an edge in ${\cal F}$ and $\pi(x,y) = \langle \nu_1,e_1,\ldots, \nu_k,e_k,\nu_{k+1}\rangle$.
Let $(A,\overline{A})$ be any $S$-mincut from 
bunch ${\mathcal C}(\nu_i,e_i)$ separating $x$ and $y$, for any $1\le i\le k$. 
\begin{enumerate}
    \item If $x\in A$ and $y\in \overline{A}$, $\pi(x)$ is a prefix and $\pi(y)$ is a suffix, and
    \item If $y\in A$ and $x\in \overline{A}$, $\pi(y)$ is a prefix and $\pi(x)$ is a suffix
\end{enumerate}
 of path $\pi(x,y)$ while traversing it in direction along $(\nu_i,\nu_{i+1})$.
\label{thm: projection-of-(x,y)-with-pi(x)-prefix-pi(y)-suffix}
\end{theorem}

\subsection{Projection of a coherent path in skeleton}
For each stretched unit $\omega$ in ${\cal F}$, there is an inherent partition on $E(\omega)$. This leads to defining a special path, called a coherent path, in ${\cal F}$ as follows. 

\begin{definition}[\cite{DBLP:conf/stoc/DinitzV94}]
Any path consisting of a single edge is a coherent path. Let $P = \langle \nu_1,e_1,\nu_{2}, \ldots, \nu_k \rangle$ be any path consisting of 2 or more edges in ${\cal F}$. $P$ is said to be a coherent path if for each $\nu_i$ with $1<i<k$, 
$\nu_i$ is a stretched unit and the two edges $e_{i-1}$ and $e_i$ lie on the opposite sides of its inherent partition.
\label{def: coherent_path}
\end{definition}

Along similar lines as the projection of an edge (see Definition \ref{def: edge-distinguished-by-a-cut}), a coherent path in ${\cal F}$ is said to be projected to a structural edge $e$ in ${\cal H}$ if there is at least one edge of the path that is distinguished by a minimal cut of ${\cal H}$ defined by $e$.

\begin{lemma}
    Let $\langle x,y,z \rangle$ be any 2-length coherent path in ${\cal F}$. $\pi(\langle x,y,z\rangle)$ is a proper path with 
    $\pi(x)$ and $\pi(z)$ as its initial and final parts respectively.
\label{lem: coherent-path-x-y-z}
\end{lemma}
\begin{proof}
Let $\pi(y)=\langle \nu_j,e_j,\nu_{j+1}, \ldots, \nu_k \rangle$.  It follows from Lemma \ref{lem: pi(x),pi(y),pi(x,y)} implies that the structural edge $e_j=(\nu_j,\nu_{j+1})$ belongs to $\pi(x,y)$ as well as $\pi(y,z)$. 
Consider the strip defined by ${\mathcal C}(\nu_{j},e_{j})$. It follows from {\textsc{Distinctness}} property (Theorem \ref{thm:distinct-nodes-in-strip}) that $y$ appears as a non-terminal in this strip. 
Furthermore, it follows from {\textsc{unique-inherent-partition}} property (Theorem \ref{thm:inherent-partition}) that edges $(x,y)$ and $(y,z)$ lie on the opposite sides of the inherent partition of $E(y)$. Without loss of generality, assume that $x$ appears on the side-{\bf s} and $z$ appears on the side-{\bf t} of $y$ in the strip defined by 
${\mathcal C}(\nu_{j},e_{j})$. Therefore, there exists a $S$-mincut $(A,\overline{A}) \in {\mathcal C}(\nu_{j},e_{j})$ such that $x\in A$ and $y\in \overline{A}$. So it follows from Theorem \ref{thm: projection-of-(x,y)-with-pi(x)-prefix-pi(y)-suffix}(1) that $\pi(y)$ is a suffix of $\pi(x,y)$ while traversing $\pi(x,y)$ in the direction along $(\nu_j,\nu_{j+1})$. Similarly, there also exists a $S$-mincut $(A',\overline{A'}) \in {\mathcal C}(\nu_{j},e_{j})$ such that $y\in A'$ and $z\in \overline{A'}$. So it follows from Theorem \ref{thm: projection-of-(x,y)-with-pi(x)-prefix-pi(y)-suffix}(2) that $\pi(y)$ is a prefix of $\pi(y,z)$  while traversing $\pi(y,z)$ in the direction along $(\nu_j,\nu_{j+1})$. Hence $\pi(\langle x,y,z\rangle)$ is a proper path with $\pi(x)$ and $\pi(z)$ as its initial and final part respectively.
\end{proof}

We can extend Lemma \ref{lem: coherent-path-x-y-z} immediately to a coherent path of any length as follows.

\begin{theorem}[Theorem 5 in  \cite{DBLP:conf/stoc/DinitzV94}]
Let $\langle \omega_1,\ldots,\omega_k\rangle$ be a coherent path between units $\omega_1$ and $\omega_k$ in ${\cal F}$. The projection of this path in ${\mathcal H}$ is a proper path with $\pi(\omega_1)$ as its initial part and $\pi(\omega_k)$ as its final part.  
\label{thm: projection-of-coherent-path}
\end{theorem}


\section{Conclusion}
We have presented the first complete, simpler, and self-contained proofs for the connectivity carcass -- a compact structure for storing all $S$-mincuts designed by Dinitz and Vainshtein \cite{DBLP:conf/stoc/DinitzV94}. We firmly believe that this manuscript will become the definitive reference for understanding the connectivity carcass in the future. 


A sensitivity oracle for all-pairs mincuts is a compact data structure for reporting mincut between any given pair of vertices upon insertion or failure of a given set of edges. It can be viewed as a dynamic structure in a very restricted setting.
Connectivity carcass has served as the foundational structure in the design of the single-edge sensitivity oracles for all-pairs mincuts \cite{DBLP:conf/soda/BaswanaP22}. So, it is natural to hope that the connectivity carcass offers a promising direction for the dynamic all-pairs mincuts. This problem has remained elusive till date, although there has been a lot of research on dynamic global mincuts (see \cite{DBLP:conf/soda/El-HayekH025} and reference therein). In our view, the simpler and lighter exposition of connectivity carcass given in this article will help the researchers pursue this direction.

\section*{Acknowledgements}
 This research work is partially supported by Tapas Mishra Memorial Chair at IIT Kanpur. This research work started when Abhyuday Pandey was a student at IIT Kanpur and Surender Baswana was at Paderborn University, on sabbatical from IIT Kanpur. Surender Baswana was supported by a fellowship from the Alexander von Humboldt Foundation during his sabbatical, and he is grateful to Friedhelm Meyer auf der Heide who, as a host for the Humboldt Fellowship at Paderborn University, provided him a highly supportive, encouraging, and free atmosphere conducive for research work.

We are grateful to Koustav Bhanja and Anupam Roy for many helpful discussions and their suggestions on improving a preliminary version of this article. We are also grateful to the anonymous reviewers for their insightful and valuable suggestions on a preliminary version of this article. We are truly indebted to Yefim Dinitz and Alek Vainshtein for their thorough review and valuable comments on a preliminary version of this article. 

\bibliography{refs.bib}

\appendix

\section{Global mincuts is a crossing families}
\label{app: global-mincuts-is-a-crossing-family}
\begin{lemma}
    The set of all global mincuts in $G$ is a crossing family.
    \label{lem:global-mincuts-are-crossing-family}
\end{lemma}
\begin{proof} 
Let $(A,\overline{A})$ and $(A',\overline{A'})$ be any pair of crossing global mincuts. 
It follows from the submodularity of cuts (Lemma \ref{lem: cor-submodularity-of-cuts}) that $A\cap A'$ as well as $\overline{A}\cap \overline{A'}$ define a global mincut. In a similar manner, if we swap $A$ with $\overline{A}$ and apply submodularity of cuts (Lemma \ref{lem: cor-submodularity-of-cuts}), it follows that $A'\setminus A$ and $A\setminus A'$ are also global mincuts. Hence the four corner sets defined by $A$ and $A'$ define global mincut as well. Thus the 1st condition for the crossing family (refer to Definition \ref{def:crossing family}) is satisfied.

Let $s$ be any vertex from $A\cap A'$ and $t$ be any vertex from $\overline{A}\cap \overline{A'}$. Observe that the cuts defined by $A$ and $A'$ are also $(s,t)$-mincuts. Therefore, it follows form Lemma \ref{lem: no edge between AA' and A'A} that there cannot be any edge between $A'\setminus A$ and $A\setminus A'$. Refer to Figure \ref{fig:proof-for-crossing-family}$(i)$. Along similar lines, one can show that there cannot be any edge between $A\cap A'$ and $\overline{A}\cap \overline{A'}$.
 \begin{figure}[ht]
 \centering  \includegraphics[width=450pt]{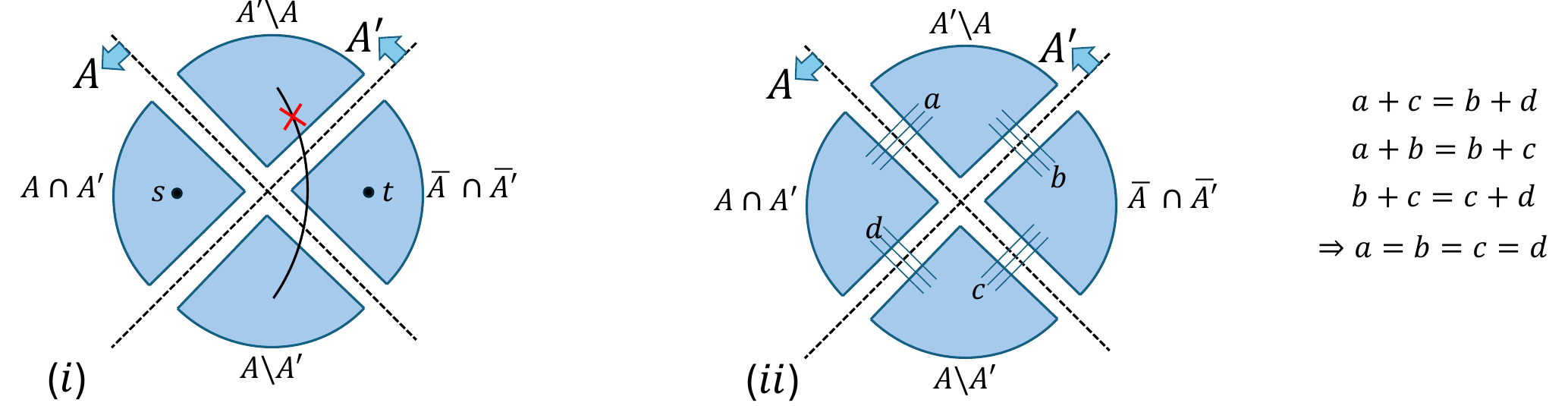} 
  \caption{($i$) No edge between $A'\backslash A$ and $A\setminus A'$, ($ii$) The edges among the 4 corner sets defined by $A$ and $A'$.}
    \label{fig:proof-for-crossing-family}
\end{figure}

So each corner set is adjacent to only two corner sets. Figure \ref{fig:proof-for-crossing-family}$(ii)$ illustrates the edges among the four corner sets defined by $A$ and $A'$. If we equate the capacity of mincuts defined by $A$ and $A'$, along with the mincuts defined by the four corner sets, we can infer that the number of edges incident on each corner set from each of its two neighbors is exactly $\lambda/2$. Now consider the cut defined by $A\setminus A'\cup A'\setminus A$. It follows from the above discussion that its capacity is $2\lambda$, hence it cannot be a global mincut. Thus the 2nd condition for the crossing family (refer to Definition \ref{def:crossing family}) is also satisfied.
\end{proof}
It follows from the proof of Lemma \ref{lem:global-mincuts-are-crossing-family} that if $\lambda$ is odd, there cannot be any pair of crossing global mincuts.


\section{A $t$-cactus for all valid cuts of \texorpdfstring{$S$}{set S}}
\label{app: cactus-construction-for-valid-cuts}
In the following section, we introduce a special family of valid cuts of $S$. This family is inspired by the {\em circular partition} defined by Nagamochi and Kameda \cite{NK94} in the context of global mincuts. This family of valid cuts will form the foundation for the $t$-cactus storing all valid cuts of $S$ that we describe in the subsequent section. 

\subsection{Circular family of valid cuts of \texorpdfstring{$S$}{set S}}

Let ${\cal A}$ be any set $\{ S_1,\ldots, S_\ell\}$ of non-empty subsets of $S$. ${\mathcal A}$ is said to be a partition of $S$ if $S_i\cap S_j = \emptyset$ for each $i\not=j$ and $\bigcup_{1\le i\le \ell} S_i=S$. Each cut of ${\mathcal A}$ also defines a cut of $S$ as follows.
If $S_{i_1},\ldots,S_{i_k}$ are any $1\le k<\ell$ elements from ${\cal A}$ defining its cut, then $\cup_{1\le j\le k}S_{i_j}$ is a proper subset of $S$, and hence defines a cut of $S$ as well. A cut of ${\mathcal A}$ is said to be a valid cut of ${\mathcal A}$ if the corresponding cut of $S$, defined in this manner, is a valid cut of $S$. We define the circular family of valid cuts as follows.


\begin{definition}[Circular family of valid cuts]
Let ${\mathcal A}  = \{S_1,\ldots,S_\ell\}$ be a partition of $S$ where each $S_i$ defines a valid cut of $S$. ${\mathcal A}$ is said to form a circular family of valid cuts of $S$ if there exists a cycle ${O}$ whose nodes are the elements of ${\mathcal A}$ such that each minimum cut of the cycle also defines a valid cut of $S$.
\label{def: circular-family}
\end{definition}

 Each minimum cut of a cycle is defined by a pair of edges of the cycle. So there are precisely $\ell(\ell-1)/2$ minimum cuts of cycle $O$. Interestingly, these and only these are the valid cuts of ${\mathcal A}$ as shown in the following lemma.

\begin{figure}[ht]
 \centering  \includegraphics[width=0.6\textwidth]{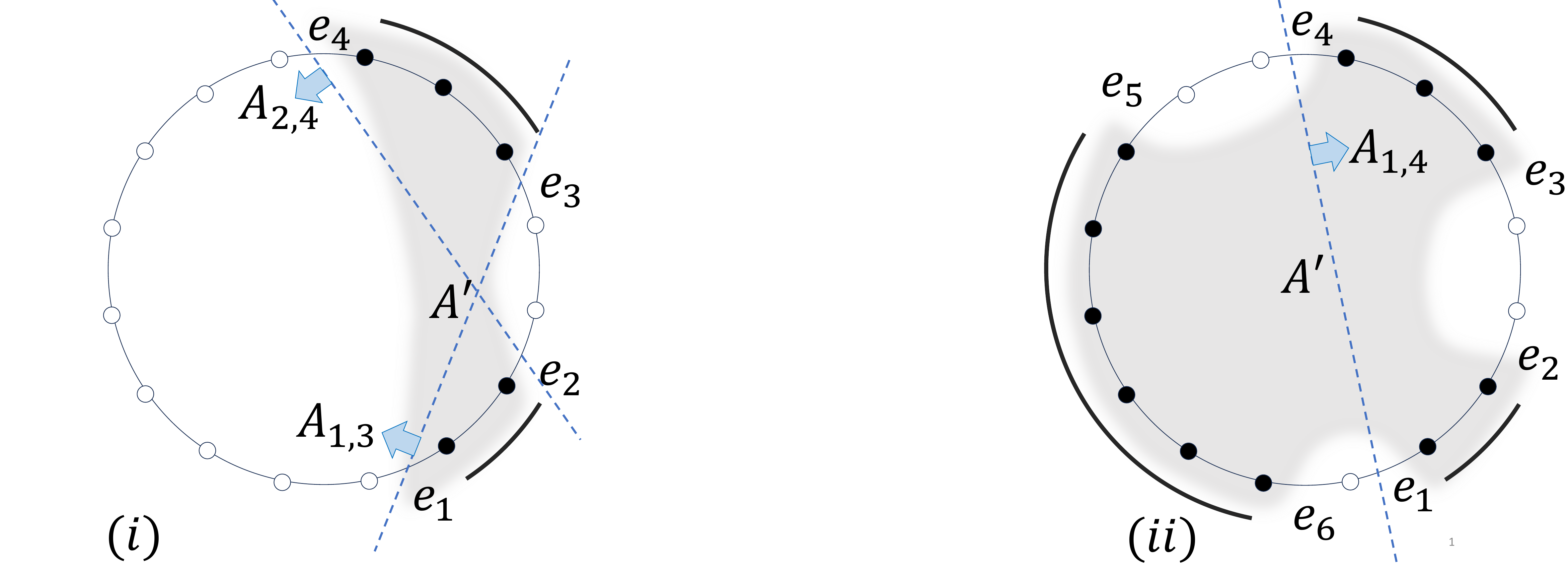} 
  \caption{($i$) $A'$ consists of 2 black segments, ($ii$) $A'$ consists of more than 2 black segments.}
    \label{fig:cyclic-valid-cuts}
\end{figure}

\begin{lemma}
   Let ${\mathcal A} = \{S_1,\ldots,S_\ell\}$ be a circular family of valid cuts of $S$, represented by a cycle $O$. Let $(A',\overline{A'})$ be any cut of ${\mathcal A}$. If  $(A',\overline{A'})$ is not a minimum cut in $O$, the corresponding cut of $S$ cannot be a valid cut of $S$.
   \label{lem:no-non-minimal-cut-of-cycle}
\end{lemma}
\begin{proof}

    Let $(A',\overline{A'})$ be a cut of ${\cal A}$ but not a minimum cut of $O$. 
    %
    Let us color the nodes of $O$ such that the nodes belonging to $A'$ are colored black and the nodes belonging to $\overline{A'}$ are colored white. Observe that $(A',\overline{A'})$ cannot be represented by a pair of edges in $O$ since it is not a minimum cut of $O$. Therefore, there will be multiple maximal segments of black nodes in cycle $O$ that define $A'$. 
    
    Suppose $A'$ is defined by precisely two black segments in cycle $O$. Let the edges defining these segments be $e_1,e_2,e_3,e_4$. Refer to Figure \ref{fig:cyclic-valid-cuts}($i$). 
    Let $(A_{1,3},\overline{A_{1,3}})$ be the cut defined by edges $e_1$ and $e_3$, and let $(A_{2,4},\overline{A_{2,4}})$ be the cut defined by edges $e_2$ and $e_4$. Each of them is a valid cut of $S$ since each of them is defined by a minimum cut of $O$. But the cut defined by $A_{1,3}$ crosses the cut defined by $A_{2,4}$. So it follows from  Definition \ref{def:crossing family} of the crossing family of cuts and Lemma \ref{lem: Valid-cuts-form-crossing-family} that $(A_{1,3} \backslash A_{2,4}) \cup  (A_{2,4} \backslash A_{1,3})$ cannot define a valid cut of $S$. But this union is set $A'$. Hence $(A',\overline{A'})$ is not a valid cut of $S$.
    
    Suppose $A'$ is defined by more than 2 black segments. So there will be more than 4 edges of the cycle that define $A'$. Let us label these edges as $e_1,e_2, ...$ in counter-clockwise order in the cycle $O$. Refer to Figure \ref{fig:cyclic-valid-cuts}($ii$). Let $({A}_{1,4},\overline{{A}_{1,4}})$ be the cut of ${\cal A}$ defined by edges $e_1$ and $e_4$. It is a valid cut of $S$ since it is a minimum cut of $O$. It can be observed that the cut defined by $A'$ crosses the cut defined by $A_{1,4}$.  If $A'$ defines a valid cut of $S$, it would follow from Lemma \ref{lem: Valid-cuts-form-crossing-family} that ${A}'\cap {A}_{1,4}$ also defines a valid cut of $S$. But that is not possible since  ${A}'\cap {A}_{1,4}$ consists of precisely 2 maximal black segments, and as explained above, it cannot be a valid cut of $S$.
\end{proof}

The following lemma can be seen as an immediate corollary of the lemma stated above.
\begin{lemma} Given a circular family of valid cuts of $S$, the cycle that represents it is unique.
\label{lem:unique-cycle-for-cyclic-family}
\end{lemma}

The following lemma states a very surprising property of a circular family of valid cuts.
\begin{lemma}
    Let ${\mathcal A} = \{S_1,\ldots,S_\ell\}$ be a partition of $S$, where the cut defined by 
    $\{S_i\}~ \forall~ 1\le i\le \ell$, is a valid cut of ${\cal A}$. 
    Moreover, these and only these are the laminar cuts in the family of all valid cuts of ${\cal A}$. 
    If there is any pair of crossing valid cuts of ${\cal A}$, ${\mathcal A}$ is a circular family of valid cuts of $S$.
    \label{lem:any-crossing-cut-implies-cyclic-family}
\end{lemma}
The rest of this section is devoted to proving Lemma \ref{lem:any-crossing-cut-implies-cyclic-family}. Let $(B',\overline{B'})$ and $(B'',\overline{B''})$ be any pair of crossing valid cuts of ${\cal A}$. We shall show that ${\cal A}$ is a circular family of valid cuts of $S$ by induction on $\ell$. If $\ell<4$, note that there cannot be any pair of crossing cuts of ${\cal A}$, so ${\cal A}$ is vacuously a crossing family of valid cuts of $S$. Therefore, we begin with $\ell=4$, which is the base case of the assertion.\\
{\em Base case:}~ $\ell=4$.  \\
In this case, $B'$ and $B''$ must have precisely 2 elements of ${\cal A}$. Without loss of generality, assume $B'=\{S_1, S_2\}$. Observe that $B''$ must have exactly one element from $B'$ and one element from $\overline{B'}$. Let us suppose $B''=\{S_2,S_3\}$. Consider the cycle $\langle S_1,S_2,S_3,S_4,S_1\rangle$. It is easy to observe using Definition \ref{def:crossing family} and Lemma \ref{lem: Valid-cuts-form-crossing-family} that
a cut of ${\cal A}$ is a valid cut if and only if it is a minimum cut in this cycle. So ${\cal A}$ is a circular family of valid cuts of $S$.\\
%
%
\noindent 
{\em Induction step:}~
Suppose the assertion holds for all $\ell < k$ for any given $k>4$. We shall now establish the assertion for the set ${\cal A}$ with $k$ elements. The proof, though long, is quite modular. We begin with its overview. First we show that there exists two subsets $C'$ and $C''$, each consisting of exactly 2 elements of ${\cal A}$, that define a pair of crossing valid cuts of ${\cal A}$. We then use these subsets and the set ${\cal A}$ to carefully construct two sets ${\cal A}_{12}$ and ${\cal A}_{23}$ each of cardinality $k-1$. We show that there exists a pair of crossing valid cuts for each of them as well. Thus, using induction, ${\cal A}_{12}$ and ${\cal A}_{23}$ are crossing families of valid cuts of $S$. We then use the cycles representing these crossing families to construct a cycle for ${\cal A}$, and show that each minimum cut in the cycle is a valid cut of ${\cal A}$. This establishes that ${\cal A}$ is a crossing family of valid cuts of $S$.

\begin{figure}[ht]
 \centering  \includegraphics[width=\textwidth]{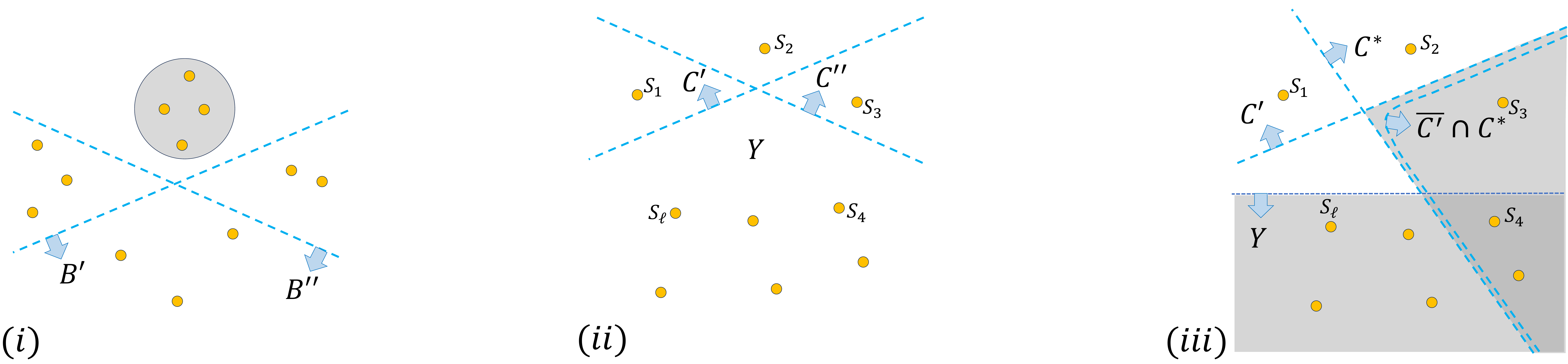} 
  \caption{($i$) $\overline{B'\cup B''}$ appears as a single element in ${\cal A}'$, ($ii$) $C'$ and $C''$ define a pair of crossing valid cuts of $S$, $(iii)$ $\overline{C'}\cap N$ and $Y$ also define a pair of crossing valid cuts of $S$.}
\label{fig:structure-of-crossing-cuts-I}
\end{figure}

If $B'$ and $B''$ have exactly 2 elements of ${\cal A}$, then $C'=B'$ and $C''=B''$. 
Otherwise, we can construct $C'$ and $C''$ as follows. The cuts defined by $B'$ and $B''$ partition ${\cal A}$ into 4 subsets. At least one of these 4 subsets must have at least 2 elements since $k>4$. Without loss of generality, let $\overline{B'\cup B''}$ be that subset. Let ${\cal A}'$ be the set obtained from ${\cal A}$ by merging all elements of $\overline{B'\cup B''}$ into a single element. Refer to Figure \ref{fig:structure-of-crossing-cuts-I}($i$). Observe that ${\cal A}'$ has $k'<k$ elements. Moreover, $B'$ and $B''$ form a pair of crossing valid cuts of ${\cal A}'$ as well. Hence using induction hypothesis, ${\cal A}'$ is a circular family of valid cuts of $S$. Without loss of generality, let the cycle representing this family be $\langle S_1,S_2,...,S_{k'},S_1\rangle$, where $S_{k'}$ represents $\overline{B'\cup B''}$.
We define $C'=\{ S_1,S_2\}$ and $C''=\{S_2,S_3\}$. Since each cut of ${\cal A}'$ is also a cut of ${\cal A}$, it can be observed that $C'$ and $C''$ together define a pair of crossing valid cuts of ${\cal A}$ as well. Figure \ref{fig:structure-of-crossing-cuts-I}($ii$) illustrates these cuts. Let $Y=\overline{C'\cup C''}$. It follows from Lemma \ref{lem: Valid-cuts-form-crossing-family} that that $Y$ defines a valid cut of ${\cal A}$.  

Let ${\cal A}_{12}$ be the set obtained from ${\cal A}$ by merging $S_1$ and $S_2$ into a single element. Likewise, let ${\cal A}_{23}$ be the set obtained from ${\cal A}$ by merging $S_2$ and $S_3$ into a single element. We shall now show that ${\mathcal A}_{12}$ as well as ${\mathcal A}_{23}$ are circular families of valid cuts of $S$. 

%
%

Let $Y=\{S_4,\ldots, S_\ell\}$. Observe that $Y$ has at least 2 elements of ${\cal A}$ since $k>4$, and $\overline{Y}$ has 3 elements of ${\cal A}$, namely $S_1,S_2,S_3$. 
Recall that each laminar cut of ${\cal A}$ has exactly one element on one of its sides. 
Therefore, $(Y,\overline{Y})$ must be a crossing valid cut of ${\cal A}$; so let $(C^*,\overline{C^*})$ be a valid cut of ${\cal A}$ that crosses it. Without loss of generality, assume that $S_3\in C^*$. Observe that either $C'\cap C^*=\emptyset$ or $(C^*,\overline{C^*})$ crosses $(C',\overline{C'})$. Figure \ref{fig:structure-of-crossing-cuts-I}$(iii)$ illustrates the latter case. It follows from Lemma \ref{lem: Valid-cuts-form-crossing-family} that  $\overline{C'}\cap C^*$ defines a valid cut of $S$. This cut crosses $(Y,\overline{Y})$ and is a cut in ${\cal A}_{12}$ as well since it keeps $S_1$ and $S_2$ on the same side. If $C'\cap C^*=\emptyset$, the cut $(C^*,\overline{C^*})$ crosses $(Y,\overline{Y})$ and is also a cut of ${\cal A}_{12}$.
So considering both the cases of $C^*$, there exists a pair of crossing valid cuts in ${\mathcal A}_{12}$. Observe that ${\mathcal A}_{12}$ has $k-1$ elements. So using induction hypothesis, it follows that ${\mathcal A}_{12}$ is a circular family of valid cuts of $S$. In a similar manner, we can show that ${\mathcal A}_{23}$ is a circular family of valid cuts of $S$. Using Lemma \ref{lem:unique-cycle-for-cyclic-family}, let $O_{12}$ and $O_{23}$ be the unique cycles that represent the circular families ${\mathcal A}_{12}$ and ${\mathcal A}_{23}$ respectively. The cycles $O_{12}$ and $O_{23}$ possess interesting structural properties as follows.

\begin{figure}[ht]
 \centering  \includegraphics[width=\textwidth]{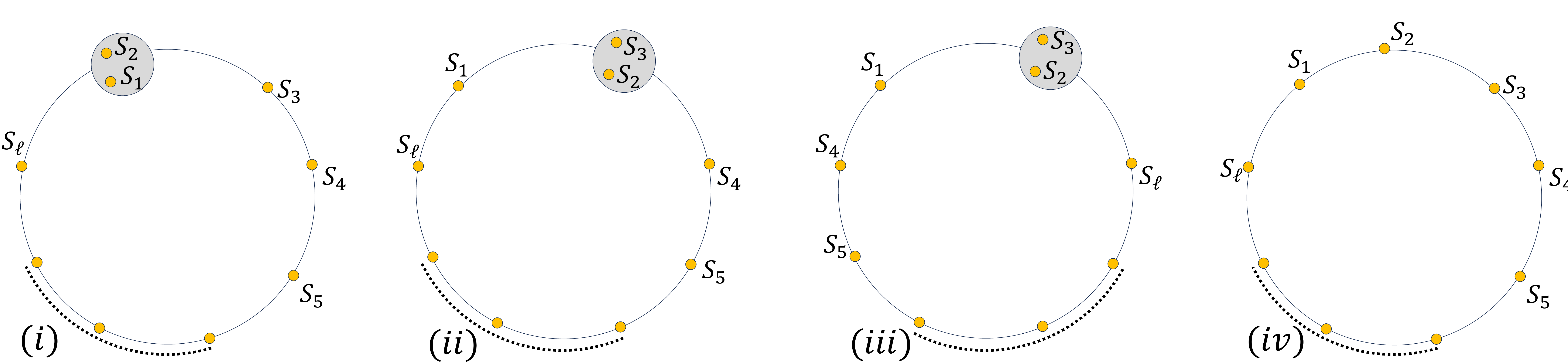} 
  \caption{($i$) The cycle $O_{12}$ representing ${\cal A}_{12}$, $(ii)$ 1st possibility of the cycle $O_{23}$, $(iii)$ 2nd possibility of the cycle $O_{23}$, $(iv)$ The cycle $O$ representing ${\cal A}$.}
  \label{fig:structure-of-crossing-cuts-II}
\end{figure}

Observe that $(Y,\overline{Y})$ is a valid cut in each of ${\mathcal A}_{12}$ and ${\mathcal A}_{23}$, so it must be defined by a pair of edges in $O_{12}$ and $O_{23}$ as implied by Lemma \ref{lem:no-non-minimal-cut-of-cycle}. Therefore, all the elements of $Y$ must appear contiguously in each of these cycles. Without loss of generality, let the cycle ${O}_{12}$ be $\langle S_3,S_4,\ldots,S_\ell,\{S_1,S_2\}, S_3\rangle$ as illustrated in Figure \ref{fig:structure-of-crossing-cuts-II}($i$). 

Consider any two elements of $Y$ that are neighboring in $O_{12}$. Let these elements be $S_i$ and $S_{i+1}$ for any $4\le i<\ell$. Observe that $\{S_i, S_{i+1}\}$ defines a valid cut of $S$ since there is a pair of edges in the cycle $O_{12}$ that defines it. So it follows from Lemma \ref{lem:no-non-minimal-cut-of-cycle} that $S_i$ and $S_{i+1}$ must appear neighboring in the cycle $O_{23}$ as well. So, just like in the cycle $O_{12}$, for any $4<i<\ell$, $S_i$ will have neighbors $S_{i-1}$ and $S_{i+1}$ in the cycle $O_{23}$. 
%
Therefore, there are only the following two possibilities for the cycle $O_{23}$ depending upon how the nodes corresponding to $S_4$ and $S_\ell$ get joined to the nodes corresponding to $S_1$ and $\{S_2,S_3\}$. 
The 1st possibility of $O_{23}$ is $\langle S_1,\{S_2,S_3\},S_4,\ldots,S_\ell,S_1\rangle$ and it is illustrated in Figure \ref{fig:structure-of-crossing-cuts-II}($ii$). The 2nd possibility of $O_{23}$ is $\langle S_1,\{S_2,S_3\},S_\ell,\ldots,S_4,S_1\rangle$ and it is illustrated in  Figure \ref{fig:structure-of-crossing-cuts-II}($iii$). We shall now rule out the 2nd possibility of $O_{23}$. 

The 2nd possibility of $O_{23}$ 
would imply that $\{S_1,S_4\}$ defines a valid cut of $S$. $C'=\{S_1,S_2\}$ already defines a valid cut of $S$ and it crosses the cut defined by $\{S_1,S_4\}$. So it follows from Lemma \ref{lem: Valid-cuts-form-crossing-family} that their union $\{S_1,S_2,S_4\}$ has to be a valid cut of $S$. But it is not possible because of the following reason. $\{S_1,S_2,S_4\}$, though defines a cut of ${\cal A}_{12}$, is represented by 4 edges in the cycle $O_{12}$; hence it cannot define a valid cut of $S$ due to Lemma \ref{lem:no-non-minimal-cut-of-cycle}. Hence $O_{23}$ will appear as shown in Figure \ref{fig:structure-of-crossing-cuts-II}($ii$).

We now construct a cycle whose nodes will be elements of ${\mathcal A}$ as follows. Consider the cycle representing the circular family ${\mathcal A}_{12}$. We replace the node corresponding to $\{S_1,S_2\}$ in this cycle by two nodes representing $S_1$ and $S_2$. We add the following set of edges -- $(S_1,S_2)$, $(S_1,S_\ell)$, and $(S_2,S_3)$. Let $O$ be the resulting cycle. Figure \ref{fig:structure-of-crossing-cuts-II}($iv$) illustrates this cycle. Notice the similarity among $O_{12}$, $O_{23}$, and $O$. In particular, $O_{12}$ (likewise $O_{23}$) can be obtained from $O$ by compressing the nodes corresponding to $S_1$ and $S_2$ (likewise $S_2$ and $S_3$) into a single node.
%
%
We use it to complete the induction step as follows. 
Consider any minimum cut of the cycle $O$. 
If the minimum cut separates $S_1$ and $S_2$, either it has only $S_2$ on one of its sides, or it is present as a minimum cut in the cycle $O_{23}$. In both the cases, it is a valid cut of $S$. If the minimum cut of $O$ does not separate $S_1$ and $S_2$, it is present as a minimum cut in the cycle 
$O_{12}$. So every minimum cut of $O$ is a valid cut of $S$.
Hence, we conclude that ${\cal A}$ is a circular family of valid cuts of $S$.

\begin{remark}
We are grateful to an anonymous reviewer for flagging incompleteness in an earlier proof of Lemma 
\ref{lem:any-crossing-cut-implies-cyclic-family} and providing the necessary arguments to make it complete.
\end{remark}

\subsection{\texorpdfstring{$t$}{t}-cactus storing all valid cuts of \texorpdfstring{$S$}{set S}}
We present the construction of a $t$-cactus ${\mathcal H}$ that compactly represents all valid cuts of $S$. 
It follows from Definition \ref{def: laminar-and-crossing} that each valid cut of $S$ is either a laminar cut or a crossing cut in the family of all valid cuts of $S$. 
While it is possible that there may be no crossing valid cut for a given set $S$, there is always at least one laminar valid cut of $S$ as stated in Lemma \ref{lem: tight-S--mincut-is-laminar}. 
Moreover, the total number of laminar valid cuts of $S$ is always ${\cal O}(|S|)$ as shown in Lemma \ref{lem: laminar-valid-cuts}.

We begin with an overview of the construction of ${\cal H}$. 
We first construct a tree $T_{L}$ whose nodes consist of (possibly empty) disjoint subsets of $S$ satisfying the following properties -- (1)~Each cut defined by an edge in $T_L$ corresponds to a laminar valid cut of ${S}$, and vice versa. (2) ~Each crossing valid cut of $S$ is {\em assigned} to a suitable empty node of $T_{L}$. In order to represent all crossing valid cuts assigned to an empty node, we implant a cycle at the node. Each crossing valid cut assigned to the node is defined by a pair of non-adjacent edges of this cycle. The resulting structure is a $t$-cactus ${\cal H}$ that stores and characterizes all valid cuts of $S$. 
%
%
%
%
%
%
%

Imagine the vertices of $S$ placed on a plane. Consider any laminar valid cut of $S$. Every
other valid cut of $S$ is bound to be parallel to it. The construction of $T_L$ exploits this property, and draws all the laminar valid cuts of $S$ on the plane, one by one. These cuts will be drawn as closed non-intersecting curves instead of infinite lines. This ensures that these cuts do not {\em cross} on the plane as well. 
We now provide below the details of the incremental construction of $T_L$.


%
%
To begin with, $T_L$ is a single node tree storing $S$. We order all the laminar valid cuts of $S$ arbitrarily. 
To include the 1st laminar valid cut of $S$ in $T_L$, we perform the following two steps. 
%

\begin{figure}[ht]
 \centering  \includegraphics[width=0.8\textwidth]{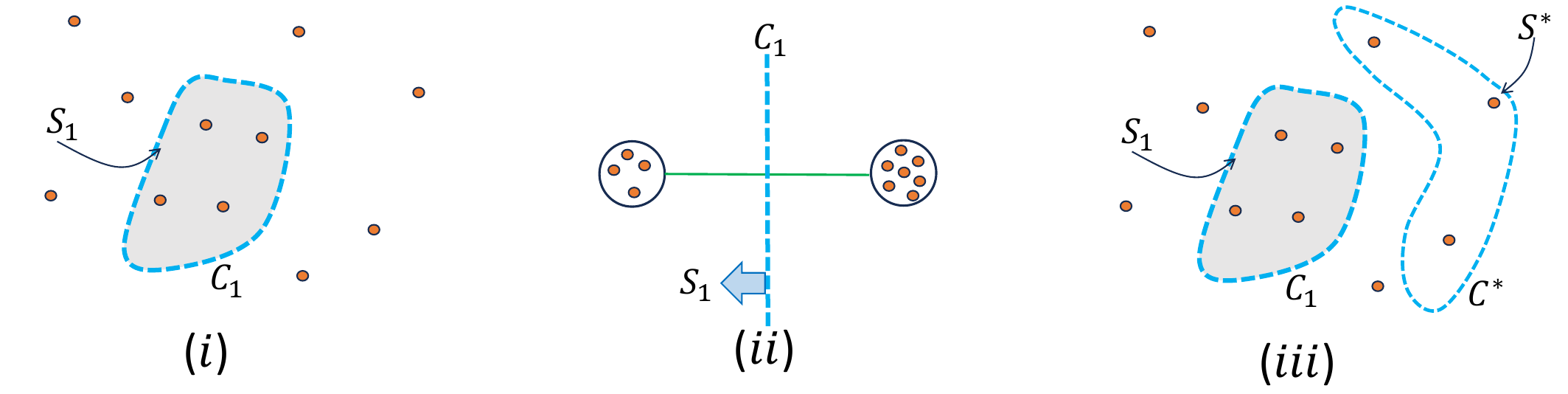} 
  \caption{($i$) Cut $C_1$ as a closed curve on the plane, ($ii$) Cut $C_1$ an edge in $T_L$, ($iii$) Cut $C^*$.}
  \label{fig:First-step-of-Tell}
\end{figure}

\begin{enumerate}
\item 
{\em Drawing the 1st laminar valid cut on plane}:~
Let $C_1=(S_1,\overline{S_1})$ be the 1st laminar valid cut of $S$. We represent $C_1$ as a closed curve on the plane keeping all vertices from $S_1$ inside the curve, and all vertices from $\overline{S_1}$ outside the curve. Refer to Figure \ref{fig:First-step-of-Tell}($i$). 
This curve splits the plane into two regions -- the region inside the curve containing $S_1$, and the region outside the curve containing $\overline{S_1}$. 
\item {\em Updating the tree:}~The single node of $T_L$ is split into two nodes, each corresponding to these two regions. One node stores $S_1$ and the other stores $\overline{S_1}$. We join these nodes by an edge, and this edge represents the cut $C_1$ in the tree. 
Refer to Figure \ref{fig:First-step-of-Tell}($ii$) for the resulting tree $T_L$.
\end{enumerate}

Observe that any other valid cut of $S$, say $C^*=(S^*,\overline{S^*})$, will split exactly one of the two subsets -- $S_1$ or $\overline{S_1}$ because $(S_1,\overline{S_1})$ is laminar. Hence, we can {\em confine} the cut $C^*$ to either the region inside or the region outside the closed curve representing $C_1$ as shown in Figure \ref{fig:First-step-of-Tell}($iii$). Accordingly in the tree $T_L$, we {\em assign} $C^*$ to the node  containing $S_1$ if $S^*\subsetneq S_1$, otherwise to the node containing $\overline{S_1}$. 

\begin{figure}[ht]
 \centering  \includegraphics[width=0.8\textwidth]{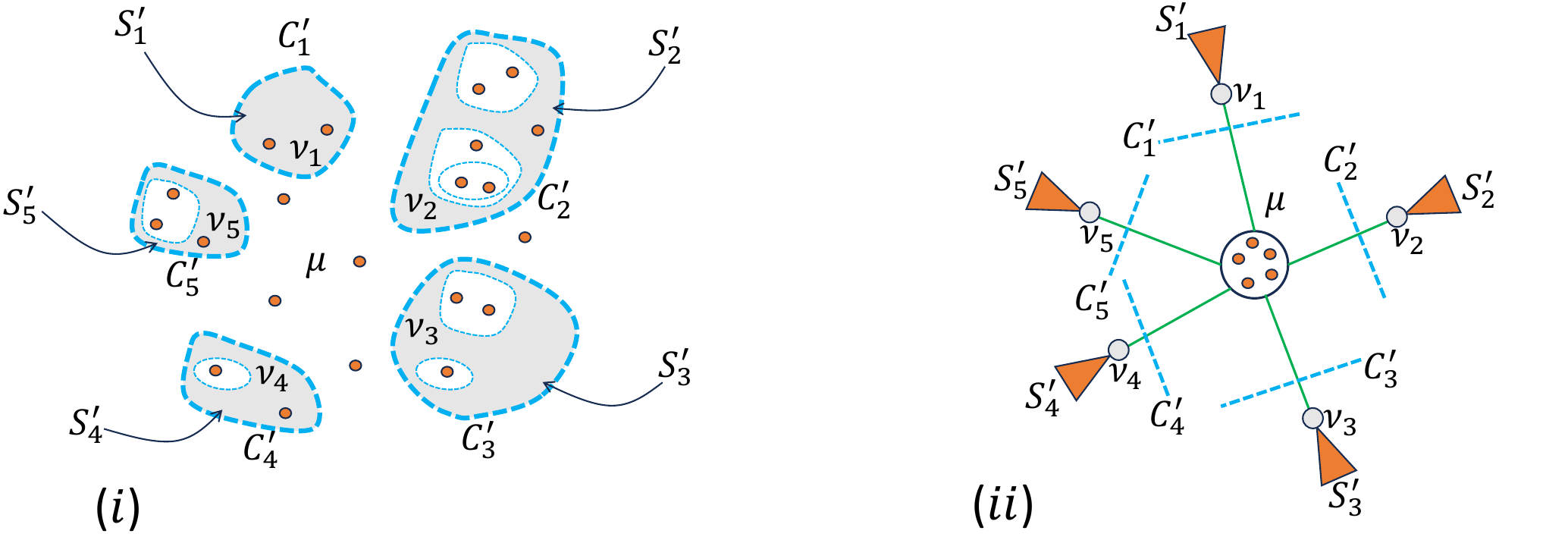} 
  \caption{($i$) The region $\mu$ defined by  laminar valid cuts, ($ii$) the tree representing these cuts.}
    \label{fig:intermediate stage of Tell}
\end{figure}

In order to describe the processing of the $(k+1)$th laminar valid cut of $S$ for any $k\ge 1$, we first describe how the first $k$ laminar valid cuts appear on the plane and in the tree. There will be $k$ closed and non-intersecting curves on the plane, each representing one of these laminar valid cuts. As a result, the plane is split into $k+1$ regions. We now provide the details of these regions and laminar  valid cuts in the plane, and how they are mapped to the nodes and edges in the tree $T_L$.\\
\noindent
{\em Regions and laminar valid cuts in the plane:}~ 
Each region on the plane is defined uniquely by a set of laminar valid cuts, and in turn, each laminar valid cut is shared by exactly two regions. Figure \ref{fig:intermediate stage of Tell}($i$) demonstrates one such region, labeled $\mu$, defined by the laminar cuts $C_1',\ldots,C_5'$, where $C_i'=(S_i',\overline{S_i'})$ for each $1\le i\le 5$.
$\nu_1,\ldots,\nu_5$ are the regions neighboring to $\mu$. These regions are shown shaded in 
Figure \ref{fig:intermediate stage of Tell}($i$).
Cut $C_i'$ is shared by region $\mu$ and region $\nu_i$.

\noindent
{\em Nodes and edges in $T_L$:}~ 
We represent each region by a unique node in the tree $T_L$. We represent each of the $k$ laminar valid cuts by an edge between two nodes; these two nodes correspond to the two regions that share the laminar valid cut. For example, the cut $C_i'$ is represented by the edge $(\mu,\nu_i)$ in $T_L$ as shown in Figure \ref{fig:intermediate stage of Tell}($ii$).
\begin{remark}
Figure \ref{fig:intermediate stage of Tell} depicts two representation of the laminar valid cuts -- one in the plane and another in the tree $T_L$. 
\end{remark}
%

\begin{figure}[ht]
 \centering  \includegraphics[width=\textwidth]{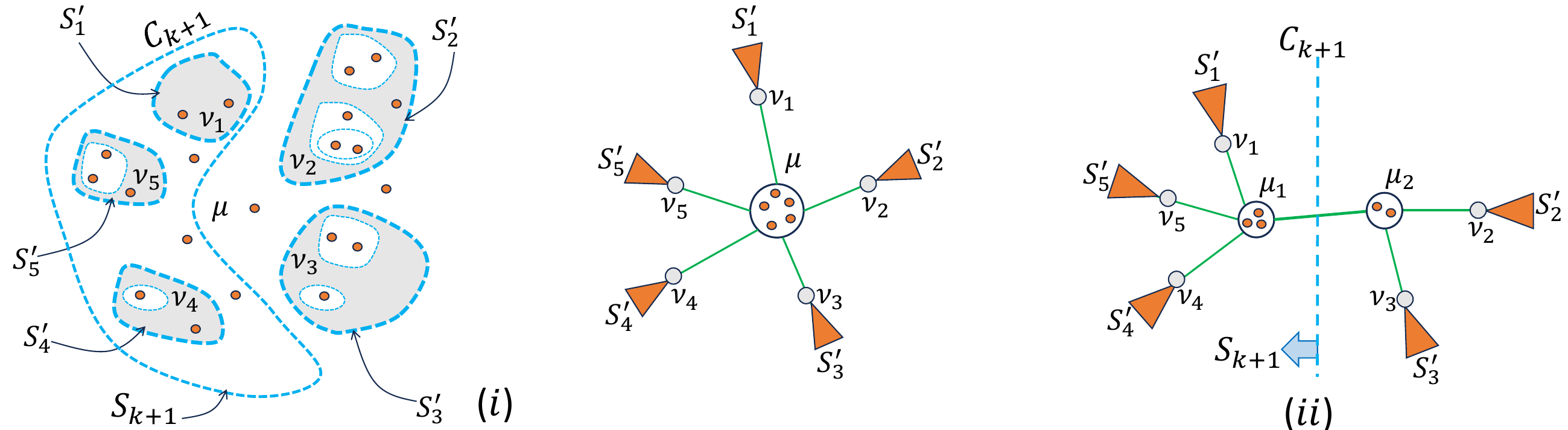} 
  \caption{($i$) The region $\mu$ defined by $5$ laminar valid cuts and the tree $T_L$,
  ($ii$) $T_L$ after introducing the cut $C_{k+1}$.}
    \label{fig:tree-for-parallel-cuts}
\end{figure}

%
%
%
Any other valid cut of $S$ is parallel to each of the $k$ laminar valid cuts. So it is confined to exactly one region, and hence  assigned to the node representing that region in $T_L$. 
To include the $(k+1)$th laminar valid cut of $S$ in $T_L$, we perform the following two steps. 
\begin{enumerate}
\item {\em Drawing the $(k+1)$th laminar valid cut on plane}:~
Let $C_{k+1}=(S_{k+1},\overline{S_{k+1}})$ be the $(k+1)$th laminar valid cut of $S$. Suppose it is assigned to the node $\mu$ of the tree $T_L$ built till now. Figure  \ref{fig:tree-for-parallel-cuts}($i$) depicts the region associated with it.  
We draw $C_{k+1}$ as a closed curve, in the region associated with $\mu$, keeping all the laminar valid cuts defining the region intact. 
As a result the region associated with $\mu$ gets split into two regions. 
\item {\em Updating the tree:}~ We split $\mu$ into 2 nodes $\mu_1$ and $\mu_2$ representing these regions. We join $\mu_1$ and $\mu_2$ by an edge that represents the cut $C_{k+1}$ and distribute the vertices assigned to $\mu$ accordingly. Let $\nu_i$ be any neighbor of $\mu$. $\nu_i$ will be joined to $\mu_1$ if $S_i'\subsetneq S_{k+1}$, otherwise $\nu_i$ will be joined to $\mu_2$. Refer to Figure \ref{fig:tree-for-parallel-cuts}($ii$). 
Each of the remaining valid cuts of $S$ assigned to $\mu$ is parallel to $C_{k+1}$ since $C_{k+1}$ is a laminar cut. So, each such cut is confined within exactly one of the two newly created regions, and so we assign it to $\mu_1$ and $\mu_2$ accordingly. 
\end{enumerate}

Let $T_L$ be the tree created after processing all laminar valid cuts of $S$, one by one, as described above. We now state a few facts about $T_L$ that follow from its construction described above. 
Each laminar valid cut of $S$ is represented by a unique edge in $T_L$, and vice versa. Note that any new region created while processing a laminar valid cut may be empty (with no vertices from $S$ belonging to it). However, it is easy to observe that each such empty region must have at least 3 laminar valid cuts defining it. Hence, the number of empty nodes in $T_L$ is less than the number of leaf nodes. This fact can be used to derive ${\cal O}(|S|)$ bound on the number of nodes in $T_L$, and hence an alternate proof for Lemma \ref{lem: laminar-valid-cuts}.
%
%

Let $u,v\in S$ be separated by one or more valid cuts of $S$. The tight valid cut from $u$ to $v$ has to be a laminar valid cut (Lemma \ref{lem: tight-S--mincut-is-laminar}). This cut has to be present in $T_L$ since $T_L$ stores all laminar valid cuts. 
The following lemma describes this cut in $T_L$; its proof is immediate from the structure of $T_L$.
\begin{lemma}
Let $u,v\in S$ be any two vertices. Let $u$ and $v$ belong to nodes $\mu$ and $\nu$ respectively in $T_L$, and let $e$ be the first edge on the path from $\mu$ to $\nu$ in $T_L$. The set of vertices mapped to the subtree containing $\mu$, after the removal of $e$ from $T_L$, defines the tight valid cut of $S$ from $u$ to $v$.
\label{lem: parallel-cut-tight-from-x}
\end{lemma}
The remaining valid cuts of $S$ are crossing cuts, and the construction described above has assigned them to the respective nodes of $T_L$. Let us consider any node $\mu$ in $T_L$ with $j$ neighbors labeled $\nu_1,\ldots,\nu_j$. It follows from the construction of $T_L$ that the laminar cuts defining $\mu$ induce a partition of $S$ into
$j+1$ sets $S_\mu, S_1,\ldots,S_j$ as follows. 
$S_\mu$ is the set of vertices from $S$ belonging to the region associated with $\mu$, and $S_\ell$ for each $1\le \ell \le j$ is the set of vertices from $S$ lying on the side of $\nu_\ell$ in the cut defined by edge $(\mu,\nu_\ell)$. The following lemma states crucial properties of $\mu$ and $S_\mu$ when there are crossing valid cuts assigned to $\mu$. 

\begin{figure}[ht]
 \centering  \includegraphics[width=\textwidth]{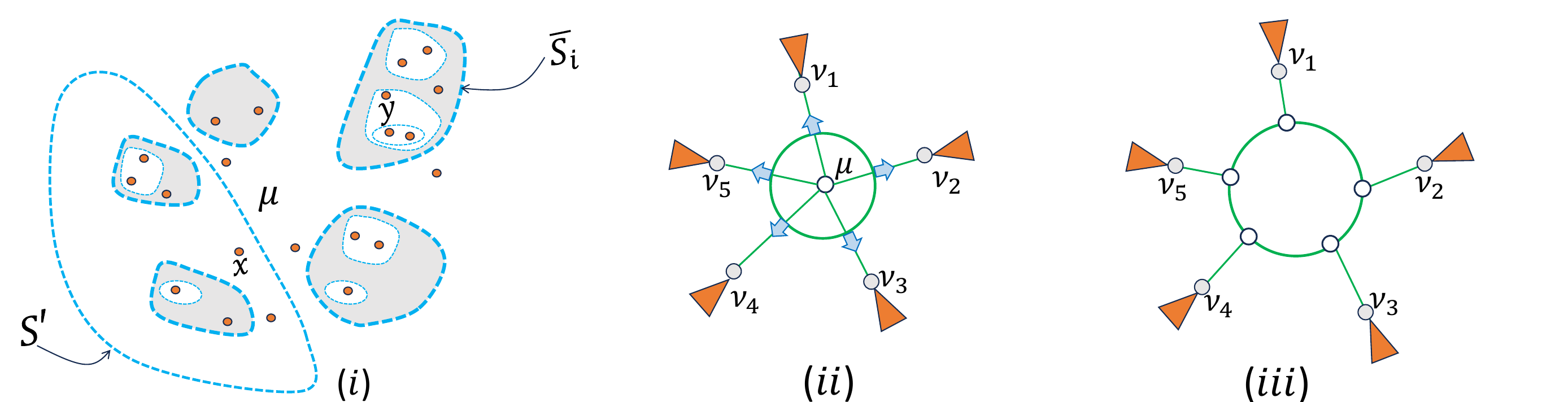} 
  \caption{($i$) Cut $(S',\overline{S'})$ in region $\mu$ is a crossing valid cut of $S$, $(ii)$ node $\mu$ having crossing valid cuts ($iii$) the cycle represents all crossing valid cuts associated with $\mu$.}
    \label{fig:cycle-for-crossing-cuts}
\end{figure}

\begin{lemma}
If there is any crossing valid cut of $S$ 
present in the region associated with $\mu$, $S_\mu$ must be empty and the degree of $\mu$ must be at least 4.
\label{lem: crossing-cuts-imply-empty-region}
\end{lemma}
\begin{proof}
Let $(S',\overline{S'})$ be any crossing valid cut of $S$ present in the region associated with $\mu$. 
Observe that the subset $S_\ell$, for each $1\le \ell \le j$, must remain intact in the cut $(S',\overline{S'})$ because $(S_\ell,\overline{S_\ell})$ is a laminar valid cut of $S$. So we may draw the cut $(S',\overline{S'})$ as a closed non-intersecting curve in the region associated with $\mu$. Refer to Figure \ref{fig:cycle-for-crossing-cuts}($i$). We shall now prove by contradiction that $S_\mu$ must be empty. 

Suppose $S_\mu$ is nonempty, and let $x$ be a vertex in $S_\mu$. Without loss of generality, assume that $x\in S'$. There are two cases.
(1) $S_i \subseteq \overline{S'}$ for some $1\le i \le j$. In this case, surely $S_i\not= \overline{S'}$ because $(S',\overline{S'})$ is a crossing valid cut of $S$. Hence $S_i\subsetneq \overline{S'}$.
(2) $S_i\subseteq S'$ for each $1\le i \le j$. So there must exist a vertex from $S_\mu$ that belongs to $\overline{S'}$ since $(S',\overline{S'})$ is a cut of $S$. In this case, we reassign $x$ to be this vertex and swap $S'$ and $\overline{S'}$. 
As a result, $S_i\subsetneq \overline{S'}$, for each $1\le i\le j$.  
Considering both the cases, we can thus conclude the following. If $S_\mu$ is nonempty, there exist a vertex $x\in S'$ and an integer $i$ satisfying $1\le i\le j$ such that $S_i\subsetneq \overline{S'}$. 
Let $y$ be any vertex belonging to $S_i$.  
Observe that $(S',\overline{S'})$ is a valid cut of $S$ with $x\in S'$ and $y\in \overline{S'}$. Refer again to Figure \ref{fig:cycle-for-crossing-cuts}($i$) for a generic description of $x,y,S_i,$ and $S'$. There is an edge between node $\mu$ and node $\nu_\ell$ in $T_L$ for each $1\le \ell \le j$. Observe that $(\mu,\nu_i)$ is the first edge on the path from $\mu$ to the node containing $y$ in $T_L$. So it follows from Lemma \ref{lem: parallel-cut-tight-from-x} that $(\overline{S_{i}},S_{i})$ is the tight valid cut from $x$ to $y$. Therefore, given that $(S',\overline{S'})$ is a valid cut of $S$ with $x\in S'$ and $y\in \overline{S'}$, we must have $\overline{S_i}\subseteq S'$, and hence $\overline{S'}\subseteq S_i$. But that would contradict $S_i\subsetneq \overline{S'}$. 
%
%

Given that $S_\mu$ is empty, the cut $(S',\overline{S'})$ must have at least 2 elements from $\{S_\ell|1\le \ell \le j\}$ on each of its sides. This is because $(S',\overline{S'})$ is a crossing valid cut of $S$. 
Thus $j\ge 4$. 
Hence the degree of $\mu$ must be at least 4.
%
%
\end{proof}

It follows from Lemma \ref{lem: crossing-cuts-imply-empty-region} that if there is any  crossing valid cut, say $(S',\overline{S'})$, present in the region associated with $\mu$, then ${\cal A}=\{S_1,\ldots,S_j\}$ defines a partition of $S$. 
The valid cut of $S$ that crosses $(S',\overline{S'})$ must keep $S_i$ intact for each $1\le i \le k$; and hence it must also be present in the region associated with $\mu$. 
This pair of crossing valid cuts stored in $\mu$ and the partition ${\cal A}=\{S_1,\ldots,S_j\}$  fulfill the conditions of Lemma \ref{lem:any-crossing-cut-implies-cyclic-family}. Therefore, $\{S_1,\ldots,S_j\}$ is a circular family of valid cuts of $S$. 
Let $O$ be the corresponding cycle that represents it. In order to represent all crossing cuts associated with $\mu$, 
we can implant cycle $O$ at node $\mu$ as follows. We first introduce an empty node at the center of each edge incident on $\mu$. Then, we join these nodes in the order specified by $O$. Finally, we remove $\mu$ and all edges incident on it. Refer to Figure \ref{fig:cycle-for-crossing-cuts} ($ii$) and $(iii)$ for the transformation of $\mu$ into a cycle. In the resulting structure, each crossing valid cut associated with $\mu$ appears as a cut defined by a pair of non-adjacent edges of the cycle. In this manner, we process each empty node of $T_L$ if there are crossing valid cuts present in the region associated with it. It follows from the construction that the resulting structure is a $t$-cactus storing all valid cuts of $S$. It is referred to as {\em skeleton} and represented by ${\cal H}$.

\subsection{An example illustrating the construction of skeleton ${\cal H}$} 
\begin{figure}[H]
 \centering  \includegraphics[width=\textwidth]{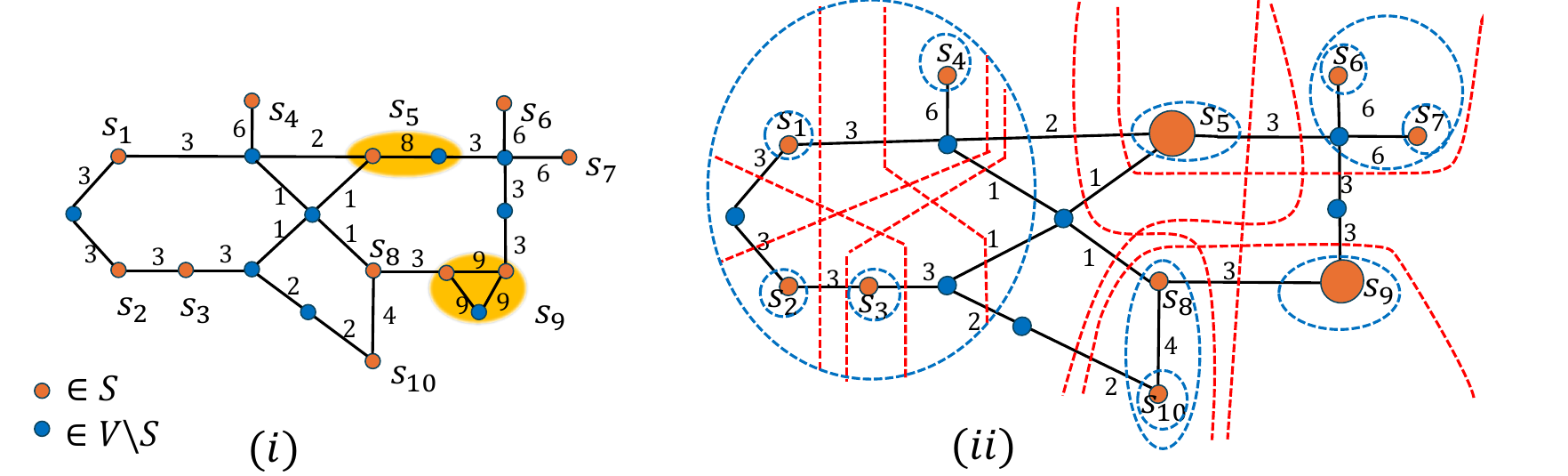} 
 \caption{The laminar valid cuts are shown using dotted curves in blue color and the crossing valid cuts are shown using dotted curves in red color.}
 \label{fig:graph-with-all-valid-cuts}
\end{figure}

We now illustrate the construction of skeleton ${\cal H}$ for a given graph based on the algorithm described in the previous section. Figure \ref{fig:graph-with-all-valid-cuts}($i$) shows the multigraph $G$; the weight on an edge represents the number of edges joining its endpoints. Observe that $\lambda$, the capacity of $S$-mincut, is 6. Figure \ref{fig:graph-with-all-valid-cuts}($ii$) shows the flesh graph ${\cal F}$ and all the valid cuts of $S$. 

\begin{figure}[ht]
 \centering  \includegraphics[width=0.9\textwidth]{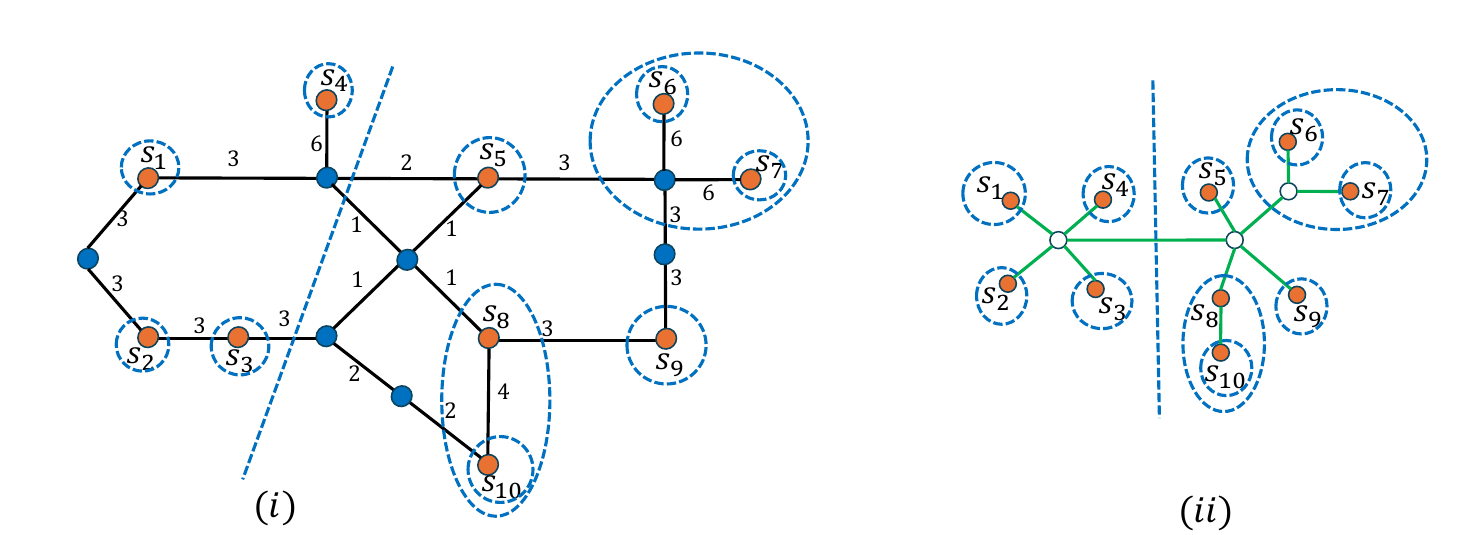} 
  \caption{($i$) Flesh ${\cal F}$ of $G$ and all laminar valid cuts ($ii$)  all laminar valid cuts in tree $T_L$.}
    \label{fig:graph-with-all-laminar-cuts}
\end{figure}

Figure \ref{fig:graph-with-all-laminar-cuts} shows the construction of the tree $T_L$ that stores all the laminar valid cuts of $S$. These cuts are shown in blue color.

\begin{figure}[H]
 \centering  \includegraphics[width=\textwidth]{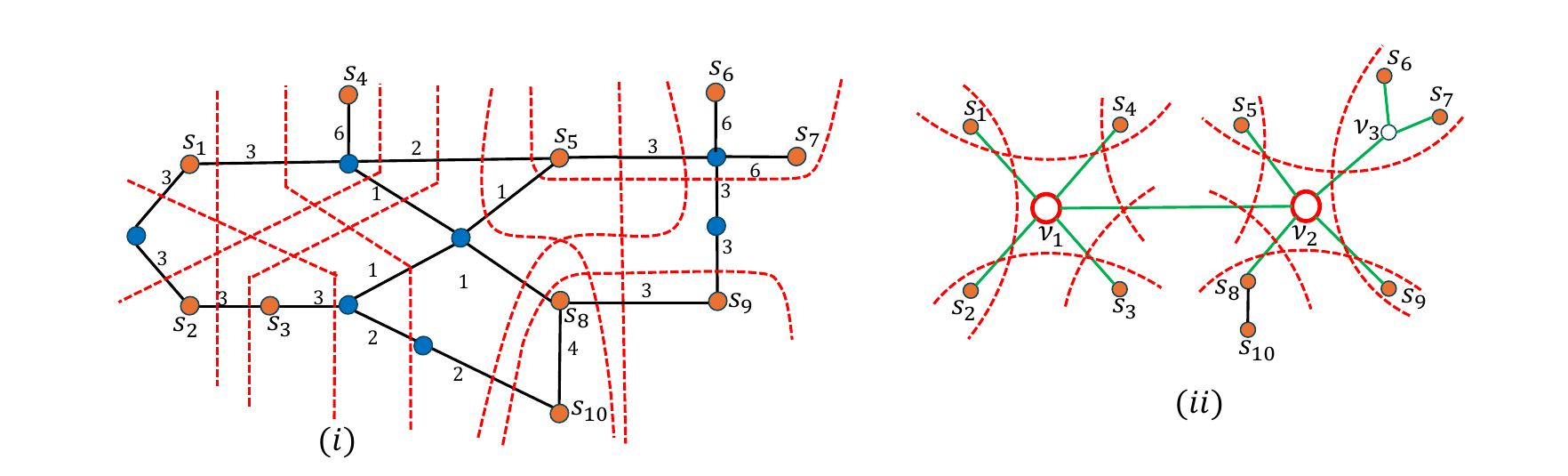} 
  \caption{($i$) Crossing  valid cuts in ${\cal F}$ ($ii$) Crossing valid cuts appearing in the regions of nodes ${\nu}_1$ and ${\nu}_2$ in $T_L$.}
    \label{fig:graph-with-crossing-valid-cuts}
\end{figure}

Lemma \ref{lem: crossing-cuts-imply-empty-region} implies that if crossing valid cuts are present in $G$, there must be empty nodes in $T_L$ such that the crossing valid cuts appear in the region corresponding to these empty nodes; nodes ${\nu}_1$ and ${\nu}_2$ in Figure \ref{fig:graph-with-crossing-valid-cuts} are such empty nodes. Although node ${\nu}_3$ is an empty node, it does not contain any crossing valid cut.

\begin{figure}[H]
 \centering  \includegraphics[width=\textwidth]{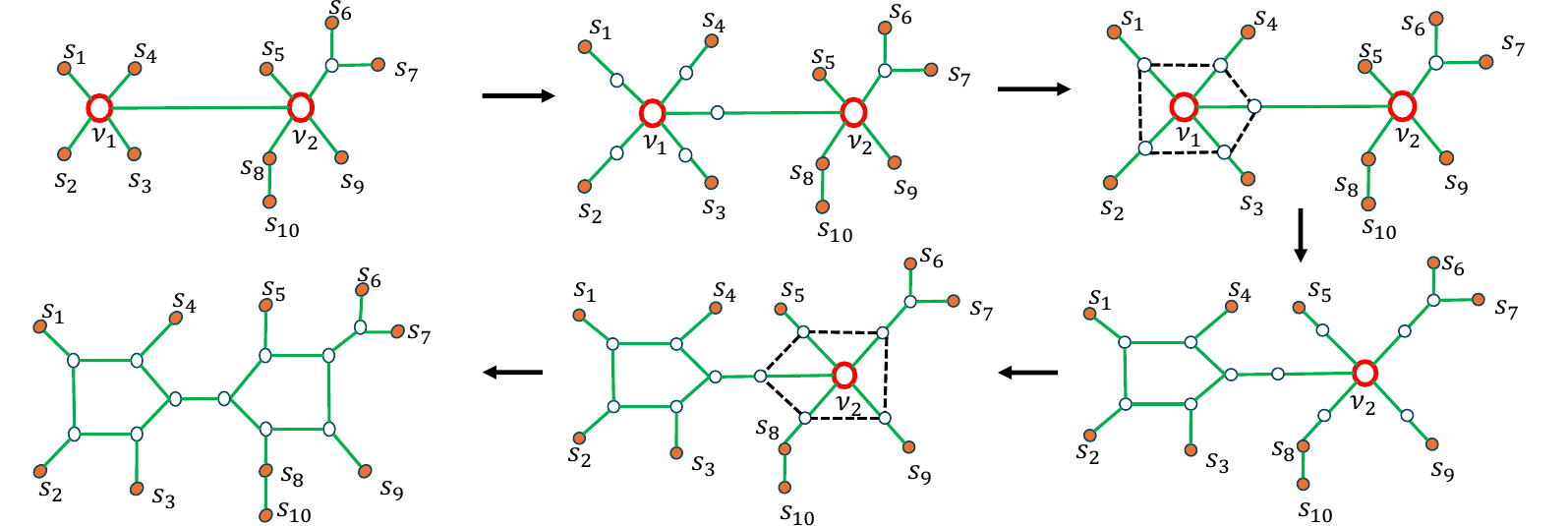} 
  \caption{Transformation of $T_L$ into $t$-cactus ${\cal H}$}
    \label{fig:transformation-into-cactus}
\end{figure}

The transformation of tree $T_L$ into ${\cal H}$ is demonstrated in Figure \ref{fig:transformation-into-cactus}. For each of the empty nodes ${\nu}_1$ and ${\nu}_2$ that have crossing cuts, we first embed an empty node in the center of each edge incident on them. Then, we join them in the order defined by the circular family associated with the node. Finally, we remove the empty node and all edges incident on it.

\end{document}